\documentclass{siamart1116}

\usepackage{lipsum}
\usepackage{amsfonts}
\usepackage{graphicx}
\usepackage{epstopdf}
\usepackage{algorithmic}
\ifpdf
  \DeclareGraphicsExtensions{.eps,.pdf,.png,.jpg}
\else
  \DeclareGraphicsExtensions{.eps,.pdf}
\fi

\usepackage{geometry}
\usepackage{booktabs}
\usepackage{mathrsfs}
\usepackage{amssymb}
\usepackage{esint}

\providecommand{\tabularnewline}{\\}

\newsiamremark{remark}{Remark}

\numberwithin{theorem}{section}
\numberwithin{equation}{section}
\numberwithin{figure}{section}

\newcommand{\TheTitle}{Strategies for Reduced-Order Models for Predicting the Statistical
Responses and Uncertainty Quantification in Complex Turbulent Dynamical
Systems}
\newcommand{\ShortTitle}{Strategies for Reduced-Order Models  in Complex Turbulent Dynamical
Systems}
\newcommand{\TheAuthors}{Andrew J. Majda, and Di Qi}

\headers{\ShortTitle}{\TheAuthors}

\title{{\TheTitle}}

\author{
  Andrew J. Majda\thanks{Department of Mathematics and Center for Atmosphere
and Ocean Science, Courant Institute of Mathematical Sciences, New
York University, New York, NY 10012
    (\email{qidi@cims.nyu.edu}, \email{jonjon@cims.nyu.edu}).}
  \and
 Di Qi\footnotemark[2]
}

\usepackage{amsopn}

\ifpdf
\hypersetup{
  pdftitle={Strategies for Reduced-Order Models},
  pdfauthor={Di Qi and Andrew Majda}
}
\fi

\usepackage{tikz}
\usetikzlibrary{shapes,arrows,snakes,backgrounds}
\usetikzlibrary{fit}
\usetikzlibrary{backgrounds}

\usepackage{color}

\usepackage{mathptmx}

\usepackage{array}
\usepackage[caption=false]{subfig}
\makeatother

\begin{document}

\maketitle

\begin{abstract}
Turbulent dynamical systems characterized by both a high-dimensional
phase space and a large number of instabilities are ubiquitous among
many complex systems in science and engineering including climate,
material, and neural science. The existence of a strange attractor
in the turbulent systems containing a large number of positive Lyapunov
exponents results in a rapid growth of small uncertainties from imperfect
modeling equations or perturbations in initial values, requiring naturally
a probabilistic characterization for the evolution of the turbulent
system. Uncertainty quantification (UQ) in turbulent dynamical systems
is a grand challenge where the goal is to obtain statistical estimates
such as the change in mean and variance for key physical quantities
in their nonlinear responses to changes in external forcing parameters
or uncertain initial data. In the development of a proper UQ scheme
for systems of high or infinite dimensionality with instabilities,
significant model errors compared with the true natural signal are
always unavoidable due to both the imperfect understanding of the
underlying physical processes and the limited computational resources
available through direct Monte-Carlo integration. One central issue
in contemporary research is the development of a systematic methodology
that can recover the crucial features of the natural system in statistical
equilibrium (\emph{model fidelity}) and improve the imperfect model
prediction skill in response to various external perturbations (\emph{model
sensitivity}).

A general mathematical framework to construct statistically
accurate reduced-order models that have skill in capturing the statistical
variability in the principal directions with largest energy of a general
class of damped and forced complex turbulent dynamical systems is discussed here. There
are generally three stages in the modeling strategy, imperfect model selection;
calibration of the imperfect model in a training phase; and prediction
of the responses with UQ to a wide class of forcing and perturbation
scenarios. The methods are developed under a universal class of turbulent
dynamical systems with quadratic nonlinearity that is representative
in many applications in applied mathematics and engineering. Several
mathematical ideas will be introduced to improve the prediction skill
of the imperfect reduced-order models. Most importantly, \emph{empirical
information theory} and \emph{statistical linear response theory}
are applied in the training phase for calibrating model errors to
achieve optimal imperfect model parameters; and \emph{total statistical
energy dynamics} are introduced to improve the model sensitivity in
the prediction phase especially when strong external perturbations
are exerted. The validity of general framework of reduced-order models
is demonstrated on instructive stochastic triad models. Recent applications
to two-layer baroclinic turbulence in the atmosphere and ocean with
combinations of turbulent jets and vortices are also surveyed. The
uncertainty quantification and statistical response for these complex
models are accurately captured by the reduced-order models with only
$2\times10^{2}$ modes in a highly turbulent system with $1\times10^{5}$
degrees of freedom. Less than $0.15\%$ of the total spectral modes
are needed in the reduced-order models.
\end{abstract}



Turbulent dynamical systems characterized by both a high-dimensional
phase space and a large number of instabilities are ubiquitous among
many complex systems in science and engineering \cite{majda2006nonlinear,majda2016introduction,vallis2006atmospheric,nicholson1983introduction}.
The existence of a strange attractor \cite{young2002srb} in turbulent
systems containing a large number of positive Lyapunov exponents results in a rapid growth of small uncertainties from imperfect modeling equations or perturbations in initial values, requiring naturally a probabilistic characterization for the evolution of the turbulent system. Uncertainty quantification (UQ) in turbulent dynamical systems is a grand challenge where the goal is to obtain statistical estimates such as the change in mean and variance for key physical quantities in their nonlinear responses to changes in external forcing parameters or uncertain initial data. One problem of practical significance in contemporary science is using UQ to understand the complexity of anisotropic turbulent processes over a wide range of spatio-temporal scales in engineering shear turbulence \cite{hinze1959turbulence,townsend1980structure,frisch1995turbulence} as well as climate atmosphere ocean science \cite{salmon1998lectures,vallis2006atmospheric,majda2006nonlinear}. This is especially important from a practical viewpoint because energy often flows intermittently from the smaller scales to affect the largest scales in such anisotropic turbulent flows.

In the development of a proper UQ scheme for systems of high
or infinite dimensionality with instabilities, the analysis and prediction of phenomena often occur through complex dynamical equations that have significant model errors compared with the true natural signal. The imperfect model errors are always unavoidable due to both the imperfect understanding of the underlying physical processes and the limited computational resources needed for repeated Monte-Carlo simulations in each different scenario of high dimensional systems. Clearly, it is important both to improve the imperfect model's capabilities to recover crucial features of the natural system and also to accurately model the sensitivities in the natural system to changes in external or internal parameters. These efforts are hampered by the fact that the actual dynamics of the natural system are unknown. Important examples with major societal impact involve the Earth's climate and climate change where climate sensitivities are studied through a suite of imperfect comprehensive computer models (\cite{emanuel2005improving,randall2007climate,majda2010quantifying}, and references therein); other examples include imperfect mesoscopic models in materials science \cite{chatterjee2007overview,katsoulakis2003coarse} and neural science \cite{rangan2009multiscale} when compared with actual observed behavior in these complex nonlinear systems.

Recently, information theory has been utilized in different ways to systematically improve model fidelity and sensitivity \cite{majda2011improving,majda2010quantifying}, to quantify the role of coarse-grained initial states in long-range forecasting \cite{giannakis2012quantifying1,giannakis2012quantifying2}, and to make an empirical link between model fidelity and forecasting skill \cite{delsole2003predictability,delsole2010model}. Imperfect models for complex systems are constrained by their capability to reproduce certain statistics in a training phase where the natural system has been observed; for example, this training phase in climate science is roughly the 60-year dataset of extensive observations of the Earth's climate system. For long-range forecasting, it is natural to guarantee statistical equilibrium fidelity for an imperfect model, and a framework using information theory is a natural way to achieve this in an unbiased fashion \cite{majda2011improving,majda2010quantifying,giannakis2012quantifying1,giannakis2012quantifying2,delsole2010model}. First, equilibrium statistical fidelity for an imperfect model depends on the choice of coarse-grained variables utilized \cite{majda2011improving,majda2010quantifying}; second, equilibrium model fidelity is a necessary but not sufficient condition to guarantee long-range forecasting skill \cite{majda2011link,giannakis2012quantifying2}. For example, Section 2.6 of \cite{majda2005information} extensively discusses three very different strongly mixing chaotic dynamical models with 40 variables and with the same Gaussian equilibrium measure, the TBH, K-Z, and IL-96 models.

One significant application of UQ through empirical information theory is quantifying uncertainty in climate change science \cite{majda2010quantifying,majda2016introduction}. The climate is an extremely complex coupled system involving multiple physical processes for the atmosphere, ocean, and land over a wide range of spatial scales from millimeters to thousands of kilometers and time scales from minutes to decades or centuries \cite{emanuel2005improving,neelin2006tropical}. Climate change science focuses on predicting the coarse-grained planetary scale long time changes in the climate system due to either changes in external forcing or internal variability such as the impact of increased carbon dioxide \cite{gritsun2007climate,giannakis2012quantifying2}. For several decades the predictions of climate change science have been carried out with some skill through comprehensive computational atmospheric and oceanic simulation (AOS) models \cite{emanuel2005improving,neelin2006tropical,randall2007climate}, which are designed to mimic the complex physical spatio-temporal patterns in nature. Such AOS models either through lack of resolution due to current computing power or through inadequate observation of nature necessarily parameterize the impact of many features of the climate system such as clouds, mesoscale and submesoscale ocean eddies, sea ice cover, etc. Thus, there are intrinsic model errors in the AOS models for the climate system and the effect of such model errors on predicting the coarse-grained large scale long time quantities is of interest. One central scientific issue in contemporary climate change science is the development of a systematic methodology that can recover the crucial features of the natural system in statistical equilibrium/climate (\emph{model fidelity}) and improve the imperfect model prediction skill in response to various external perturbations like climate change and mitigation scenarios (\emph{model sensitivity}) \cite{abramov2012low,bell1980climate,delsole2010model,majda2011link,majda2016introduction}.

Here we discuss a general mathematical framework to construct statistically accurate reduced-order models that have the skill in capturing the statistical variability in the principal directions with largest energy of a general class of damped and forced complex turbulent dynamical systems. Low-order truncation methods is especially important for UQ with practical impact since \emph{the curse of ensemble size} \cite{bengtsson2008curse,majda2016introduction} forbids to run Monte-Carlo simulations for all possible uncertain forcing scenarios in order to do attribution studies. Thus reduced-order models (ROM) are needed on a low-dimensional subspace where key physical significant quantities are characterized by the degrees of freedom that carry the largest energy or variance. In general, there are three stages in the modeling procedure, \emph{imperfect
model selection}; \emph{calibration of the imperfect model} in a training
phase using only data in the low-order perfect statistics; and \emph{prediction of the responses} to a wide class of forcing and perturbation scenarios. The methods are developed for a universal class of turbulent dynamical systems with quadratic nonlinearity that is representative in many applications in applied mathematics and engineering \cite{majda2012lessons,majda2006nonlinear,majda2016introduction}.
Several mathematical ideas will be introduced to help improve the
prediction skill of the imperfect reduced-order models. Most importantly,
\emph{empirical information theory} \cite{majda2005information} and
\emph{statistical linear response theory} \cite{majda2011link} are
applied in the training phase for calibrating model errors to achieve
optimal imperfect model parameters; and \emph{total statistical energy
dynamics} \cite{majda2015statistical} are introduced to improve the
model sensitivity in the prediction phase especially when strong external
perturbations are exerted. The validity of general framework of reduced-order
models has been verified by testing the methods on a series of representative
turbulent dynamical models ranging from the 40-dimensional Lorenz
'96 model \cite{majda2016improving}, one-layer barotropic turbulence
with topography \cite{qi2016}, and finally the two-layer baroclinic
turbulence with internal instability \cite{qi2016low}.

In the following parts of the paper, we first display the general
formulation of the turbulent system with quadratic nonlinearity, and
its statistical moment dynamical equations in Section 1. The skill and limitation of many previous low-order modeling ideas are also discussed. Theoretical
toolkits that are useful for the development of reduced-order models
are introduced in Section 2, where a general strategy to improve imperfect
model sensitivity is described using empirical information theory
and a general total statistical energy dynamics. Section 3 discusses
the construction of reduced-order models in detail under this general
framework with these various theoretical tools. In Section 4, we illustrate
all these procedures and algorithms for the reduced-order models for
some simple but instructive systems of triad stochastic equations
with several novel features. In Section 5, we give examples of the
skill of the procedures and algorithms on two-layer
baroclinic models for both atmosphere and ocean regimes with turbulent
jets and vortices with roughly $1\times10^{5}$ degrees of freedom
and direct and inverse turbulent cascades. In these very tough regimes,
the reduced-order strategies show skill in capturing the response
to changes in external forcing using only $200$ modes, less than $0.15\%$ of the modes in the original system.

\section{General Formulation of Turbulent Dynamical Systems with Nonlinearity}

One representative feature in many turbulent dynamical systems from
nature is the quadratic energy conserving nonlinear interaction that
transfers energy from the unstable modes to stable ones where the energy
is dissipated resulting in a statistical steady state in equilibrium.
We consider the following abstract formulation of the turbulent dynamical
systems about state variables $\mathbf{u}\in\mathbb{R}^{N}$ in a
high-dimensional phase space
\begin{equation}
\frac{d\mathbf{u}}{dt}=\left(\mathcal{L+D}\right)\mathbf{u}+B\left(\mathbf{u},\mathbf{u}\right)+\mathbf{F}\left(t\right)+\boldsymbol{\sigma}\left(t\right)\dot{\mathbf{W}}\left(t;\omega\right).\label{eq:abs_formu}
\end{equation}
On the right hand side of the above equation (\ref{eq:abs_formu}),
the first two components, $\left(\mathcal{L+D}\right)\mathbf{u}$,
represent linear dispersion and dissipation effects so that\addtocounter{equation}{0}\begin{subequations}\label{prop}
\begin{equation}
\mathcal{L}^{*}=-\mathcal{L},\:\mathrm{skew-symmetric};\quad\mathcal{D}^{*}=\mathcal{D}<0,\:\mathrm{negative-definite},\label{eq:prop1}
\end{equation}
where the superscript star `$*$' represents conjugate transpose of
the matrix. The nonlinear effect in the dynamical system is introduced
through a quadratic form, $B\left(\mathbf{u},\mathbf{u}\right)$,
about the state variables $\mathbf{u}$ that conserves energy when
linear operators and all forcing in (\ref{eq:abs_formu}) are ignored,
such that,
\begin{equation}
\mathbf{u}\cdot B\left(\mathbf{u},\mathbf{u}\right)=\sum_{j=1}^{N}u_{j}B_{j}\left(\mathbf{u},\mathbf{u}\right)\equiv0,\quad\mathrm{Energy\:Conservation},\label{eq:prop2}
\end{equation}
\end{subequations}where the dot on the left hand side denotes the
inner product under a proper metric according to the conserved quantity
\cite{majda2015statistical,majda2016introduction}. Besides, the system
is forced by external forcing effects that are decomposed into a deterministic
component, $\mathbf{F}\left(t\right)$, and a stochastic component
usually represented by a Gaussian random process, $\boldsymbol{\sigma}\left(t\right)\dot{\mathbf{W}}\left(t;\omega\right)$.
It needs to be noticed that $\mathbf{F}\left(t\right)$ might be inhomogeneous and
introduce anisotropic structure into the system, and $\boldsymbol{\sigma}\left(t\right)\dot{\mathbf{W}}\left(t;\omega\right)$
might further alter the energy structure in the fluctuation modes.

Many complex turbulent dynamical systems can be categorized into this
abstract mathematical structure in (\ref{eq:abs_formu}) satisfying
the properties (\ref{eq:prop1}) and (\ref{eq:prop2}), including the 
(truncated) Navier-Stokes equation \cite{pope2001turbulent} as well
as basic geophysical models for the atmosphere, ocean, and the climate
systems with rotation, stratification, and topography \cite{salmon1998lectures,majda2006nonlinear,majda2016introduction}.
The main goal of the remainder of this paper is to provide a survey about the development of a consistent mathematical framework for systems like (\ref{eq:abs_formu}) and illustrate emerging applications of turbulent dynamical systems with model error and the curse of ensemble size.

\subsection{Exact statistical moment equations for the abstract formulation}

We use a finite-dimensional representation of the stochastic field
$\mathbf{u}$ consisting of a fixed-in-time, $N$-dimensional, orthonormal
basis $\left\{ \mathbf{e}_{i}\right\} _{i=1}^{N}$
\begin{equation}
\mathbf{u}\left(t\right)=\bar{\mathbf{u}}\left(t\right)+\sum_{i=1}^{N}Z_{i}\left(t;\omega\right)\mathbf{e}_{i},\label{eq:spec_expansion}
\end{equation}
where $\bar{\mathbf{u}}\left(t\right)=\left\langle \mathbf{u}\left(t\right)\right\rangle $
represents the ensemble average of the model state variable response
(we use angled bracket to represent ensemble average), i.e. the mean
field, and $Z_{i}\left(t;\omega\right)$ are stochastic coefficients
measuring the fluctuation processes along the direction $\mathbf{e}_{i}$.

By taking the statistical (ensemble) average over the original equation
(\ref{eq:abs_formu}) and using the mean-fluctuation decomposition
(\ref{eq:spec_expansion}), the \textbf{evolution equation of the
mean state $\bar{\mathbf{u}}=\left\langle \mathbf{u}\right\rangle $}
is given by the following dynamical equation
\begin{equation}
\frac{d\bar{\mathbf{u}}}{dt}=\left(L+D\right)\bar{\mathbf{u}}+B\left(\bar{\mathbf{u}},\bar{\mathbf{u}}\right)+\sum_{i,j}R_{ij}B\left(\mathbf{e}_{i},\mathbf{e}_{j}\right)+\mathbf{F},\label{eq:mean_dyn}
\end{equation}
with $R=\left\langle \mathbf{Z}\mathbf{Z}^{*}\right\rangle $ the
second-order covariance matrix of the stochastic coefficients $\mathbf{Z}=\left\{ Z_{i}\right\} _{i=1}^{N}$.
The term $B\left(\bar{\mathbf{u}},\bar{\mathbf{u}}\right)$ represents
the nonlinear interactions between the mean state, and $R_{ij}B\left(\mathbf{e}_{i},\mathbf{e}_{j}\right)$
is the higher-order feedbacks from the fluctuation modes to the mean
state dynamics. Moreover the random fluctuation component of the solution,
$\mathbf{u}^{\prime}=\sum_{i}Z_{i}\left(t;\omega\right)\mathbf{e}_{i}$
satisfies
\[
\frac{d\mathbf{u}^{\prime}}{dt}=\left(L+D\right)\mathbf{u}^{\prime}+B\left(\bar{\mathbf{u}},\mathbf{u}^{\prime}\right)+B\left(\mathbf{u}^{\prime},\bar{\mathbf{u}}\right)+B\left(\mathbf{u}^{\prime},\mathbf{u}^{\prime}\right)-\left\langle B\left(\mathbf{u}^{\prime},\mathbf{u}^{\prime}\right)\right\rangle +\sigma\left(t\right)\dot{\mathbf{W}}\left(t;\omega\right).
\]
By projecting the above equation to each orthonormal basis element
$\mathbf{e}_{i}$ we obtain
\[
\frac{dZ_{i}}{dt}=Z_{j}\left[\left(L+D\right)\mathbf{e}_{j}+B\left(\bar{\mathbf{u}},\mathbf{e}_{j}\right)+B\left(\mathbf{e}_{j},\bar{\mathbf{u}}\right)\right]\cdot\mathbf{e}_{i}+\left[B\left(\mathbf{u}^{\prime},\mathbf{u}^{\prime}\right)-\left\langle B\left(\mathbf{u}^{\prime},\mathbf{u}^{\prime}\right)\right\rangle \right]\cdot\mathbf{e}_{i}+\sigma\left(t\right)\dot{\mathbf{W}}\left(t;\omega\right)\cdot\mathbf{e}_{i}.
\]
From the last equation we directly obtain the exact \textbf{evolution
equation of the covariance matrix} $R=\left\langle \mathbf{Z}\mathbf{Z}^{*}\right\rangle $
by multiplying $Z_{j}^{*}$ on both sides of the equation and taking
ensemble statistical average
\begin{equation}
\frac{dR}{dt}=L_{v}\left(\bar{\mathbf{u}}\right)R+RL_{v}^{*}\left(\bar{\mathbf{u}}\right)+Q_{F}+Q_{\sigma},\label{eq:cov_dyn}
\end{equation}
where we have:\addtocounter{equation}{0}\begin{subequations}
\begin{description}
\item [{i)}] the linear dynamical operator $L_{v}\left(\bar{\mathbf{u}}\right)$
expresses energy transfers between the mean field and the stochastic
modes (effect due to $B$), as well as energy dissipation (effect
due to $\mathcal{D}$) and non-normal dynamics (effect due to $\mathcal{L}$)
\begin{equation}
\left\{ L_{v}\right\} _{ij}=\left[\left(\mathcal{L}+\mathcal{D}\right)\mathbf{e}_{j}+B\left(\bar{\mathbf{u}},\mathbf{e}_{j}\right)+B\left(\mathbf{e}_{j},\bar{\mathbf{u}}\right)\right]\cdot\mathbf{e}_{i};\label{eq:lin_operator}
\end{equation}

\item [{ii)}] the positive definite operator $Q_{\sigma}$ expresses energy
transfer due to the external stochastic forcing 
\begin{equation}
\left\{ Q_{\sigma}\right\} _{ij}=\sum_{k}\left(\mathbf{e}_{i}\cdot\sigma_{k}\right)\left(\sigma_{k}\cdot\mathbf{e}_{j}\right);\label{eq:sto_operator}
\end{equation}

\item [{iii)}] as well as the energy flux $Q_{F}$ expresses nonlinear
energy transfer between different modes due to non-Gaussian statistics
(or nonlinear terms) modeled through third-order moments
\begin{equation}
\left\{ Q_{F}\right\} _{ij}=\sum_{m,n}\left\langle Z_{m}Z_{n}Z_{j}\right\rangle B\left(\mathbf{e}_{m},\mathbf{e}_{n}\right)\cdot\mathbf{e}_{i}+\left\langle Z_{m}Z_{n}Z_{i}\right\rangle B\left(\mathbf{e}_{m},\mathbf{e}_{n}\right)\cdot\mathbf{e}_{j}.\label{eq:nonlinear_flux}
\end{equation}

One important property to notice is that the energy conservation property of the quadratic operator
$B$ is inherited in the statistical equations by the matrix $Q_{F}$
since
\begin{equation}
\mathrm{tr}\left(Q_{F}\right)=2\sum_{i}\sum_{m,n}\left\langle Z_{m}Z_{n}Z_{i}\right\rangle B\left(\mathbf{e}_{m},\mathbf{e}_{n}\right)\cdot\mathbf{e}_{i}=2B\left(\mathbf{u}^{\prime},\mathbf{u}^{\prime}\right)\cdot\mathbf{u}^{\prime}\equiv0.\label{eq:energy_flux_conserving}
\end{equation}
\end{description}\end{subequations}The above exact statistical equations for the state
of the mean (\ref{eq:mean_dyn}) and covariance matrix (\ref{eq:cov_dyn})
will be the starting point for the developments in the reduced-order
models on UQ methods.

Note that the statistical dynamics for the mean (\ref{eq:mean_dyn})
and covariance (\ref{eq:cov_dyn}) are still not closed due to the
inclusion of third-order moments through the nonlinear interactions
in $Q_{F}$ in (\ref{eq:nonlinear_flux}). The basic idea in the general development of reduced-order schemes concerns about proper approximation about this
energy flux term $Q_{F}$ in a simple and efficient manner so that
the energy mechanism can be modeled properly in the reduced-order
schemes \cite{lesieur2012turbulence,salmon1998lectures,sapsis2013statistically1}.

\subsubsection{Low-order truncation methods for UQ and their limitations}
Next we briefly discuss some popular low-order truncation methods for closing the statistical equations (\ref{eq:mean_dyn}) and (\ref{eq:cov_dyn}) and their limitations. Low-order truncation models for UQ include projection of the dynamics on leading order empirical orthogonal functions (EOF’s) \cite{holmes1998turbulence}, truncated polynomial chaos (PC) expansions \cite{hou2006wiener,knio2001stochastic,najm2009uncertainty}, and dynamically orthogonal (DO) truncations \cite{sapsis2013attractor,sapsis2009dynamically}. Then ideas about closing the low-order truncated system within the resolved modes need to be proposed. A pioneering statistical prediction strategy \cite{epstein1969stochastic,epstein1971depicting} overcomes the curse of ensemble size for moderate size turbulent dynamical systems by simply neglecting the third-order moments by setting $Q_{F}\equiv0$ in the covariance equations (\ref{eq:cov_dyn}). This \emph{Gaussian closure method} has been applied to short time statistical prediction for truncated geophysical models like the one-layer geophysical models in (\ref{eq:onelayer}) with some success \cite{epstein1971depicting,srinivasan2012zonostrophic}. Based on the  similar idea of neglecting third-order moments, the \emph{eddy-damped quasi-normal Markovian approximation} (EDM) \cite{salmon1998lectures,lesieur2012turbulence} is another approximation to the moment hierarchy (\ref{eq:mean_dyn}) and (\ref{eq:cov_dyn}) that closes the second moments with (inconsistent) Gaussian approximation in the higher order equations. With a much larger  \emph{eddy-damped} parameters, the EDM equations are realizable in a stochastic model.

Moreover concise mathematical models and analysis reveal fundamental limitations in truncated EOF expansions \cite{aubry1993preserving,crommelin2004strategies}, PC expansions \cite{branicki2013fundamental,majda2012lessons}, and DO truncations \cite{sapsis2013blended}, due to different manifestations of the fact that in many turbulent dynamical systems, modes that carry small variance on average can have important, highly intermittent dynamical effects on the large variance modes. Furthermore, the large dimension of the active variables in turbulent dynamical systems makes direct UQ by large ensemble Monte-Carlo simulations impossible in the foreseeable future while once again, concise mathematical models \cite{majda2012lessons} point to the limitations of using moderately large yet statistically too small ensemble sizes. Other important methods for UQ involve the linear statistical response to change in external forcing or initial data through the fluctuation dissipation theorem (FDT) which only requires the measurement of suitable time correlations in the unperturbed system \cite{abramov2007blended,gritsun2007climate,gritsun2008climate,hairer2010simple,majda2010low}.  Despite some significant success with this approach for turbulent dynamical systems \cite{abramov2007blended,gritsun2007climate,gritsun2008climate,majda2010low}, the method is hampered by the need to measure suitable approximations to the exact correlations for long time series as well as the fundamental limitation to parameter regimes with a linear statistical response. All the limitations above imply the need of a more careful treatment for the higher-order statistics in $Q_{F}$ in the exact equations for mean and covariance (\ref{eq:mean_dyn}) and (\ref{eq:cov_dyn}).

\subsection{The overall prediction strategy for the development of reduced-order statistical models}

Before preceding to the details about developing the reduced-order
statistical model framework, we illustrate the basic ideas in the
modeling process as a general overview. Overall, this can serve as
a generic procedure where rigorous mathematical theories and various
computational strategies are combined to get a crucial improvement
for understanding turbulent dynamical systems. In general, we can
decompose the reduced-order statistical modeling strategy into three
stages: i) imperfect model selection according to the complexity of
the problem; ii) model calibration in the training phase using equilibrium
data; and iii) model prediction with the optimized model parameters
for various responses to external perturbations. The overall prediction
strategy is summarized in a diagram in Figure \ref{fig:The-general-strategy}.
The basic procedure for developing statistical models illustrates
a representative example where various mathematical theories and numerical
methods interact and are combined for achieving a better understanding
about the natural system.

\subsubsection{Ergodicity and non-trivial invariant measure for the true turbulent dynamical systems}

In the first place, the best reduced-order approximation strategy
can only be achieved through a good understanding about the true turbulent
dynamical system. Several important mathematical theories are especially
useful for characterizing the statistical structure of the turbulent
system. Under special damping and random noise forms without the deterministic
forcing $\mathbf{F}\equiv0$, a Gaussian invariant measure can be
generated in the statistical steady state, whereas this Gaussian distribution
from equilibrium statistical mechanics can be only derived from special
damping and noise terms \cite{majda2016introduction,qi2016}. One
more generalized situation with importance in many realistic applications
is when no stochastic forcing in the damped and forced dynamical system
(\ref{eq:abs_formu}), so that the deterministic system with $\sigma\equiv0$
has non-trivial long-time dynamics. The uncertainty in such deterministic
systems is measured by the unstable sub-phase space with a number
of positive Lyapunov exponents, thus a nontrivial global attractor
is generated through the strong interaction and exchange of energy
\cite{robinson2001infinite}. This scenario is similar to the Sinai-Ruelle-Bowen
(SRB) measure problem \cite{young2002srb,ruelle1997differentiation}.
In that case, a unique distinguished invariant measure $p_{\mathrm{eq}}$,
the SRB measure, is the one selected by the vanishing noise limit
with appropriate assumptions on the system and noise. This distinguished
invariant measure forms up a stationary statistical solution $p_{\mathrm{eq}}$
in equilibrium, so that
\begin{equation}
p_{\mathrm{eq}}\left(\left(\Phi^{t}\right)^{-1}\left(\Omega\right)\right)=p_{\mathrm{eq}}\left(\Omega\right),\quad\mathrm{for\:any}\:t>0\;\mathrm{and}\;\Omega\subset\mathbb{R}^{N},\label{eq:inv_measure}
\end{equation}
with $\Phi^{t}:\mathbb{R}^{N}\rightarrow\mathbb{R}^{N}$ as the flow
map. This invariant measure (\ref{eq:inv_measure}) provides a mechanism
for explaining how local instability on attractors can produce coherent
statistics for orbits starting from large sets in the basin. The statistical
ensemble behaviour in equilibrium such as the mean state and covariance
can be deduced by taking averages with respect to the invariant measure.

Ergodicity is then one important property for the turbulent dynamical
system with uncertainty, and means that there exists a unique invariant
measure in statistical equilibrium which attracts all statistical
initial data. Geometric ergodicity for finite dimensional Garlerkin
truncation models (for example, the two or three dimensional Navier-Stokes
equations) with minimal stochastic forcing is an important research
topic \cite{weinan2001gibbsian,majda2016introduction,majda2015ergodicity}.
With proper ergodicity assumption about the abstract system (\ref{eq:abs_formu})
and rigorously justified for the system with minimal stochastic forcing \cite{majda2016introduction,majda2015ergodicity},
the statistical expectation of any functionals about the state variables
can be calculated through averaging the time-series in steady state,
that is,
\begin{equation}
\left\langle g\left(\mathbf{u}\right)\right\rangle =\int_{\mathbb{R}^{N}}g\left(\mathbf{u}\right)p_{\mathrm{eq}}\left(\mathbf{u}\right)d\mathbf{u}=\lim_{T\rightarrow\infty}\frac{1}{T}\int_{t_{0}}^{t_{0}+T}g\left[\mathbf{u}\left(t\right)\right]dt,\label{eq:ergodicity}
\end{equation}
where $g$ is any functional about the state variables $\mathbf{u}$,
and $p_{\mathrm{eq}}$ is the invariant measure (\ref{eq:inv_measure}) in statistical equilibrium.
Taking the ensemble averages from the first equality of (\ref{eq:ergodicity})
is usually an extremely challenging problem, while the average along
a trajectory over a long time as the right hand side of (\ref{eq:ergodicity})
forms a more practical approach. Ergodicity is crucial in this prediction
strategy for achieving accurate perfect model statistics, and will
be assumed throughout the following discussions.

\subsubsection{Model selection, model calibration, and prediction with optimized imperfect model}

The ergodic theory and invariant measure enable us to get access to
the model equilibrium statistical structures in steady state. Still
the major goal in this investigation is to find the model sensitivity
in response to various external perturbations. Especially for the
turbulent dynamical systems with instability like (\ref{eq:abs_formu}), nonlinearity forms
the key mechanism in the complex chaotic behaviour, and even small
perturbations may drive the system away from its original equilibrium
state. Furthermore, strong non-Gaussianity due to the strange attractor
from the SRB measure is another characteristic feature in these turbulent
systems with non-Gaussian measures even in equilibrium. The reduced-order
statistical modeling procedure aims at capturing these nonlinear non-Gaussian
statistical responses in the principal directions in the system in
an accurate and efficient way.

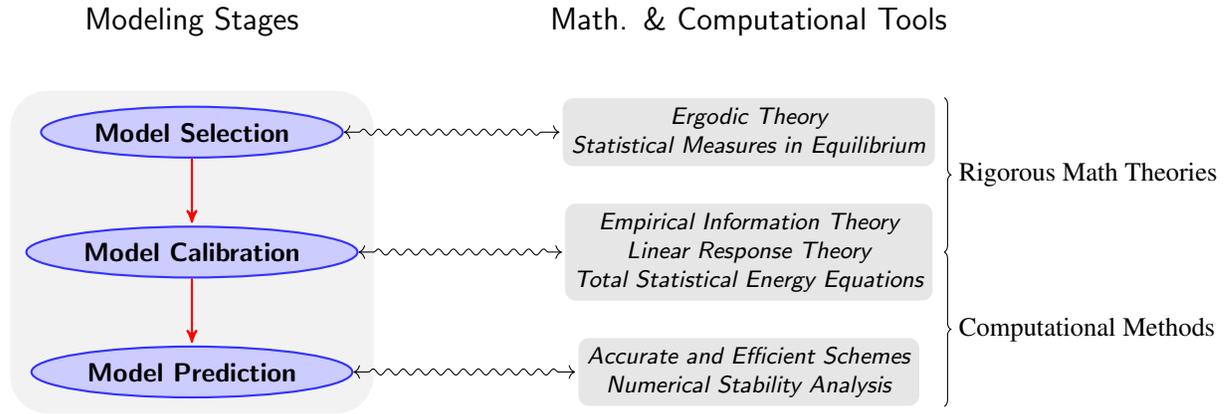
\begin{figure}
\usetikzlibrary{matrix,arrows.meta}
\usetikzlibrary{calc,positioning,decorations.pathreplacing,decorations.pathmorphing,shapes }
\tikzset{   
	centered/.style = { align=center, anchor=center },      
	empty/.style = { font=\sffamily\large, text width=20em, minimum height=3em, text centered },        
	box/.style = { font=\sffamily, ellipse, minimum width=80pt, align=center, thick, draw=blue!80, fill=blue!20 },     
	result/.style = { font=\sffamily\small, fill=black!10, align=center, rounded corners },      
	arrow/.style = { thick, color=red, ->, shorten >=1pt, >=stealth'}, 
  arrowsnake/.style = { <->,snake=snake, segment amplitude=.4mm, segment length=2mm, shorten >=1pt, line before snake=2mm, line after snake=2mm}, 
	decoration={brace},   tuborg/.style={decorate},   tubnode/.style={midway, right=2pt},} 
\newcommand*{\nothing}{}
\tikzstyle{background}=[rectangle, fill=gray!10, inner sep=0.2cm, rounded corners=5mm]

\begin{tikzpicture}[scale=1]

	\matrix (m)[row sep=0.5cm, column sep=0.4em] {             
	\node (model)    [empty]{Modeling Stages};                  &               
	\node (theory)   [empty]{Math. \&  Computational Tools};    \\
	\node (select1)  [box]{\textbf {Model Selection}};                  &               
	\node (select2)  [result]{\em Ergodic Theory \\ \em Statistical Measures in Equilibrium };    \\
	\node (cali1)  [box]{\textbf {Model Calibration}};                  &               
	\node (cali2)  [result]{\em Empirical Information Theory \\ \em Linear Response Theory \\ \em Total Statistical Energy Equations };    \\
	\node (pred1)  [box]{\textbf {Model Prediction}};                  &               
	\node (pred2)  [result]{\em Accurate and Efficient  Schemes \\ \em Numerical Stability Analysis };    \\
    };

	\path[->]         
		(select1) edge[arrowsnake] (select2)
		(cali1) edge[arrowsnake] (cali2)
		(pred1) edge[arrowsnake] (pred2)
		(select1) edge[arrow] (cali1)
		(cali1) edge[arrow] (pred1)
	;

	\begin{pgfonlayer}{background}         
		\node [background, fit=(select1) (cali1) (pred1)] {};
	\end{pgfonlayer}

	\draw[tuborg, decoration={brace}] let \p1=(select2.north), \p2=(cali2.center) in ($(6.3, \y1)$) -- ($(6.3, \y2)$) node[tubnode] {Rigorous Math Theories};
	\draw[tuborg, decoration={brace}] let \p1=(cali2.center), \p2=(pred2.south) in ($(6.3, \y1)$) -- ($(6.3, \y2)$) node[tubnode] {Computational Methods};
	
\end{tikzpicture}

\caption{The general strategy for the development of reduced-order statistical
models. Three sequential stages are required to carry out the reduced-order
statistical model, and rigorous mathematical theories are combined
with numerical analysis to calibrate model errors and improve the
imperfect model prediction skill.\label{fig:The-general-strategy}}
\end{figure}

As illustrated in Figure \ref{fig:The-general-strategy} for the general
strategy, the modeling procedure begins with the model selection stage
where proper approximation method is adopted through a careful analysis
about the statistical theories. Specifically in the reduced-order
models to be developed here, usually additional damping and
random noise corrections are introduced for the unresolved higher-order statistics.
The equilibrium invariant measure and ergodic theory \cite{majda2015ergodicity,sapsis2013statistically1}
can help determine the optimal Galerkin truncation wavenumber for
the reduced-order model and the proper basis that can cover the most
important directions in the system. Especially, non-Gaussian statistics
in the unperturbed equilibrium state would also become important and
require careful consideration in the model calibration.

The model calibration procedure is usually carried out in a training
phase before the prediction, so that the optimal imperfect model parameters
can be achieved through a careful calibration about the true higher-order
statistics. The ideal way is to find a unified systematic strategy
where various external perturbations can be predicted from the same
set of optimal parameters through this training phase. To achieve
this, various statistical theories and numerical strategies need to
be blended together in a judicious fashion. Most importantly, we need
to consider the linear statistical response theory to calibrate the
model responses in mean and variances \cite{majda2005information,majda2010linear,hairer2010simple};
and use empirical information theory \cite{majda2011improving,majda2011link,majda2006nonlinear}
to get a balanced measure for the error in the leading order moments.
In the final model prediction stage, the optimized imperfect model parameters
are applied for the forecast of various model responses to perturbations.
In the construction about numerical models, numerical issues also
need be taken into account to make sure numerical stability and accuracy.
Especially, proper schemes with accuracy order consistent with the
reduced model approximation error should be proposed to ensure optimal
performance.

\subsection{Low-order models illustrating model selection, calibration, and prediction in UQ}
Here we provide a brief discussion of some instructive quantitative and qualitative low-order models where the above strategy for improved prediction and UQ is displayed. The test models as in nature often exhibit intermittency \cite{frisch1995turbulence,pope2001turbulent} where some components of a turbulent dynamical system have low amplitude phases followed by irregular large amplitude bursts of extreme events. Intermittency is an important physical phenomena. Exactly solvable test models as a test bed for the prediction and UQ strategy \cite{majda2016introduction,majda2005information,majda2010quantifying} including information barriers are discussed extensively in models ranging from linear stochastic models to nonlinear models with intermittency in the research expository article \cite{majda2012lessons} as well as in \cite{branicki2013non,branicki2012quantifying}. Some more sophisticated applications are mentioned next in Section 1.4.

Turbulent diffusion in exactly solvable models is a rich source of highly nontrivial spatiotemporal multi-scale models to test the strategies in \emph{empirical information theory} and \emph{kicked statistical response theory} in a more complex setting \cite{gershgorin2012quantifying,majda2011improving,majda2011link,majda2013elementary}. Even though these models have no positive Lyapunov exponents, they have been shown rigorously to exhibit intermittency and extreme events \cite{majda2015intermittency}. Calibration strategies for imperfect models using information theory have been developed recently to yield statistical accurate prediction of these extreme events by imperfect inexpensive linear stochastic models for the velocity field \cite{qi2015predicting}. This topic merits much more attention by other modern applied mathematicians \cite{mohamad2015probabilistic,mohamad2016probabilistic}.

\subsubsection{Nonlinear regression models for time series}
A central issue in contemporary science is the development of data driven statistical dynamical models for the time series of a partial set of observed variables which arise from suitable observations from nature (see \cite{cressie2015statistics} and references therein); examples are multi-level linear autoregressive models as well as \emph{ad hoc} quadratic nonlinear regression models. It has been established recently \cite{majda2012fundamental} that \emph{ad hoc} quadratic multi-level regression models can have finite time blow up of statistical solutions and pathological behavior of their invariant measure even though they match the data with high precision. A new class of physics-constrained multi-level nonlinear regression models was developed which involve both memory effects in time as well as physics-constrained energy conserving nonlinear interactions \cite{harlim2014ensemble,majda2012physics}, which completely avoid the above pathological behavior with full mathematical rigor.

A striking application of these ideas combined with information calibration to the predictability limits of tropical intraseasonal variability such as the Madden-Julian oscillation (MJO) and the monsoon has been developed in a series of papers \cite{chen2014predicting,chen2015predicting1,chen2015predicting}. They yield an interesting class of low-order turbulent dynamical systems with extreme events and intermittency. The nonlinear low-order stochastic model (see Section 4.2 of \cite{majda2016introduction}) has been shown to have significant skill for determining the predictability limits of the large-scale cloud patterns of the boreal winter MJO \cite{chen2014predicting} and the summer monsoon \cite{chen2015predicting}.  It is an interesting open problem to rigorously describe the intermittency and other mathematical features in these low-order turbulent dynamical systems.

\subsection{Examples of complex turbulent dynamical systems}

Here we list some typical prototype models of complex turbulent dynamical
systems with the structure in (\ref{eq:abs_formu}). These qualitative
and quantitative models with increasing complexity form a desirable
set of testing models for prediction, UQ, and state estimation \cite{majda2016introduction}.
We will finally test the reduced-order modeling strategies on all
these typical models as a thorough discussion about the effectiveness
and limitations of the model reduction ideas including a complete
new treatment for the triad example.
\begin{description}
\item [{(A)}] \emph{The triad system with quadratic energy transfer}. The
triad model \cite{majda2002priori,majda2016introduction} is the elementary
building block of complex turbulent systems with energy conserving
nonlinear interactions. It is a 3-dimensional ODE system with inhomogeneous
damping and both deterministic and stochastic forcing terms\addtocounter{equation}{0}\begin{subequations}
\begin{equation}
\begin{aligned}\frac{du_{1}}{dt}= & L_{2}u_{3}-L_{3}u_{2}-d_{1}u_{1}+B_{1}u_{2}u_{3}+F_{1}+\sigma_{1}\dot{W}_{1},\\
\frac{du_{2}}{dt}= & L_{3}u_{1}-L_{1}u_{3}-d_{2}u_{2}+B_{2}u_{3}u_{1}+F_{2}+\sigma_{2}\dot{W}_{2},\\
\frac{du_{3}}{dt}= & L_{1}u_{2}-L_{2}u_{1}-d_{3}u_{3}+B_{3}u_{1}u_{2}+F_{3}+\sigma_{3}\dot{W_{3}}.
\end{aligned}
\label{eq:triad}
\end{equation}
The triad system is an instructive test model for the reduced-order strategies. A self-contained pedagogical discussion about the triad system is shown in Section 4.

\item [{(B)}] \emph{40-dimensional Lorenz '96 model}. The Lorenz '96 model
\cite{lorenz1996predictability,majda2016improving,majda2016introduction,majda2006nonlinear}
is a 40-dimensional turbulent dynamics defined with periodic boundary
condition which mimics weather waves of the mid-latitude atmosphere.
Various representative statistical features can be generated by changing
the external forcing values in $F$
\begin{equation}
\frac{du_{j}}{dt}=\left(u_{j+1}-u_{j-2}\right)u_{j-1}-d_{j}u_{j}+F_{j},\quad u_{0}=u_{J},\;j=0,\cdots,J-1,\quad J=40.\label{eq:L96}
\end{equation}
See \cite{majda2016improving} for the detailed reduced-order modeling
strategy.
\item [{(C)}] \emph{One-layer barotropic model with topography}. The one-layer
barotropic system \cite{majda2006nonlinear,majda2016introduction,qi2016}
is a basic and simple geophysical model for the atmosphere or ocean
with the essential geophysical effects of rotation, topography, and
both deterministic and random forcing.
\begin{equation}
\begin{aligned}\frac{\partial q}{\partial t}+\nabla^{\bot}\psi\cdot\nabla q+U\frac{\partial q}{\partial x} & =-\mathcal{D}\left(\Delta\right)\psi+\mathcal{F}\left(\mathbf{x},t\right)+\varSigma\left(\mathbf{x}\right)\dot{W}\left(t\right),\\
\frac{dU}{dt}+\fint\frac{\partial h}{\partial x}\psi\left(t\right) & =-\mathcal{D}_{0}U+\mathcal{F}_{0}\left(t\right)+\varSigma_{0}\dot{W}_{0}\left(t\right),\\
q & =\Delta\psi+h+\beta y.
\end{aligned}
\label{eq:onelayer}
\end{equation}
See \cite{qi2016} for the detailed reduced-order modeling strategy.
\item [{(D)}] \emph{Two-layer quasi-geostrophic model with baroclinic instability}.
The two-layer quasi-geostrophic model with baroclinic instability
in a two-dimensional periodic domain \cite{salmon1998lectures,vallis2006atmospheric,qi2016low}
is one fully nonlinear fluid model, and is quite capable in capturing
the essential physics of the relevant internal variability despite
its relatively simple dynamical structure.
\begin{equation}
\begin{aligned}\frac{\partial q_{\psi}}{\partial t}+J\left(\psi,q_{\psi}\right)+J\left(\tau,q_{\tau}\right)+\beta\frac{\partial\psi}{\partial x}+U\frac{\partial}{\partial x}\Delta\tau & =-\frac{\kappa}{2}\Delta\left(\psi-\tau\right)-\nu\Delta^{s}q_{\psi}+\mathcal{F}_{\psi}\left(\mathbf{x},t\right),\\
\frac{\partial q_{\tau}}{\partial t}+J\left(\psi,q_{\tau}\right)+J\left(\tau,q_{\psi}\right)+\beta\frac{\partial\tau}{\partial x}+U\frac{\partial}{\partial x}\left(\Delta\psi+k_{d}^{2}\psi\right) & =\frac{\kappa}{2}\Delta\left(\psi-\tau\right)-\nu\Delta^{s}q_{\tau}+\mathcal{F}_{\tau}\left(\mathbf{x},t\right).
\end{aligned}
\label{eq:twolayer}
\end{equation}
\end{subequations}See \cite{qi2016low} and discussions in Section
5 for the reduced-order modeling strategy.
\end{description}
\

\section{Statistical Theory Toolkits for Improving Model Prediction Skill}

In this section we introduce the general theoretical toolkits that
are useful for capturing the key statistical features in turbulent
systems like (\ref{eq:abs_formu}) and improving imperfect model prediction
skill. Despite the complex model statistical responses in each component
as the turbulent dynamical system gets perturbed, there exists a simple
and exact statistical energy conservation principle for the total
statistical energy of the system describing the overall (inhomogeneous)
statistical structure in the system through a simple scalar dynamical
equation \cite{majda2015statistical,majda2016introduction}. The theory
is briefly described in Section 2.1. Then the construction about the
imperfect reduced-order models concerns about the consistency in equilibrium
(climate fidelity) and the responses to perturbations (model sensitivity).
Equilibrium statistical fidelity should be guaranteed in the first
place so that the reduced-order model will converge to the true unperturbed
equilibrium statistics. To further calibrate the detailed model sensitivity
to perturbations in each statistical component, the linear response
theory can offer useful quantities to measure for quantifying the
crucial statistics in the model structure. Combining with the relative
entropy under empirical information theory, a general information-theoretical
framework can be proposed to tune the imperfect model parameters in
a training phase, thus optimal model parameters can be used for model
prediction in various dynamical regimes. We will describe the basic
statistical theories in this section.

\subsection{A statistical energy conservation principle}

Despite the fact that the exact equations for the statistical mean
(\ref{eq:mean_dyn}) and the covariance fluctuations (\ref{eq:cov_dyn})
are not closed equations, there is suitable statistical symmetry so
that the energy of the mean plus the trace of the covariance matrix
satisfies an energy conservation principle even with general deterministic
and random forcing. Here we briefly introduce the theory developed
in \cite{majda2015statistical,majda2016introduction} about a total
statistical energy dynamics for the abstract system (\ref{eq:abs_formu}).
This total statistical energy offers a general description about the
total responses in the perturbed system and will be shown useful for
the construction of reduced-order models.

Consider the statistical mean energy, $\bar{E}=\frac{1}{2}\left|\bar{\mathbf{u}}\right|^{2}=\frac{1}{2}\bar{\mathbf{u}}\cdot\bar{\mathbf{u}}$,
and the statistical fluctuation energy, $E^{\prime}=\frac{1}{2}\left\langle \mathbf{u}^{\prime}\cdot\mathbf{u}^{\prime}\right\rangle =\frac{1}{2}\mathrm{tr}R$.
Assume the following symmetries involving the nonlinear interaction
operator $B$ under the orthonormal basis $\left\{ \mathbf{e}_{i}\right\} $:
\begin{description}
\item [{A)}] The self interactions vanish in the quadratic interaction,
\addtocounter{equation}{0}\begin{subequations}\label{prop_assp}
\begin{equation}
B\left(\mathbf{e}_{i},\mathbf{e}_{i}\right)\equiv0,\quad1\leq i\leq N;\label{eq:prop_assp1}
\end{equation}

\item [{B)}] The dyad interaction coefficients vanish through the symmetry,
\begin{equation}
\mathbf{e}_{i}\cdot\left[B\left(\mathbf{e}_{j},\mathbf{e}_{i}\right)+B\left(\mathbf{e}_{i},\mathbf{e}_{j}\right)\right]=0,\quad\mathrm{for\,any\,}i,j.\label{eq:prop_assp2}
\end{equation}
\end{subequations}
\end{description}
Therefore the \emph{detailed triad symmetry} guarantees that the nonlinear
interaction $B\left(\mathbf{u},\mathbf{u}\right)$ will not alter
the total statistical energy structure in the system (though the state
of the mean and covariance may both change due to the nonlinear term
in (\ref{eq:mean_dyn}) and (\ref{eq:cov_dyn})). So we have the following
theorem \cite{majda2015statistical,majda2016introduction}:
\begin{theorem}
\label{thm:thm1}(\textbf{Statistical Energy Conservation Principle})
Under the structural assumptions (\ref{eq:prop_assp1}), (\ref{eq:prop_assp2})
on the basis $\mathbf{e}_{i}$, for any turbulent dynamical systems
in (\ref{eq:abs_formu}), the total statistical energy, $E=\bar{E}+E^{\prime}=\frac{1}{2}\bar{\mathbf{u}}\cdot\bar{\mathbf{u}}+\frac{1}{2}\mathrm{tr}R$,
satisfies
\begin{equation}
\frac{dE}{dt}=\bar{\mathbf{u}}\cdot D\bar{\mathbf{u}}+\bar{\mathbf{u}}\cdot\mathbf{F}+\mathrm{tr}\left(DR\right)+\frac{1}{2}\mathrm{tr}Q_{\sigma},\label{eq:energy_conservation}
\end{equation}
where $R$ satisfies the exact covariance equation in (\ref{eq:cov_dyn}).
Matrix $Q_{\sigma}$ expresses energy transfer due to external stochastic
forcing, and is a diagonal matrix with entries, $Q_{\sigma,kk}=\left|\sigma_{k}\right|^{2}$.
\end{theorem}
For most practical dynamical systems, for example, the systems we
have illustrated in (\ref{eq:triad})-(\ref{eq:twolayer}), the symmetries
in (\ref{prop_assp}) are usually satisfied. Also a generalization
allowing both dyads and triads in the statistical energy conservation
principle is in \cite{majda2016introduction}. Thus the statistical
energy conservation principle can always be applied. Notice that especially
under the homogeneous dissipation case, $D=-dI$, the right hand side
of the statistical energy equation (\ref{eq:energy_conservation})
will become a linear damping term for the total energy, $-dE$, plus
the deterministic forcing applying on the mean state and the stochastic
forcing contribution. This implies that the total energy structure
(and thus the total variance in all the modes) can be determined from
the statistical mean state by solving the scalar equation above.

\subsection{Statistical equilibrium fidelity in approximation models}

Here we consider the statistical energy of the dynamical system in
each individual (spectral) mode. Statistical equilibrium fidelity
concerns the convergence to the true equilibrium statistics in statistical
steady state in the reduced-order models. Recall the true second-order
statistical equation (\ref{eq:cov_dyn}) about the covariance matrix
\[
\frac{dR}{dt}=L_{v}R+RL_{v}^{*}+Q_{F}+Q_{\sigma}.
\]
The most difficult and expensive part in solving the above system
comes from evaluating the nonlinear flux term $Q_{F}$ where higher
order statistics are involved, that is,
\[
\left\{ Q_{F}\right\} _{ij}=\sum_{m,n}\left\langle Z_{m}Z_{n}Z_{j}\right\rangle B\left(\mathbf{e}_{m},\mathbf{e}_{n}\right)\cdot\mathbf{e}_{i}+\left\langle Z_{m}Z_{n}Z_{i}\right\rangle B\left(\mathbf{e}_{m},\mathbf{e}_{n}\right)\cdot\mathbf{e}_{j}.
\]
Note that the third-order moments always include triad interactions
of modes $\left\{ Z_{m},Z_{n},Z_{j}\right\} $ between different scales,
where nonlinear energy forward-cascade and backward-cascade along
the energy spectrum can be induced. Thus the central issue in developing
closure models becomes to find proper approximation about the nonlinear
flux term $Q_{F}^{M}\sim Q_{F}$ which can offer a statistically consistent
estimation. First of all, it is important to remember the conservation
of the total nonlinear flux $\mathrm{tr}Q_{F}\equiv0$ from (\ref{eq:energy_flux_conserving}).
This equality implies that the nonlinear interactions will not introduce
additional energy source or sink into the system. Thus the same constraint
should be maintained in designing the approximation models, $\mathrm{tr}Q_{F}^{M}=0$.
Consideration about accuracy and computational efficiency should be
balanced in determining the explicit form of $Q_{F}^{M}$ in the implementation
of reduced methods. Here we first display some theoretical principles
about the equilibrium nonlinear flux $Q_{F,\mathrm{eq}}$ that can
be used as guidelines for determining the values in $Q_{F}^{M}$.

\subsubsection{Calibration about higher-order statistics in full phase space}

In the prediction of model responses it is most important to find
the variability along each principal direction. In general, the nonlinear
flux $Q_{F}$ illustrates the nonlinear energy transfer between modes
with different scales. In fact, we can decompose the matrix $Q_{F}=Q_{F}^{+}+Q_{F}^{-}$
by singular value decomposition into a positive-definite and negative-definite
component. The positive definite part $Q_{F}^{+}$ illustrates the
additional energy that is injected into this mode from other scales,
while the negative definite part $Q_{F}^{-}$ shows the extraction
of energy through nonlinear transfer to other scales. Thus the accurate
approximation about the nonlinear flux $Q_{F,ij}$ in each (spectral)
component becomes important. On the other hand, this approximation
requires the calibration about the third-order moments $\left\langle Z_{m}Z_{n}Z_{j}\right\rangle $
and $\left\langle Z_{m}Z_{n}Z_{i}\right\rangle $, and will always
include the interactions between the (resolved) large-scale modes
and (unresolved) smaller-scale fluctuations. Direct simulation would
require ensemble averages for the third-order moments, where large
numerical errors and high computational loads are almost unavoidable.

Instead, from the statistical dynamics for the covariance equation
(\ref{eq:cov_dyn}) in statistical steady state, the temporal derivative
on the left hand side vanishes, $\frac{d}{dt}R_{\mathrm{eq}}\equiv0$,
thus the equilibrium solution $\left(\bar{\mathbf{u}}_{\mathrm{eq}},R_{\mathrm{eq}}\right)$
necessarily satisfies the steady state equation
\[
0=L_{v}\left(\bar{\mathbf{u}}_{\mathrm{eq}}\right)R_{\mathrm{eq}}+R_{\mathrm{eq}}L_{v}^{*}\left(\bar{\mathbf{u}}_{\mathrm{eq}}\right)+Q_{F,\mathrm{eq}}+Q_{\sigma},
\]
where $Q_{F,\mathrm{eq}}$ includes the third-order moments evaluated
at the statistical steady state. Therefore we can get the measurements
about equilibrium third-order nonlinear flux through the lower order
steady state solution of the mean, $\bar{\mathbf{u}}_{\mathrm{eq}}$,
and the covariance, $R_{\mathrm{eq}}$, so that
\begin{equation}
Q_{F,\mathrm{eq}}=-L_{v}\left(\bar{\mathbf{u}}_{\mathrm{eq}}\right)R_{\mathrm{eq}}-R_{\mathrm{eq}}L_{v}^{*}\left(\bar{\mathbf{u}}_{\mathrm{eq}}\right)-Q_{\sigma}.\label{eq:thrd_equili}
\end{equation}
The quasi-linear operator $L_{v}\left(\bar{\mathbf{u}}_{\mathrm{eq}}\right)$
is defined through (\ref{eq:lin_operator}) containing the interactions
between mean state. Especially, the non-trivial third moments play
a crucial dynamical role in the statistical closure models. As an
example in the case without random forcing $Q_{\sigma}\equiv0$, the
necessary and sufficient condition for a non-Gaussian statistical
steady state \cite{majda2016introduction} requires that 
\[
L_{v}\left(\bar{\mathbf{u}}_{\mathrm{eq}}\right)R_{\mathrm{eq}}+R_{\mathrm{eq}}L_{v}^{*}\left(\bar{\mathbf{u}}_{\mathrm{eq}}\right)\neq0,
\]
so the above matrix has non-zero entries. This is an important constraint
that needs to be considered first in the construction about reduced-order
models in the next section.

\subsubsection{Calibration of higher-order statistics in the reduced subspace}

Despite the above exact model calibration for higher-order statistics
(\ref{eq:thrd_equili}) using equilibrium mean and covariance, in
many realistic problems, resolving the entire covariance matrix $R\in\mathbb{C}^{N\times N}$
of order $O\left(N^{2}\right)$ is still expensive and unnecessary
especially for high dimensional systems $N\gg1$ with strong interactions
between small and large scales. Often the key physical significant
quantities are characterized by the degrees of freedom which carry
the largest energy (or variance). Thus, for most cases we are mostly
interested in the model variability in a low-dimensional subspace
along the principal directions spanned by the subspatial basis
\[
P=\left[\mathbf{v}_{1},\cdots,\mathbf{v}_{s}\right],\quad s\ll N.
\]
One simplest proposal to get the low-order basis $\left\{ \mathbf{v}_{j}\right\} $
is through the leading order EOFs or energy based proper orthogonal decomposition \cite{holmes1998turbulence,aubry1993preserving}.
The reduced-order third-order nonlinear flux can be calculated through
a more efficient way using only the mean state, $\bar{\mathbf{u}}_{\mathrm{eq}}$,
and covariance in the subspace of interest, $C=P^{*}RP\in\mathbb{\mathbb{C}}^{s\times s}$.
By projecting the original nonlinear flux formulation (\ref{eq:thrd_equili})
onto the subspace, we have the reduced-order formulation
\begin{equation}
Q_{F,\mathrm{eq}}^{\mathrm{red}}\equiv P^{*}Q_{F,\mathrm{eq}}P=-L_{v}^{\mathrm{red}}\left(\bar{\mathbf{u}}_{\mathrm{eq}}\right)C_{\mathrm{eq}}-C_{\mathrm{eq}}L_{v}^{\mathrm{red}*}\left(\bar{\mathbf{u}}_{\mathrm{eq}}\right)-Q_{\sigma}^{\mathrm{red}},\label{eq:thrd_equili_red}
\end{equation}
where the reduced-order quasi-linear operator and reduced-order noise
can also be calculated efficiently only using information in the subspace
with resolved leading order statistics
\[
L_{v,ij}^{\mathrm{red}}\equiv\left\{ P^{*}L_{v}P\right\} _{ij}=\left[\left(\mathcal{L}+\mathcal{D}\right)\mathbf{v}_{j}+B\left(\bar{\mathbf{u}},\mathbf{v}_{j}\right)+B\left(\mathbf{v}_{j},\bar{\mathbf{u}}\right)\right]\cdot\mathbf{v}_{i},\quad Q_{\sigma}^{\mathrm{red}}=P^{*}Q_{\sigma}P.
\]
Thus even though $Q_{F}^{\mathrm{red}}$ may still include many third
moments between the low-wavenumber resolved modes and high-wavenumber
modes that are not calculated explicitly in the reduced-order equations
only for $C$, we can still achieve the equilibrium nonlinear flux
constrained in the resolved subspace of interest by using only the
mean and covariances $\left(\bar{\mathbf{u}}_{\mathrm{eq}},C_{\mathrm{eq}}\right)$
along the resolved directions $\left\{ \mathbf{v}_{1},\cdots,\mathbf{v}_{s}\right\} $.

In general, the first two moments in equilibrium can be achieved through
the ergodicity (\ref{eq:ergodicity}) by averaging the variables of
interest along one solution trajectory, thus we can get the calibration
about the third-order moment feedbacks in the second-order dynamics
by solving the equation (\ref{eq:thrd_equili}) or (\ref{eq:thrd_equili_red}).
Besides, we also find one necessary condition for confirming equilibrium
fidelity for the reduced-order models for the construction of nonlinear
flux term, so that consistent nonlinear flux $Q_{F}$ is guaranteed
in the final steady state
\begin{equation}
Q_{F}^{M}\rightarrow Q_{F,\mathrm{eq}},\quad\mathrm{as}\;t\rightarrow\infty.\label{eq:flux_consis}
\end{equation}
Actually, the idea of estimating the higher-order statistics through
low-order moments has been exploited for several specific models in
\cite{sapsis2013statistically1,sapsis2013statistically,majda2016improving}.
The equilibrium statistics from (\ref{eq:thrd_equili_red}) can efficiently
calibrate the model nonlinear energy transfer mechanism along each
resolved principal direction. However, as external perturbations are
exerted, nonlinear responses will take place with large deviation
from the original equilibrium statistical data calculated in $Q_{F,\mathrm{eq}}$.
Next we will discuss the strategy to calibrate the model sensitivity
to perturbations in a unified way.

\subsection{Linear response theory and kicked responses}

The linear response theory as well as fluctuation-dissipation theorem
(FDT) offers a convenient way to get leading-order statistical linear
approximation about model responses to perturbations \cite{carnevale1991fluctuation,majda2005information,marconi2008fluctuation,majda2011link}.
Consider the general unperturbed system (\ref{eq:abs_formu}), $\delta\mathbf{F}=\mathbf{0}$,
with invariant measure $p_{\mathrm{eq}}\left(\mathbf{u}\right)$,
and an external forcing perturbation in separation with temporal and
spatial variables, 
\[
\delta\mathbf{F}\left(\mathbf{u},t\right)=\mathbf{w}\left(\mathbf{u}\right)\delta f\left(t\right).
\]
Therefore the resulting perturbed probability density $p^{\delta}$
can be asymptotically expanded as the equilibrium and the fluctuation
correction \cite{majda2005information}
\begin{equation}
p^{\delta}\left(t\right)=p_{\mathrm{eq}}+\delta p^{\prime}\left(t\right),\quad\int_{\mathbb{R}^{N}}p_{\mathrm{eq}}\left(\mathbf{u}\right)d\mathbf{u}=1,\;\int_{\mathbb{R}^{N}}\delta p^{\prime}\left(\mathbf{u}\right)d\mathbf{u}=0.\label{eq:pert_dist}
\end{equation}
The equilibrium statistics and leading-order correction to the perturbation
of some functional about the state variable $A\left(\mathbf{u}\right)$
can be formulated as an asymptotic expansion, $\left\langle A\left(\mathbf{u}\right)\right\rangle =\left\langle A\left(\mathbf{u}\right)\right\rangle _{\mathrm{eq}}+\delta\left\langle A\left(\mathbf{u}\right)\right\rangle \left(t\right)+O\left(\delta^{2}\right)$
according to the measure (\ref{eq:pert_dist}) with $\left\langle A\left(\mathbf{u}\right)\right\rangle _{\mathrm{eq}}=\int A\left(\mathbf{u}\right)p_{\mathrm{eq}}\left(\mathbf{u}\right)$
the expectation of $A$ according to equilibrium distribution $p_{\mathrm{eq}}$,
while $\delta\left\langle A\left(\mathbf{u}\right)\right\rangle =\int A\left(\mathbf{u}\right)\delta p^{\prime}\left(\mathbf{u}\right)$
according to $\delta p^{\prime}$. Therefore we get the leading order
responses from
\begin{equation}
\left\langle A\left(\mathbf{u}\right)\right\rangle _{\mathrm{eq}}=\int_{\mathbb{R}^{N}}A\left(\mathbf{u}\right)p_{\mathrm{eq}}\left(\mathbf{u}\right)d\mathbf{u},\quad\delta\left\langle A\left(\mathbf{u}\right)\right\rangle \left(t\right)=\int_{0}^{t}\mathcal{R}_{A}\left(t-s\right)\delta f\left(s\right)ds.\label{eq:LRT}
\end{equation}
Above the pointed-bracket denotes the statistical average under the
solution from Fokker-Planck equation. $\mathcal{R}_{A}\left(t\right)$
is the \emph{linear response operator} corresponding to the functional
$A$, which is calculated through correlation functions in the unperturbed
statistical equilibrium (climate) only
\begin{equation}
\mathcal{R}_{A}\left(t\right)=\left\langle A\left[\mathbf{u}\left(t\right)\right]B\left[\mathbf{u}\left(0\right)\right]\right\rangle _{\mathrm{eq}},\quad B\left(\mathbf{u}\right)=-\frac{\mathrm{div}_{\mathbf{u}}\left(\mathbf{w}p_{\mathrm{eq}}\right)}{p_{\mathrm{eq}}}.\label{eq:LRO}
\end{equation}
The noise in the equations is not needed for FDT to be valid, but
is required to generate the smooth equilibrium measure $p_{\mathrm{eq}}$
for the linear response operator $\mathcal{R}_{A}$. There is even
a rigorous proof of the validity of FDT in this context \cite{hairer2010simple}.
Note that even though in general the linear response operator is difficult
to calculate considering the complicated and unaccessible equilibrium
distribution, a variety of Gaussian approximations for $p_{\mathrm{eq}}$
and improved algorithms have been developed for response via FDT \cite{leith1975climate,majda2005information,majda2010low,majda2010linear}.
FDT can have high skill for the mean response and some skill for the
variance response for a wide variety of turbulent dynamical systems
\cite{majda2010high,abramov2007blended,majda2010low,abramov2012low,lutsko2015applying,gritsun2008climate}.

\subsubsection{Calculate linear response operators through initial kicked responses}

The problem in calculating the leading order response using (\ref{eq:LRO})
is that the equilibrium distribution $p_{\mathrm{eq}}$ is expensive
to calculate for general systems with non-Gaussian features in a high
dimensional phase space. One strategy to approximate the linear response
operator which avoids direct evaluation of $p_{\mathrm{eq}}$ through
the FDT formula but still includes important non-Gaussian statistics
is through the \emph{kicked response} of an unperturbed system to
a perturbation $\delta\mathbf{u}$ of the initial state from the equilibrium
measure, that is, to set the initial distribution with the same variance
but a perturbation in the mean state
\begin{equation}
p\mid_{t=0}=p_{\mathrm{eq}}\left(\mathbf{u}-\delta\mathbf{u}\right)=p_{\mathrm{eq}}-\delta\mathbf{u}\cdot\nabla p_{\mathrm{eq}}+O\left(\delta^{2}\right).\label{eq:init_pert}
\end{equation}
One important advantage of adopting this kicked response strategy
is that higher-order statistics due to nonlinear dynamics will not
be ignored (compared with the other linearized strategy using only
Gaussian statistics \cite{majda2010high}). Then the kicked response
theory gives the following proposition \cite{majda2005information,majda2011improving}
for calculating the linear response operator:
\begin{proposition}\label{prop1}
For $\delta$ small enough, the linear response operator $\mathcal{R}_{A}\left(t\right)$
can be calculated by solving the unperturbed system (\ref{eq:abs_formu})
with a perturbed initial distribution in (\ref{eq:init_pert}). Therefore,
the linear response operator can be achieved through
\begin{equation}
\delta\mathcal{R}_{A}\left(t\right)\equiv\delta\mathbf{u}\cdot\mathcal{R}_{A}=\int A\left(\mathbf{u}\right)\delta p^{\prime}+O\left(\delta^{2}\right).\label{eq:kicked_resp}
\end{equation}
Here $\delta p^{\prime}$ is the resulting leading order expansion
of the transient density function from unperturbed dynamics using
initial value perturbation. From the formula in (\ref{eq:kicked_resp}),
the response operators for the mean and variance can be achieved from
the perturbation part of the probability density $\delta p^{\prime}$.
And this density function can also be used to measure the information
distance between the truth and imperfect models in the training phase.
\end{proposition}
The proof of the above Proposition \ref{prop1} is a direct application of \emph{Duhamel's
principle} to the corresponding Fokker-Planck equation with forcing
perturbations \cite{majda2005information}. Thus the variability in
the external forcing can be transferred to the perturbations in initial
values. More importantly, the kicked response formulation (\ref{eq:kicked_resp})
with initial mean state perturbation (\ref{eq:init_pert}) is independent
of the specific perturbation forms. Thus the operator $\mathcal{R}_{A}$
describes the inherent dynamical mechanisms of the system. We summarize
the practical strategies to calculate the kicked response operators
for the mean and variance from (\ref{eq:kicked_resp}) in Appendix
A.

\subsection{Empirical information theory for measuring imperfect model errors}

\subsubsection{Empirical information theory for leading order statistics}

The empirical information theory \cite{jaynes1957information,majda2005information}
builds the least biased probability measure consistent with the leading
order measurements of the true perfect system. Information theory
is often used in statistical science for imperfect model selection
\cite{burnham2003model}. A natural way to measure the lack of information
in one probability density from the imperfect model, $p^{M}$, compared
with the true probability density, $p$, is through the \emph{relative
entropy} or \emph{information distance} \cite{kullback1951information,majda2005information},
given by 
\begin{equation}
\mathcal{P}\left(p,p^{M}\right)=\int p\log\frac{p}{p^{M}}.\label{eq:rela_entropy}
\end{equation}
Despite the lack of symmetry in its arguments (that is, $\mathcal{P}\left(p_{1},p_{2}\right)\neq\mathcal{P}\left(p_{2},p_{1}\right)$
in general), the relative entropy, $\mathcal{P}\left(p,p^{M}\right)$
provides an attractive framework for assessing model error like a
probabilistic metric. Importantly, the following two crucial features
are satisfied in the relative entropy: (i) $\mathcal{P}\left(p,p^{M}\right)\geq0$,
and the equality holds if and only if $p=p^{M}$; and (ii) it is invariant
under any invertible change of variables. The most practical setup
for utilizing the framework of empirical information theory arises
when only the Gaussian statistics of the distributions are considered.
By only comparing the first two moments of the density functions,
we get the following fact \cite{majda2002mathematical}:
\begin{proposition}\label{prop2}
If the probability density functions $p$, $p^{M}$ contain only the
first two moments, that is, $p\sim\mathcal{N}\left(\bar{\mathbf{u}},R\right)$
and $p^{M}\sim\mathcal{N}\left(\bar{\mathbf{u}}_{M},R_{M}\right)$,
the relative entropy in (\ref{eq:rela_entropy}) has the explicit
formula
\begin{equation}
\mathcal{P}\left(p,p^{M}\right)=\frac{1}{2}\left(\bar{\mathbf{u}}-\bar{\mathbf{u}}_{M}\right)^{T}R_{M}^{-1}\left(\bar{\mathbf{u}}-\bar{\mathbf{u}}_{M}\right)+\frac{1}{2}\left(\mathrm{tr}\left(RR_{M}^{-1}\right)-N-\log\det\left(RR_{M}^{-1}\right)\right).\label{eq:entro_Gau}
\end{equation}
The first term on the right hand side of (\ref{eq:entro_Gau}) is
called the \emph{signal}, reflecting the model error in the mean but
weighted by the inverse of the model variance $R_{M}$; whereas the
second term is the \emph{dispersion,} involving only the model error
covariance ratio $RR_{M}^{-1}$, measuring the differences in the
covariance matrices. 
\end{proposition}
Above usually we will use $p$ to denote the probability distribution
of the perfect model, which is actually unknown. Nevertheless, we
can construct the measure of the perfect model $p_{L}$ using $L$
measurements of the true system. Consider the imperfect model prediction
with its associated probability density $p_{L}^{M}$, the definition
of relative entropy (\ref{eq:rela_entropy}) facilitates the practical
calculation \cite{kleeman2002quantifying,majda2011improving,majda2011link,majda2016introduction,majda2012lessons}
\begin{eqnarray*}
\mathcal{P}\left(p,p^{M}\right) & = & \mathcal{P}\left(p,p_{L}\right)+\mathcal{P}\left(p_{L},p_{L}^{M}\right)\\
 & = & \left[\mathcal{S}\left(p_{L}\right)-\mathcal{S}\left(p\right)\right]+\mathcal{P}\left(p_{L},p_{L}^{M}\right).
\end{eqnarray*}
The entropy difference $\mathcal{S}\left(p_{L}\right)-\mathcal{S}\left(p\right)$
precisely measures an intrinsic error from $L$ measurements of the
perfect system, and this is a simple example of an information barrier
for any imperfect model based on $L$ measurements for calibration.
With the measurements $L$ representing the first two moments, the
Gaussian approximation (\ref{eq:entro_Gau}) can be used to estimate
the information error $\mathcal{P}\left(p_{L},p_{L}^{M}\right)$ considering
only the first $L$ statistical measurements (in practice, it is usually
the measurements about the statistical mean and covariance).

\subsubsection{Climate information barrier in single point statistics in homogeneous systems}

Here as one example, we briefly illustrate the inherent information
barrier in special\emph{ homogeneous systems} like the L-96 model
in (\ref{eq:L96}) (see \cite{majda2016improving}) with \emph{uniform
damping and forcing} using the above relative entropy metric. Of particular
interest in both theory and applications, the statistical mean and
variance at each individual grid point \cite{majda2006nonlinear,majda2016improving,majda2003systematic,delsole2010model}
play an important role as key statistical quantities to predict. In
climate science, these might be the mean and variance of the surface
temperature at every grid point. The single point mean $\bar{u}_{\mathrm{1pt}}$
and single point variance $r_{\mathrm{1pt}}$ can be defined by averaging
each grid component with presumed homogeneity, that is,
\begin{equation}
\bar{u}_{\mathrm{1pt}}=\frac{1}{N}\sum_{j=1}^{N}\bar{u}_{j},\quad r_{\mathrm{1pt}}=\frac{1}{N}\mathrm{tr}R.\label{eq:one-point}
\end{equation}
For simplicity in representation, we assume homogeneous damping, $D=-dI$,
and forcing, $F=fI$, in the energy dynamics (\ref{eq:energy_conservation}),
thus the total statistical energy equations for the true model $E$
and reduced-order model approximation $E^{M}$ become
\begin{eqnarray*}
\frac{dE}{dt} & = & -2dE+f\bar{u}_{\mathrm{1pt}}+\frac{1}{2}\mathrm{tr}Q_{\sigma},\\
\frac{dE^{M}}{dt} & = & -2dE^{M}+f\bar{u}_{\mathrm{1pt}}^{M}+\frac{1}{2}\mathrm{tr}Q_{F}^{M}+\frac{1}{2}\mathrm{tr}Q_{\sigma}.
\end{eqnarray*}
Above the statistical energy can be defined through the single point
statistics as 
\[
E=\frac{N}{2}\left(\bar{u}_{\mathrm{1pt}}^{2}+r_{\mathrm{1pt}}\right),\quad E^{M}=\frac{N}{2}\left(\left(\bar{u}_{\mathrm{1pt}}^{M}\right)^{2}+r_{\mathrm{1pt}}^{M}\right),
\]
where we assume homogeneity in the first two moments. The last part
on the right hand side of $E^{M}$ equation comes from the error in
the approximation for nonlinear flux $Q_{F}^{M}$. Taking the difference
of the above two equations for $E$ and $E^{M}$ and using Gronwall's
inequality gives the error in the total statistical energy $\delta E=E-E^{M}$
\[
\left\Vert \delta E\right\Vert \leq\tilde{C}_{0}\left\Vert \delta\bar{u}_{\mathrm{1pt}}\right\Vert +C_{1}\left\Vert \mathrm{tr}Q_{F}^{M}\right\Vert ,
\]
and the definition about the statistical energy offers the estimation
\[
\left\Vert \delta E\right\Vert \geq\left\Vert \delta r_{\mathrm{1pt}}\right\Vert -N\left\Vert \bar{u}_{\mathrm{1pt}}\right\Vert \left\Vert \delta\bar{u}_{\mathrm{1pt}}\right\Vert .
\]
The error estimation for the single point variance through the error
from the mean and nonlinear flux by combining the above two inequalities
\begin{equation}
\left\Vert \delta r_{\mathrm{1pt}}\right\Vert \leq C_{0}\left\Vert \delta\bar{u}_{\mathrm{1pt}}\right\Vert +C_{1}\left\Vert \mathrm{tr}Q_{F}^{M}\right\Vert ,\label{eq:single_consistency}
\end{equation}
with $C_{0},C_{1}$ constants. The inequality in (\ref{eq:single_consistency})
illustrates that the error in the second-order statistics $\delta r_{\mathrm{1pt}}$
can be controlled by the error in the first-order mean with a good
approximation for the nonlinear flux term $Q_{F}^{M}$. Similar special
results for the 40-dimensional L-96 model are described in \cite{majda2016improving}.

\begin{figure}
\centering
\subfloat{\includegraphics[scale=0.32]{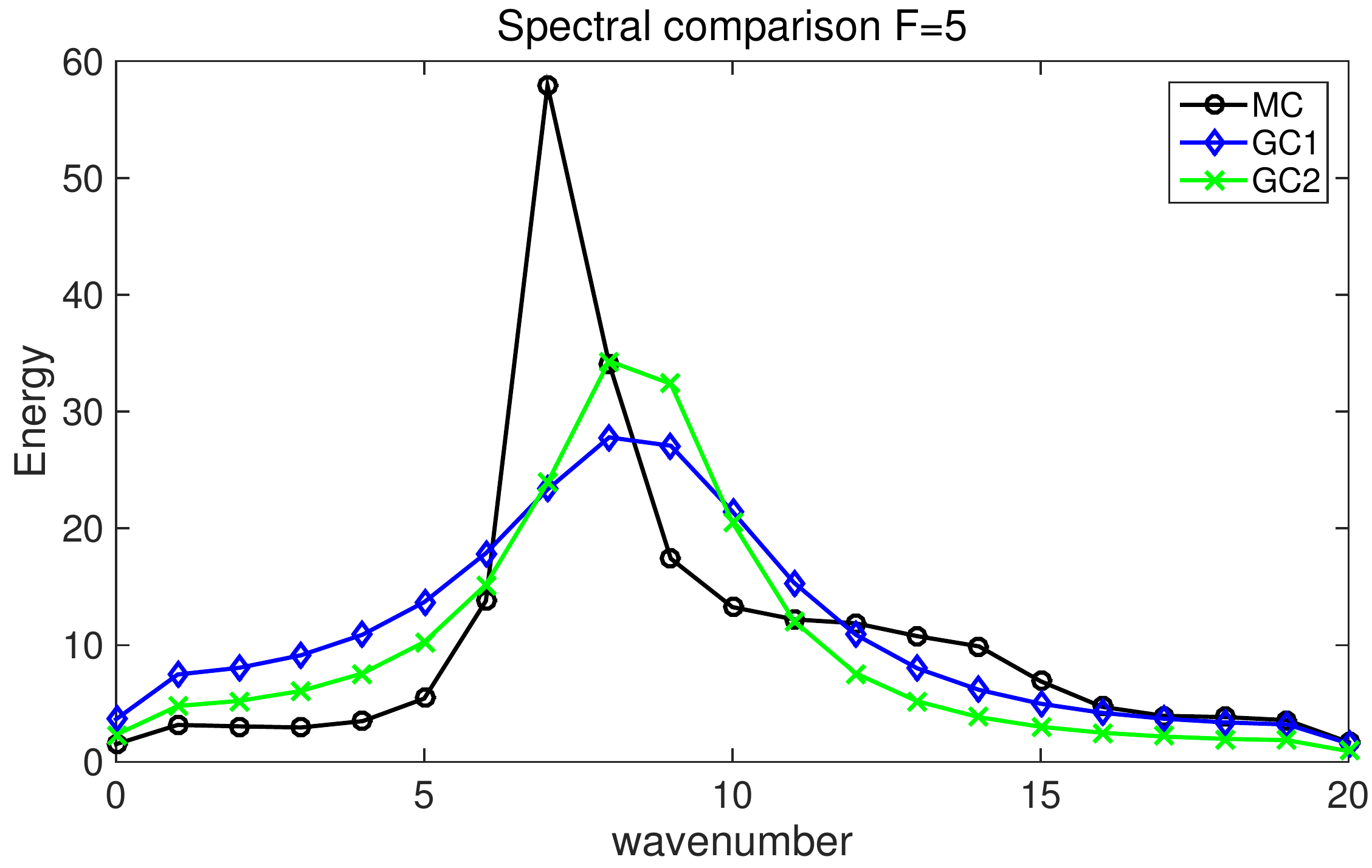}}\subfloat{\includegraphics[scale=0.32]{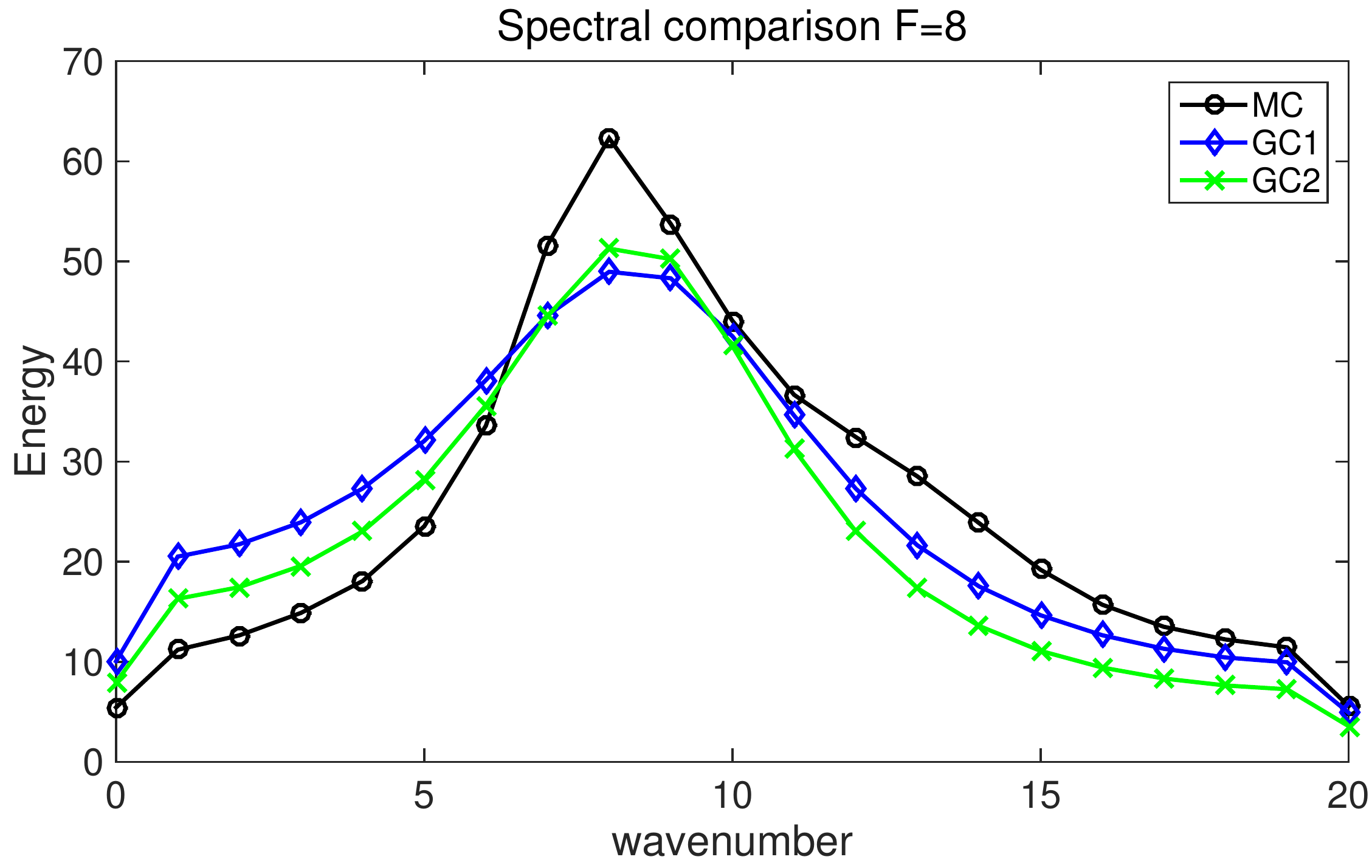}}

\caption{Information barriers for imperfect closure models with only consistent
equilibrium single point statistics $\left(\bar{u}_{\mathrm{1pt}},r_{\mathrm{1pt}}\right)$
in the L-96 system. The steady state variances under Fourier basis
from the two imperfect model results are compared with the truth from
Monte-Carlo simulations in two typical dynamical regimes $F=5$ and
$F=8$.\label{fig:spec_1pt}}
\end{figure}

With the help of the relative entropy, we can first illustrate the
inherent information barrier with the single-point statistics approximation
(\ref{eq:one-point}). It will be shown even with consistent single-point
statistics in $\left(\bar{u}_{\mathrm{1pt}},r_{\mathrm{1pt}}\right)$,
large errors may still appear due to the lack of consideration in
the covariances between different modes. Through the definition in
(\ref{eq:rela_entropy}) (and referring to Proposition 4.1 of \cite{majda2002mathematical}),
the relative entropy between the truth $p$ and imperfect model single-point
statistics $p_{\mathrm{1pt}}^{M}$ has the form
\begin{equation}
\mathcal{P}\left(p,p_{\mathrm{1pt}}^{M}\right)=\left[\mathcal{S}\left(p_{G}\right)-\mathcal{S}\left(p\right)\right]+\mathcal{P}\left(p_{G},p_{\mathrm{1pt}}^{G}\right)+\mathcal{P}\left(p_{\mathrm{1pt}}^{G},p_{\mathrm{1pt}}^{M}\right).\label{eq:info_expan}
\end{equation}
Above, $p_{G}$ is the Gaussian fit for the original probability distribution
$p$ with the same mean and covariance from the truth; $p_{\mathrm{1pt}}^{G}=\mathscr{N}\left(\bar{u}_{\mathrm{1pt}},r_{\mathrm{1pt}}\right)$
is the single-point approximation for the true system, and $p_{\mathrm{1pt}}^{M}=\mathscr{N}\left(\bar{u}_{M},r_{M}\right)$
is the reduced-order model prediction from the imperfect model with
consistent single-point statistics. The first part on the right hand
side of (\ref{eq:info_expan}) is the intrinsic information barrier
in Gaussian approximation. And the third part with homogeneous assumption
of the system will vanish (or at least be minimized) due to the single-point
statistics fidelity from (\ref{eq:single_consistency}). The error
from single point approximation (and ignoring the cross-covariance)
then comes only from the information barrier in marginal approximation
$\mathcal{P}\left(p_{G},p_{\mathrm{1pt}}^{G}\right)$ as shown in
the second part on the right hand side of (\ref{eq:info_expan}).
Simple calculation using the formula (\ref{eq:entro_Gau}) and Jensen\textquoteright s
inequality \cite{majda2016improving} yields the estimation for the
information barrier in single-point approximation
\begin{equation}
\mathcal{P}\left(p_{G},p_{\mathrm{1pt}}^{G}\right)=N\log\left[\frac{\left(\sum_{j=1}^{N}r_{j}\right)/N}{\left(\prod_{j=1}^{N}r_{j}\right)^{1/N}}\right]\sim r_{\mathrm{1pt}}^{-1}\left(\sigma_{\max}-\sigma_{\min}\right)^{2},\label{eq:info_barrier}
\end{equation}
where $r_{j}$ is the variance in the spectral modes, and $\sigma_{\max}^{2}=\max\left\{ r_{j}\right\} ,\sigma_{\min}^{2}=\min\left\{ r_{j}\right\} $
are the largest and smallest variance. In Figure \ref{fig:spec_1pt},
we demonstrate this information barrier for imperfect models with
exact one-point statistics $\left(\bar{u}_{\mathrm{1pt}},r_{\mathrm{1pt}}\right)$
consistency for the L-96 model with $F=5,8$. Large errors in the
statistical steady state spectra (thus information barrier for these
models) exist for each individual mode for both dynamical regimes
$F=5$ (weakly chaotic) and $F=8$ (strongly chaotic), consistent
with what we have calculated from (\ref{eq:info_barrier}) for single
point statistics.

The barrier from (\ref{eq:info_barrier}) could become significant
considering the gap between the largest and smallest variances due
to common decaying energy spectra in turbulent systems. See \cite{majda2016improving}
for a detailed example for the L-96 model. This information barrier
can only be overcome by introducing more careful calibration about
the dynamics in each eigen-direction of the system individually.

\subsubsection{Dynamical calibration for imperfect model improvement}

The prediction skill of imperfect models can be improved by comparing
the information distance through the linear response operator with
the true model. The following fact offers a convenient way to measure
the lack of information in the perturbed imperfect model requiring
only knowledge of linear responses for the mean and variance $\delta\bar{\mathbf{u}}\equiv\delta\mathcal{R}_{\mathbf{u}},\delta R\equiv\delta\mathcal{R}_{\left(\mathbf{u}-\bar{\mathbf{u}}\right)^{2}}$.
For this result, it is important to tune the imperfect model to satisfy
equilibrium model fidelity, 
\[
\mathcal{P}\left(p_{G}\left(\mathbf{u}\right),p_{G}^{M}\left(\mathbf{u}\right)\right)=0,
\]
in the first place. Statistical equilibrium fidelity is a natural
necessary condition to tune the mean and variance of the imperfect
model to match those of the perfect model; it is far from a sufficient
condition \cite{majda2016introduction,majda2011improving,majda2011link}.
Using simplified assumptions with block-diagonal covariance matrices
$R=\mathrm{diag}\left(R_{k}\right)$ and equilibrium model fidelity
$\mathcal{P}\left(p_{G},p_{G}^{M}\right)=0$, the relative entropy
in (\ref{eq:rela_entropy}) between the true perturbed density $p_{\delta}$
and the perturbed model density $p_{\delta}^{M}$ with small perturbation
$\delta$ can be expanded componentwisely as the following proposition:
\begin{proposition}\label{prop3}
Under assumptions with block-diagonal covariance matrices $R=\mathrm{diag}\left(R_{k}\right)$
and equilibrium model fidelity $\mathcal{P}\left(p_{G},p_{G}^{M}\right)=0$,
the relative entropy in (\ref{eq:entro_Gau}) between perturbed model
density $p_{\delta}^{M}$ and the true perturbed density $p_{\delta}$
with small perturbation $\delta$ can be expanded componentwisely
as
\begin{eqnarray}
\mathcal{P}\left(p_{\delta},p_{\delta}^{M}\right) & = & \mathcal{S}\left(p_{G,\delta}\right)-\mathcal{S}\left(p_{\delta}\right)\nonumber \\
 &  & +\frac{1}{2}\sum_{k}\left(\delta\bar{u}_{k}-\delta\bar{u}_{M,k}\right)R_{k}^{-1}\left(\delta\bar{u}_{k}-\delta\bar{u}_{M,k}\right)\nonumber \\
 &  & +\frac{1}{4}\sum_{k}R_{k}^{-2}\left(\delta R_{k}-\delta R_{M,k}\right)^{2}+O\left(\delta^{3}\right).\label{eq:entro_pert}
\end{eqnarray}
Here in the first line $\mathcal{S}\left(p_{G,\delta}\right)-\mathcal{S}\left(p_{\delta}\right)$
is the intrinsic error from Gaussian approximation of the system.
$R_{k}$ is the equilibrium variance in $k$-th component, and $\delta\bar{u}_{k}$
and $\delta R_{k}$ are the linear response operators for the mean
and variance in $k$-th component.
\end{proposition}
Detailed derivation about this result is shown in \cite{majda2011link}.
The inherent information error from the first row of (\ref{eq:entro_pert})
is due to the measurement in only first two order of moments, and
is independent of the specific imperfect model structures. As a result,
this component, $\mathcal{S}\left(p_{G,\delta}\right)-\mathcal{S}\left(p_{\delta}\right)$,
can be viewed as a constant and does not need to be calculated in
the optimization procedure. The second row of the information distance
(\ref{eq:entro_pert}) illustrates the \emph{signal error} from the
estimation about the mean responses, while the third row is the \emph{dispersion
error} for the errors from the variance responses.

\

The above Proposition \ref{prop3} about empirical information theory and linear
response theory together provides a convenient and unambiguous way
of improving the performance of imperfect models in terms of increasing
their model sensitivity regardless of the specific form of external
perturbations $\delta\mathbf{f}^{\prime}$. The formula (\ref{eq:kicked_resp})
in Proposition \ref{prop1} as well as (\ref{eq:LRT}) illustrates that the skill
of an imperfect model in predicting forced changes to perturbations
with general external forcing is directly linked to the model's skill
in estimating the linear response operators $\mathcal{R}_{A}$ for
the mean and variances (that is, use the functional $A=\mathbf{u},\left(\mathbf{u}-\bar{\mathbf{u}}\right)^{2}$
in calculating the linear response operators) in a suitably weighted
fashion as dictated by information theory (\ref{eq:entro_pert}).
This offers us useful hints of training imperfect models for optimal
responses for the mean and variance in a universal sense. From the
linear response theory, it shows that the system's responses to various
external perturbations can be approximated by a convolution with the
linear response operator $\mathcal{R}_{A}$ (which is only related
to the statistics in the unperturbed equilibrium steady state). It
is reasonable to claim that an imperfect model with precise prediction
of this linear response operator should possess uniformly good sensitivity
to different kinds of perturbations. On the other hand, the response
operator can be calculated easily by the transient state distribution
density function using the kicked response formula as in (\ref{eq:kicked_resp}).
Considering all these good features of the linear response operator,
the information barrier due to model sensitivity to perturbations
can be overcome by minimizing the information error in the imperfect
model kicked response distribution relative to the true response from
observation data. 

To summarize, consider a class of imperfect models, $\mathcal{M}$.
The optimal model $M^{*}\in\mathcal{M}$ that ensures best information
consistent responses to various kinds of perturbations is characterized
with the smallest additional information in the linear response operator
$\mathcal{R}_{A}$ among all the imperfect models, such that
\begin{equation}
\left\Vert \mathcal{P}\left(p_{\delta},p_{\delta}^{M^{*}}\right)\right\Vert _{L^{1}\left(\left[0,T\right]\right)}=\min_{M\in\mathcal{M}}\left\Vert \mathcal{P}\left(p_{\delta},p_{\delta}^{M}\right)\right\Vert _{L^{1}\left(\left[0,T\right]\right)},\label{eq:opt_model}
\end{equation}
where $p_{\delta}^{M}$ can be achieved through a kicked response
procedure (\ref{eq:kicked_resp}) in the training phase compared with
the actual observed data $p_{\delta}$ in nature, and the information
distance between perturbed responses $\mathcal{P}\left(p_{\delta},p_{\delta}^{M}\right)$
can be calculated with ease through the expansion formula (\ref{eq:entro_pert}).
The information distance $\mathcal{P}\left(p_{\delta}\left(t\right),p_{\delta}^{M}\left(t\right)\right)$
is measured at each time instant, so the entire error is averaged
under the $L^{1}$-norm inside a proper time window $\left[0,T\right]$
before the linear response function decays back to zero.

\section{Reduced-Order Statistical Models for the Turbulent Systems}

Previously in Section 2, the general idea about finding the optimal
imperfect model is proposed according to the statistical theories
and information distance metric. And we have shown the basic theoretical
tools that can help construct the reduced-order statistical approximations
and illustrate the information barriers due to these approximations.
Then it is important to construct the explicit forms of the reduced-order
models according to the exact dynamics for the mean and covariance
in (\ref{eq:mean_dyn}) and (\ref{eq:cov_dyn}). Generally the statistical
model for the leading two moments can be formulated in the full phase
space as\addtocounter{equation}{0}\begin{subequations}
\begin{eqnarray}
\frac{d\bar{\mathbf{u}}_{M}}{dt} & = & \left(L+D\right)\bar{\mathbf{u}}_{M}+B\left(\bar{\mathbf{u}}_{M},\bar{\mathbf{u}}_{M}\right)+R_{M,ij}B\left(\mathbf{e}_{i},\mathbf{e}_{j}\right)+\mathbf{F},\label{eq:mean_red}\\
\frac{dR_{M}}{dt} & = & L_{v}\left(\bar{\mathbf{u}}_{M}\right)R_{M}+R_{M}L_{v}^{*}\left(\bar{\mathbf{u}}_{M}\right)+Q_{F}^{M}+Q_{\sigma},\label{eq:cov_red}
\end{eqnarray}
\end{subequations}where $\bar{\mathbf{u}}_{M}\in\mathbb{R}^{N}$
is the model approximated mean, and $R_{M}$ is the $N\times N$ full
order covariance matrix about the fluctuation state variable $\mathbf{u}^{\prime}\in\mathbb{R}^{N}$.
Comparing with the original statistical dynamics (\ref{eq:mean_dyn})
and (\ref{eq:cov_dyn}), the most expensive but crucial part comes
from the nonlinear flux term $Q_{F}$ in
(\ref{eq:nonlinear_flux}) where important third-order
moments are included representing the nonlinear interactions between
different modes. Therefore the key issue in this section is to
construct a judicious estimation about this nonlinear interaction
term $Q_{F}^{M}$ in the statistical closure models. Here the basic
idea is to start with the simplest possible imperfect model and
compare the advantages and limitations of different levels of imperfect
models due to different degrees of approximation and model calibration,
and finally check how the theories from previous sections can help with improving
the model prediction skill, especially the model sensitivity to various
perturbations.

\subsection{A hierarchy of statistical reduced-order modeling ideas based on stochastic models}

We may consider the statistical closure ideas by taking another look
at the dynamics for stochastic coefficients
\[
\frac{dZ_{i}}{dt}=Z_{j}\left[\left(L+D\right)\mathbf{e}_{j}+B\left(\bar{\mathbf{u}},\mathbf{e}_{j}\right)+B\left(\mathbf{e}_{j},\bar{\mathbf{u}}\right)\right]\cdot\mathbf{e}_{i}+B\left(\mathbf{u}^{\prime},\mathbf{u}^{\prime}\right)\cdot\mathbf{e}_{i}+\sigma\left(t\right)\dot{\mathbf{W}}\left(t;\omega\right)\cdot\mathbf{e}_{i}.
\]
Major nonlinearity comes from the term above representing interactions
between different fluctuation modes $B\left(\mathbf{u}^{\prime},\mathbf{u}^{\prime}\right)\cdot\mathbf{e}_{i}$.
The first idea here is to model the effect of the nonlinear energy
transfers on each mode by adding additional damping $d_{M,i}$ balancing
the linearly unstable character of these modes, and adding additional
(white) stochastic excitation with standard deviation $\sigma_{M,i}$
which will model the energy received by the stable modes. We want
to constrain ourselves to second order models concentrating on the
mean and variance and maintaining the computational expense in a low
level, hence the additional parts $d_{M,i},\sigma_{M,i}$ only include
statistics up to second order moments. Specifically we replace this
high-order nonlinear term by
\[
B\left(\mathbf{u}^{\prime},\mathbf{u}^{\prime}\right)\cdot\mathbf{e}_{i}\equiv\sum_{m,n}Z_{m}Z_{n}B\left(\mathbf{e}_{m},\mathbf{e}_{n}\right)\cdot\mathbf{e}_{i}\rightarrow-d_{M,i}\left(\mathrm{tr}R\right)Z_{i}+\sigma_{M,i}\left(\mathrm{tr}R\right)\dot{W}_{i},
\]
with $\mathrm{tr}R=\sum_{j}\left\langle Z_{j}Z_{j}^{*}\right\rangle $
for measuring the total energy (variance) structure in the system.
Corresponding to the statistical equations, the nonlinear flux $Q_{F}$ representing the higher-order interactions
is replaced by 
\begin{equation}
Q_{F}^{M}=Q_{F-}^{M}+Q_{F+}^{M}=-D_{M}\left(R\right)R_{M}-R_{M}D_{M}^{*}\left(R\right)+\Sigma_{M}\left(R\right).\label{eq:numer_flux}
\end{equation}
In (\ref{eq:numer_flux}), $\left(D_{M},\Sigma_{M}\right)$ are $N\times N$
matrices that replace the original nonlinear unstable and stable effects
from the original dynamics. Here $Q_{F-}^{M}=-D_{M}\left(R\right)R_{M}-R_{M}D_{M}^{*}\left(R\right)$
represents the additional damping effect to stabilize the unstable
modes with positive Lyapunov coefficients, while $Q_{F+}^{M}=\Sigma_{M,k}\left(R\right)$
is the positive-definite additional noise to compensate for the overdamped
modes. Now the problem is converted to finding expressions for $D_{M}$
and $\Sigma_{M}$. In the following by gradually adding more detailed
characterization about the statistical dynamical model we display
the general procedure of constructing a hierarchy of the closure methods
step by step. Below is a review about several model closure ideas
\cite{majda2016introduction,sapsis2013statistically,majda2016improving,qi2016}
with increasing complexity:
\begin{enumerate}
\item \emph{Quasilinear Gaussian closure model}: The simplest approximation
for the closure methods at the first stage should be simply neglecting
the nonlinear part entirely \cite{epstein1969stochastic,fleming1971stochastic,srinivasan2012zonostrophic}.
That is, set
\begin{equation}
D_{M}\left(R\right)\equiv0,\quad\Sigma_{M}\left(R\right)\equiv0,\qquad Q_{F}^{\mathrm{QG}}\equiv0.\label{eq:qGC}
\end{equation}
Thus the nonlinear energy transfer mechanism will be entirely neglected
in this Gaussian closure model. This is the similar idea in the \emph{eddy-damped Markovian model} where the moment hierarchy is closed at the level of second moments with Gaussian assumption and a much larger \emph{eddy-damped} parameter is introduced to replace the molecular viscosity (see Chapter 5 of \cite{salmon1998lectures} and \cite{lesieur2012turbulence} for details). Obviously this crude Gaussian approximation
will not work well in general due to the cutoff of the energy flow
when strong nonlinear interactions between modes occur. Actually,
the deficiency of this crude approximation have been shown under the
L-96 framework, and in final equilibrium state there exists only one
active mode with critical wavenumber \cite{sapsis2013statistically1,majda2016improving}.
Such closures are only useful in the weakly nonlinear case where the
quasi-linear effects are dominant.
\item \emph{Models with consistent equilibrium statistics}: Next the strategy is to construct the simplest closure model with consistent equilibrium
statistics. So the direct way is to choose constant damping and noise
term at most scaled with the total variance. We propose two possible
choices as in \cite{majda2016improving} for the damping and noise in (\ref{eq:numer_flux}) below.\\
\textbf{Gaussian closure 1 (GC1)}: let
\begin{equation}
D_{M}\left(R\right)=\epsilon_{M}I_{N}\equiv\mathrm{const.},\quad\Sigma_{M}\left(R\right)=\sigma_{M}^{2}I_{N}\equiv\mathrm{const.},\qquad Q_{F}^{\mathrm{GC1}}=-\left(\epsilon_{M}R+R\epsilon_{M}\right)+\sigma_{M}^{2}I_{N};\label{eq:GC1}
\end{equation}
\\
\textbf{Gaussian closure 2 (GC2)}: let
\begin{equation}
D_{M}\left(R\right)=\epsilon_{M}\left(\frac{\mathrm{tr}R}{\mathrm{tr}R_{\mathrm{eq}}}\right)^{1/2}I_{N},\quad\Sigma_{M}\left(R\right)=\sigma_{M}^{2}\left(\frac{\mathrm{tr}R}{\mathrm{tr}R_{\mathrm{eq}}}\right)^{3/2}I_{N},\quad Q_{F}^{\mathrm{GC2}}=-\left(\frac{\mathrm{tr}R}{\mathrm{tr}R_{\mathrm{eq}}}\right)^{1/2}\left(\epsilon_{M}R+R\epsilon_{M}\right)+\sigma_{M}^{2}\left(\frac{\mathrm{tr}R}{\mathrm{tr}R_{\mathrm{eq}}}\right)^{3/2}I_{N}.\label{eq:GC2}
\end{equation}
Above only two scalar model parameters $\left(\epsilon_{M},\sigma_{M}\right)$ are introduced,
and $I_{N}$ represents the $N\times N$ identity matrix. GC1 is the
familiar strategy of adding constant damping and white noise forcing
to represent nonlinear interaction; GC2 scales with the total variance
$\mathrm{tr}R$ (or total statistical energy) so that the model sensitivity
can be further improved as the system is perturbed. From both GC1
and GC2, we introduce uniform additional damping rate for each spectral
mode controlled by a single scalar parameter $\epsilon_{M}$; while
the additional noise with variance $\sigma_{M}^{2}$ is added to make
sure climate fidelity in equilibrium (we leave the detailed discussion
for climate fidelity in Section 3.2.1).\\
The statistical model closure $Q_{F}^{M}$ is used to approximate
the third-order moments in the true dynamics, thus the exponents of
the total energy $\mathrm{tr}R$ in GC2 should be consistent in scaling
dimension. In the positive-definite part $Q_{F+}^{M}$, it calibrates
the rate of energy injected into the spectral mode due to nonlinear
effect in the order $\left|u^{\prime}\right|^{3}$.
The factor scales with the total energy with exponent $3/2$ so that
the corrections keep consistent with the third-order moment approximations;
In the negative damping rate $Q_{F-}^{M}$, the scaling function is
used to characterize the amount of energy that flows out the spectral
mode due to nonlinear interactions. Scaling factor with a square-root
of the total energy with exponent $1/2$ is applied for this damping
rate multiplying the variance in order $\left|u^{\prime}\right|^{2}$
to make it consistent in scaling dimension with third moments.
\item \emph{Modified quasi-Gaussian closure with equilibrium statistics}:
In this modified quasi-Gaussian closure model originally proposed
in \cite{sapsis2013statistically1,sapsis2013statistically}, we exploit
more about the true nonlinear energy transfer mechanism from the equilibrium
statistical information. Thus the additional damping and noise proposed
like before are calibrated through the equilibrium nonlinear flux
by letting
\begin{equation}
D_{M}\left(R\right)=-N_{M,\mathrm{eq}},\quad\Sigma_{M}\left(R\right)=Q_{F,\mathrm{eq}}^{+},\qquad Q_{F}^{\mathrm{MQG}}=-\left(N_{M}R+RN_{M}^{*}\right)+Q_{F}^{+};.\label{eq:MQG}
\end{equation}
$N_{M,\mathrm{eq}}$ is the effective damping from equilibrium, and
$Q_{F,\mathrm{eq}}^{+}$ is the effective noise from the positive-definite
component. Unperturbed equilibrium statistics in the nonlinear flux
$Q_{F,\mathrm{eq}}$ are used to calibrate the higher-order moments
as additional energy sink and source. The true equilibrium higher-order
flux can be calculated without error from first and second order moments
in $\left(\bar{\mathbf{u}}_{\mathrm{eq}},R_{\mathrm{eq}}\right)$
from the unperturbed true dynamics (\ref{eq:cov_dyn}) in steady state
following the steady state statistical solution relation (\ref{eq:thrd_equili})
as discussed in Section 2.2
\begin{equation}
Q_{F,\mathrm{eq}}=Q_{F,\mathrm{eq}}^{-}+Q_{F,\mathrm{eq}}^{+}=-L_{v}\left(\bar{\mathbf{u}}_{\mathrm{eq}}\right)R_{\mathrm{eq}}-R_{\mathrm{eq}}L_{v}^{*}\left(\bar{\mathbf{u}}_{\mathrm{eq}}\right)-Q_{\sigma},\quad N_{M,\mathrm{eq}}=\frac{1}{2}Q_{F,\mathrm{eq}}^{-}R_{\mathrm{eq}}^{-1}.\label{eq:MQG_cali}
\end{equation}
$Q_{F,\mathrm{eq}}^{-},Q_{F,\mathrm{eq}}^{+}$ are the negative and
positive definite components in the unperturbed equilibrium nonlinear
flux $Q_{F,\mathrm{eq}}$. Since exact model statistics are used in
the imperfect model approximations, the true mechanism in the nonlinear
energy transfer can be modeled under this first correction form. This
is the similar idea used for measuring higher-order interactions in
\cite{sapsis2013statistically}, where more sophisticated and expensive
calibrations are required to make that model work there.
\end{enumerate}

\subsection{A reduced-order statistical energy model with optimal consistency and sensitivity}

The above closure model ideas, especially (\ref{eq:GC1}), (\ref{eq:GC2}),
and (\ref{eq:MQG}), have advantages of their own. Models in (\ref{eq:GC1})
and (\ref{eq:GC2}) are simple and efficient to construct with consistent
equilibrium consistency, while (\ref{eq:MQG}) involves the true information
about the higher-order statistics in equilibrium so that the energy
mechanism can be characterized well. The validity of these approaches
has been tested and compared from several papers \cite{sapsis2013statistically1,sapsis2013statistically,majda2016improving}
using the simplified triad model and L-96 model. Still when it comes
to the more complicated and realistic flow systems like the quasi-geostrophic
equations, more detailed calibration for model consistency and sensitivity
is required to achieve the optimal performance. A preferred approach
for the nonlinear flux $Q_{F}^{M}$ combining both the detailed model
energy mechanism and control over model sensitivity is proposed in
the form
\begin{equation}
Q_{F}^{M}=Q_{F}^{M,-}+Q_{F}^{M,+}=f_{1}\left(R\right)\left[-\left(N_{M,\mathrm{eq}}+d_{M}I_{N}\right)R_{M}\right]+f_{2}\left(R\right)\left[Q_{F,\mathrm{eq}}^{+}+\Sigma_{M}\right].\label{eq:model_nonflux}
\end{equation}
The closure form (\ref{eq:model_nonflux}) consists of three indispensable
components:
\begin{description}
\item [{i)}] \emph{Higher-order corrections from equilibrium statistics}:
In the first part of the correction using the damping and noise operator
as $\left(N_{M,\mathrm{eq}},Q_{F,\mathrm{eq}}^{+}\right)$, unperturbed
equilibrium statistics in the nonlinear flux $Q_{F,\mathrm{eq}}$
are used to calibrate the higher-order moments as additional energy
sink and source following the procedure in (\ref{eq:MQG}). Therefore
the equilibrium statistics can be guaranteed to be consistent with
the truth, and the true energy mechanism can be restored;
\item [{ii)}] \emph{Additional damping and noise to model changes in nonlinear
flux}: The above corrections in step i) by using equilibrium information
for nonlinear flux is found to be insufficient for accurate prediction
in the reduced-order methods since the scheme is only marginally stable
and the energy transferring mechanism may change with large deviation
from the equilibrium case when external perturbations are applied.
Thus we also introduce the additional damping and noise $\left(d_{M},\Sigma_{M}\right)$
as from (\ref{eq:GC1}). $d_{M}$ is just a constant scalar parameter
to add uniform dissipation on each mode, and $\Sigma_{M}$ is the
further correction as an additional energy source to maintain climate
fidelity;
\item [{iii)}] \emph{Statistical energy scaling to improve model sensitivity}:
Still note that these additional parameters are added regardless of the
true nonlinear perturbed energy mechanism where only unperturbed equilibrium
statistics are used. To capture the responses to a specific perturbation
forcing, it is better to make the imperfect model parameters change
adaptively according to the total energy structure. Considering this,
the additional damping and noise corrections are scaled with factors
$f_{1}\left(R\right),f_{2}\left(R\right)$ related with the total
statistical variance $\mathrm{tr}R$ as
\begin{equation}
f_{1}\left(R\right)=\left(\frac{\mathrm{tr}R}{\mathrm{tr}R_{\mathrm{eq}}}\right)^{1/2},\quad f_{2}\left(R\right)=\left(\frac{\mathrm{tr}R}{\mathrm{tr}R_{\mathrm{eq}}}\right)^{3/2}.\label{eq:scaling}
\end{equation}

\end{description}
Note that in the full model formulation (\ref{eq:mean_red}) and (\ref{eq:cov_red})
with the entire covariance matrix $R$ resolved, the total variance
structure $\mathrm{tr}R$ is easy to achieve. However in the low-order
models with only the variances in the principal modes resolved explicitly
as will be discussed in the following subsection, $\mathrm{tr}R$
is generally not available directly. This is where the statistical
energy dynamics (\ref{eq:energy_conservation}) can play an important
role and help the development of reduced order models. Besides, a further skew-symmetric correction for dispersion effects in addition to the scalar model damping $d_{M}$ in the reduced-order models might be useful in some situations as the following remark.

\begin{remark*}
In the additional damping correction in (\ref{eq:model_nonflux}), only a scalar damping parameter $d_{M}$ is considered. A little more detailed calibration about the nonlinear exchange of energy is to also introduce an imaginary skew-symmetric operator $i\Omega_{M}$ applied on the covariance $R_{M}$, that is,
\[
Q_F^{M,-}=\left(-d_{M}I_{N}+i\Omega_{M}\right)R_{M}
\]
This term will not alter the entire energy structure of the system due to skew symmetry but can offer correction for the dispersion relation in this imperfect model. However, we may have the additional difficulty in fitting the $N$ by $N$ parameter matrix in the general case. In practical applications, instead we can exploit the physical structure of the specific model and introduce only one additional dispersion parameter $i\omega_{M}$; see \cite{qi2016} and \cite{qi2016low} for two examples of adding the dispersion correction to effectively improve model prediction skill under the barotropic and baroclinic models.
\end{remark*}

Next we discuss the detailed calibrations about the nonlinear flux
approximations. Two steps of model calibration should be considered
as from the general framework described in Section 1.2: i) \emph{the
equilibrium consistency} that the reduced model must converge to the
true equilibrium statistical state as no perturbations are added;
ii) \emph{model sensitivity} by blending statistical response and
information theory so that the imperfect model can capture the responses
to various kinds of perturbations as the system is perturbed. The
construction in (\ref{eq:MQG}) guarantees equilibrium consistency
using the true equilibrium model nonlinear flux structure. On the
other hand, to improve model sensitivity, the linear response operators
with information distance metric are used to find optimal parameters
from the correction part in (\ref{eq:GC1}) or (\ref{eq:GC2}).

\subsubsection{Equilibrium statistical fidelity through the additional damping and noise}
In designing the reduced-order models, equilibrium fidelity for consistent
statistics should be guaranteed in the first place in the unperturbed
climate. That is, the same final unperturbed statistical equilibrium
$R_{\mathrm{eq}}$ should be recovered from the closure models $R_{M}$
in each component. Comparing the true statistical equation (\ref{eq:cov_dyn})
with the reduced-order model (\ref{eq:cov_red}), time derivatives
about the statistics on the left hand sides vanish in statistical
steady state, thus climate consistency can be achieved if we have
exact recovery of the estimation in the nonlinear flux term. Specifically,
it requires that the model nonlinear flux correction term (\ref{eq:model_nonflux})
converges to the truth, $Q_{M}\rightarrow Q_{F,\mathrm{eq}}$, when
no external perturbation is added. Under this condition in steady
state the closure model covariance equation (\ref{eq:cov_red}) goes
to the true unperturbed statistics, the equilibrium statistical relation
(\ref{eq:thrd_equili}) implies the relation 
\[
0=L_{v}\left(\bar{\mathbf{u}}_{\mathrm{eq}}\right)R_{M,\mathrm{eq}}+R_{M,\mathrm{eq}}L_{v}^{*}\left(\bar{\mathbf{u}}_{\mathrm{eq}}\right)+Q_{F,\mathrm{eq}}^{M}+Q_{\sigma}\;\rightarrow\;R_{M,\mathrm{eq}}=R_{\mathrm{eq}}.
\]
In construction the first component $\left(N_{M,\mathrm{eq}},Q_{F,\mathrm{eq}}^{+}\right)$
comes from the true equilibrium statistics, and in equilibrium state
it will guarantee the consistency with the truth that
\[
-\left(N_{M,\mathrm{eq}}R_{\mathrm{eq}}+R_{\mathrm{eq}}N_{M,\mathrm{eq}}^{*}\right)+Q_{F,\mathrm{eq}}^{+}=Q_{F,\mathrm{eq}}.
\]
This part will be automatically equal to the true nonlinear flux in
equilibrium. On the other hand climate consistency requires that the
second component correction due to the parameters $\left(d_{M}I_{N},\Sigma_{M}\right)$
adds no additional energy source or sink in the unperturbed system,
and no further correction in the scaling functionals. That is, we
need $\Sigma_{M}$ to satisfy
\begin{equation}
\Sigma_{M}=\frac{1}{2}d_{M}R_{\mathrm{eq}},\quad f_{1}\left(\mathrm{tr}R_{\mathrm{eq}}\right)=1,\;f_{2}\left(\mathrm{tr}R_{\mathrm{eq}}\right)=1.\label{eq:param_clim}
\end{equation}
Note again $f_{1}$ and $f_{2}$ in (\ref{eq:param_clim}) calibrate
the model sensitivity to perturbations according to the total energy
structure $\mathrm{tr}R$. Thus it is natural to assume no additional
correction in the unperturbed case.

By choosing parameters according to (\ref{eq:param_clim}), the climate
consistency for the imperfect reduced-order models in (\ref{eq:cov_red})
in the unperturbed equilibrium is guaranteed. In addition, we still
leave one controlling parameter $d_{M}$ for the freedom to tune the
imperfect model performance, considering that climate consistency
is only the necessary but not sufficient condition for good model
prediction \cite{majda2011link}.

\subsubsection{Model calibration blending statistical response and information theory}

The above methods (\ref{eq:GC1}), (\ref{eq:GC2}), (\ref{eq:MQG}),
as well as (\ref{eq:model_nonflux}) construct statistical approximation
models with consistent equilibrium statistics. Still equilibrium fidelity
of imperfect models is a necessary but not sufficient condition for
model prediction skill with many examples \cite{majda2016introduction,majda2011link,majda2016improving}.
In order to get precise forecasts for various forced responses, it
is also crucial to seek models that can correctly reflect the system's
`memory' to its previous states. From Section 2.2, it shows that the
linear response operator $\mathcal{R}_{A}$ represents the lagged-covariance
of certain functions (and thus can describe the `memory' of the system
to previous states). We try to find a unified way to achieve the optimal
model parameters $d_{M}$ such that the imperfect models can maintain
high performance for various kinds of external perturbations. Adopting
the general strategy suggested in Section 2.3, we can improve model
sensitivity through tuning imperfect models in a training phase before
the prediction step. Thus the optimal model parameter can be selected
through minimizing the information distance in the linear response
operators in (\ref{eq:opt_model}) between the imperfect closure model
and the truth.

\paragraph*{Information-theoretical framework to measure the linear responses in the training phase}

In this training phase, we try to find the optimal model parameters
$d_{M}$ by comparing the linear response operators from the true
system and imperfect approximation model. The true model linear response
operator and the reduced-order model response operator can be calculated
from (\ref{eq:LRO}), following the procedure from the kicked response
strategy with detailed procedure shown in Appendix A. The distance
between these two operators can be calculated through the information
metric (\ref{eq:entro_pert}) which offers an unbiased and invariant
measure for model distributions 
\begin{eqnarray*}
\mathcal{P}\left(p_{\delta},p_{\delta}^{M}\right) & = & \frac{1}{2}\sum_{k}\left(\delta\bar{u}_{k}-\delta\bar{u}_{M,k}\right)R_{k}^{-1}\left(\delta\bar{u}_{k}-\delta\bar{u}_{M,k}\right)\\
 &  & +\frac{1}{4}\sum_{k}R_{k}^{-2}\left(\delta R_{k}-\delta R_{M,k}\right)^{2}+O\left(\delta^{3}\right).
\end{eqnarray*}
The first row above is the signal error due to the estimation about
the mean; and the second row is the dispersion error for calibrating
the linear responses in the first two order of moments, $\delta R_{\mathbf{k}}$.
The intrinsic error due to second-order closure $\mathcal{S}\left(p_{G,\delta}\right)-\mathcal{S}\left(p_{\delta}\right)$
is independent of the specific forms of the reduced-order models and
is not included in this metric. The optimization principle in (\ref{eq:opt_model})
is then performed over the parameter $d_{M}$.

\subsubsection{Comparisons with stochastic modeling about the mean and fluctuations and realizability}

To achieve a better understanding about the statistical models, it
is useful to compare the reduced-order statistical energy model (\ref{eq:mean_red})
and (\ref{eq:cov_red}) with its stochastic correspondences. In the
stochastic formulation, we consider the separation with a deterministic
mean state and the stochastic fluctuations 
\[
\mathbf{u}_{M}=\bar{\mathbf{u}}_{M}\left(t\right)+\sum_{j}Z_{j}\left(t\right)\mathbf{e}_{j},
\]
where $\bar{\mathbf{u}}_{M}=\left\langle \mathbf{u}_{M}\right\rangle $
is the statistical mean state following the same dynamical mean equation
as before together with the stochastic dynamics for the fluctuation
modes\addtocounter{equation}{0}\begin{subequations}
\begin{eqnarray}
\frac{d\bar{\mathbf{u}}_{M}}{dt} & = & \left(L+D\right)\bar{\mathbf{u}}_{M}+B\left(\bar{\mathbf{u}}_{M},\bar{\mathbf{u}}_{M}\right)+\left\langle Z_{i}Z_{j}^{*}\right\rangle B\left(\mathbf{e}_{i},\mathbf{e}_{j}\right)+\mathbf{F},\label{eq:stoc_mean}\\
\frac{dZ_{i}}{dt} & = & Z_{j}\left[\left(L+D\right)\mathbf{e}_{j}+B\left(\bar{\mathbf{u}}_{M},\mathbf{e}_{j}\right)+B\left(\mathbf{e}_{j},\bar{\mathbf{u}}_{M}\right)\right]\cdot\mathbf{e}_{i}\:-d_{M,i}\left(\left\langle \mathbf{ZZ^{*}}\right\rangle \right)Z_{i}+\Sigma_{M,ij}\left(\left\langle \mathbf{ZZ^{*}}\right\rangle \right)\dot{W}_{ij}.\label{eq:stoc_cov}
\end{eqnarray}
\end{subequations}Above the effective damping and noise $\left(d_{M}I_{N},\Sigma_{M}\right)$
are added in the same way as constructed in (\ref{eq:model_nonflux}).
The mean dynamics (\ref{eq:stoc_mean}) get the small scale feedbacks
from the nonlinear statistical interaction $\left\langle Z_{i}Z_{j}^{*}\right\rangle B\left(\mathbf{e}_{i},\mathbf{e}_{j}\right)$,
while the fluctuation stochastic dynamics are linked with the mean
state through the quasilinear interactions. By direct comparison with
the statistical equations (\ref{eq:mean_red}) and (\ref{eq:cov_red}),
we see that the mean equation is identical while the equation for
the stochastic fluctuations differs in the nonlinear term. The constructed
set of closed stochastic equations is a representative of a new class
of stochastic systems where the evolution of each stochastic realization
depends on the global statistics, i.e. on the collective or statistical
behavior of all the realizations due to $\left\langle \mathbf{ZZ^{*}}\right\rangle $.
In particular, the associated formal Fokker-Planck equation becomes
nonlinear and nonlocal. This guarantees realizability of the reduced-order
models. These novel stochastic equations deserve further mathematical
study as a complex version of McKean-Vlasov systems \cite{mckean1966class}.

\subsection{Reduced-order statistical model for principal modes}

Here for genuinely high-dimensional systems, the computational cost
for the full covariance matrix $R_{M}$ is still unaffordable even
with the first two moment closure \cite{bengtsson2008curse,majda2012lessons};
for example, climate systems usually have the dimensionality at least
of order $10^{3}$ . On the other hand, in many situations, we are
mostly interested in the variability in the statistics of the first
most energetic principal directions. Therefore one alternative practical
strategy is to develop reduced-order methods that only calculate variances
in a low-dimensional subspace spanned by primary EOFs $\left\{ \mathbf{v}_{1},\cdots,\mathbf{v}_{s}\right\} $
with $s\ll N$ ($N$ the dimensionality of the full system). The corresponding
reduced-order representation of the state variables under these resolved
basis becomes $\mathbf{u}=\bar{\mathbf{u}}+\sum_{k=1}^{s}Y_{k}\mathbf{v}_{k}$.
To see the possibility of achieving this, first note that the dynamical
equations for variances (\ref{eq:cov_red}) in each mode $r_{k}=\left\langle Y_{k}Y_{k}^{*}\right\rangle $
are rather independent with each other according to the previous closure
strategies with higher-order interactions replaced by additional damping
and noise terms. Thus it is realizable to restrict the variance equations
inside the chosen subspace. Actually following the same strategy by
replacing the high-order interaction terms by proper damping and noise,
the equivalent counterpart of the closure models can be formulated
as a low-order stochastic system\addtocounter{equation}{0}\begin{subequations}\label{red_model}
\begin{eqnarray}
\frac{d\bar{\mathbf{u}}_{M}}{dt} & = & \left(L+D\right)\bar{\mathbf{u}}_{M}+B\left(\bar{\mathbf{u}}_{M},\bar{\mathbf{u}}_{M}\right)+\sum_{i,j\leq s}C_{M,ij}B\left(\mathbf{v}_{i},\mathbf{v}_{j}\right)+\mathbf{F}+\mathbf{G},\label{eq:mean_red1}\\
\frac{dC_{M}}{dt} & = & L_{v}^{\mathrm{red}}C_{M}+C_{M}L_{v}^{\mathrm{red}*}+Q_{F,M}^{\mathrm{red}}+Q_{\sigma}^{\mathrm{red}},\quad C_{M}\in\mathbb{C}^{s\times s}.\label{eq:cov_red1}
\end{eqnarray}
\end{subequations}The mean dynamics (\ref{eq:mean_red1}) is the
same as the previous closure model (\ref{eq:mean_red}) with an additional
correction term $\mathbf{G}$ to compensate the unresolved modes.
$C_{M}$ is the reduced-order $s\times s$ covariance matrix where
only the leading primary modes are resolved, that is,
\[
C_{M}=P^{*}R_{M}P,
\]
where $P=\left[\mathbf{v}_{1},\mathbf{v}_{2},\cdots,\mathbf{v}_{s}\right]\in\mathbb{C}^{N\times s}$
projects the modes to the subspace.

Through proper choice of the parameters according to GC1 (\ref{eq:GC1}),
GC2 (\ref{eq:GC2}), MQG (\ref{eq:MQG}), or the blended method (\ref{eq:model_nonflux})
as before but concentrating on the resolved subspace, these reduced
system should converge to the same first two order statistics with
the moment closure model. Still several new problems need to be taken
care of for the above model reduction process: i) How to ensure correct
modeling about the true statistics in the mean dynamics (\ref{eq:mean_red1})
due to the many unresolved directions in the covariance $C_{M}$;
ii) How to include the nonlocal scale factor (which always includes
the total energy $\mathrm{tr}R=\sum_{k=1}^{N}r_{k}$) in (\ref{eq:scaling})
in the nonlinear flux approximation $Q_{F,M}^{\mathrm{red}}$ if only
subspace variances are resolved. Here in the general strategy to do
this, we follow the ideas in \cite{majda2016improving,qi2016,qi2016low}.

\subsubsection{Correction for the mean dynamics}

Still the simplest way of estimating the unresolved parts in the mean
dynamics is through the statistical equilibrium information $\mathbf{G}_{\mathrm{eq}}$.
The value of the additional forcing $\mathbf{G_{\mathrm{eq}}}$ is determined
using statistical steady state information for the covariance $C_{\mathrm{eq}}$
and the mean $\bar{\mathbf{u}}_{\mathrm{eq}}$. In particular we have
the equilibrium equation through the steady state mean dynamics where
$\frac{d}{dt}\bar{\mathbf{u}}_{\mathrm{eq}}\equiv0$
\begin{equation}
\mathbf{G_{\mathrm{eq}}}=-\left(L+D\right)\bar{\mathbf{u}}_{\mathrm{eq}}+B\left(\bar{\mathbf{u}}_{\mathrm{eq}},\bar{\mathbf{u}}_{\mathrm{eq}}\right)-\sum_{i,j\leq s}C_{\mathrm{eq},ij}B\left(\mathbf{v}_{i},\mathbf{v}_{j}\right)-\mathbf{F}_{\mathrm{eq}}.\label{eq:mean_corr}
\end{equation}
Similar as in the estimation about the nonlinear flux, the mean dynamics
correction term can also be scaled with the total variance in the
system, so that,
\begin{equation}
\mathbf{G}=\frac{\mathrm{tr}R}{\mathrm{tr}R_{\mathrm{eq}}}\mathbf{G}_{\mathrm{eq}}.
\end{equation}
In this way, the mean dynamics (\ref{eq:mean_red1}) become consistent
in the statistical equilibrium state, and the corrections $\mathbf{G}$
can change sensitively according to the total energy structure through
$\mathrm{tr}R$.
\begin{remark*}
One further optional correction for the unresolved modes in the reduced-order
mean equations is to make further use of the linear response theory
predictions. To estimate the values for unresolved modes, we can improve
it from the equilibrium statistics by introducing finer approximation
making use of the linear response operator (\ref{eq:LRT})
\begin{equation}
r_{k,\mathrm{un}}\sim r_{k,\mathrm{eq}}+\delta r_{k}^{\prime}=r_{k,\mathrm{eq}}+\int_{0}^{t}\mathcal{R}_{r_{k}}\left(t-s\right)\delta F^{\prime}\left(s\right)ds,\quad k>s.\label{eq:model_linresp}
\end{equation}
Therefore these first-order predictions for the unresolved variances
$r_{k,\mathrm{un}}$ can also be used in (\ref{eq:mean_red1}) for
estimating the unresolved modes. This idea is applied to the L-96
system with improvements shown in \cite{majda2016improving}.
\end{remark*}

\subsubsection{Correction through total statistical energy}

In the reduced-order covariance dynamics (\ref{eq:cov_red1}) using
the closure form (\ref{eq:model_nonflux}), two additional scaling
factors, $f_{1},f_{2}$, are introduced to further quantify the nonlinear
energy flux in and out the spectral modes due to the nonlinear interactions.
We propose the dynamical corrections with the total statistical energy
$\mathrm{tr}R$ as in the forms (\ref{eq:scaling}). This total energy
correction introduces global information into each spectral mode so
the nonlinear energy transfer can be better characterized in the imperfect
model, while solving only one additional scalar equation is the only
additional cost in computation. The scaling factor from $\mathrm{tr}R$
introduces nonlinear global effect into the additional damping and
noise corrections in each mode. This can be solved efficiently by
introducing one additional scalar equation as described in (\ref{eq:energy_conservation})
\[
\frac{dE}{dt}=\bar{\mathbf{u}}\cdot D\bar{\mathbf{u}}+\bar{\mathbf{u}}\cdot\mathbf{F}+\mathrm{tr}\left(DR\right)+\frac{1}{2}\mathrm{tr}Q_{\sigma}.
\]
Then $\mathrm{tr}R$ can be achieved by solving $E=\frac{1}{2}\bar{\mathbf{u}}^{2}+\frac{1}{2}\mathrm{tr}R$. 

Especially in uniform damping case, $D=-dI$, the above statistical
energy equation can be simplified as
\begin{equation}
\frac{dE}{dt}=-dE+\bar{\mathbf{u}}\cdot\mathbf{F}+\frac{1}{2}\mathrm{tr}Q_{\sigma}.\label{eq:homo_ene}
\end{equation}
Note that on the right hand side of (\ref{eq:homo_ene}), only mean
state information (which can be fully resolved in the reduced mean
dynamics in (\ref{eq:mean_red1})) needs to be calculated to get the
total statistical energy $E$. In this way, the total second order
moments $\mathrm{tr}R$ can be entirely determined only through the
first order mean $\bar{\mathbf{u}}$ and the scalar statistical energy
equation (\ref{eq:homo_ene}).

The final issue in the reduced-order model construction is about the
tuning process in the training phase. Still the same kicked-response
strategy (\ref{eq:kicked_resp}) can be applied to the reduced-order
formulation (\ref{red_model}). Importantly, the relative entropy
expansion for the responses in (\ref{eq:entro_pert}) is decomposed
into each component. Thus it can be directly applied to the reduced-order
case by calculating only the signal and dispersion error in the resolved
subspace. Therefore through the same procedure as the previous case,
we can find the optimal model parameter in the training phase for
the reduced-order model, and then apply the optimal model for prediction
with various forcing perturbations.

\subsection{Summary of the \emph{Reduced-Order Statistical Energy Closure} algorithm}

We summarize the low-dimensional reduced-order statistical closure
algorithm with calibrations from total statistical energy and linear
response theory. The general reduced-order model algorithm is split
into the separated steps of a \emph{training phase} and a \emph{prediction
phase} after a proper \emph{imperfect model selection} step according
to the problem. The training phase is used to improve model sensitivity
by tuning the imperfect model parameter using only unperturbed equilibrium
statistics for the linear response operator. Then the optimal parameter
can be applied for predicting model responses to different kinds of
external perturbations.
\begin{algorithm*}
\caption{Reduced-order statistical closure model for general turbulent systems}
\begin{itemize}
\item Model selection stage:
\begin{itemize}
\item Decide the low-dimensional subspace spanned by orthonormal basis $\left\{ \mathbf{v}_{k}\right\} _{k=-s}^{s}$
covering the directions with largest variances (energy) among the
spectrum. 
\item Set up statistical dynamical equations (\ref{red_model}) by Galerkin
projecting the original equations to the resolved subspace for modes
with wavenumbers $1\leq\left|k\right|\leq s$, as well as the statistical
energy equation (\ref{eq:homo_ene}) to get the total statistical
energy $E_{M}$ in the system. 
\end{itemize}

\item Model calibration stage:
\begin{itemize}
\item Construct low-order approximation of the nonlinear flux $Q_{F}^{M}$
in the statistical equations using the statistical energy closure
proposed in (\ref{eq:model_nonflux}) consistent with the equilibrium
first two moments;
\item Compute the true linear response operator from the unperturbed equilibrium
statistics, and calculate the imperfect model predicted linear response
operator from the kicked response strategy;
\item Determine the imperfect model parameter value through minimizing the
information distance in (\ref{eq:entro_pert}) and (\ref{eq:opt_model})
between linear response operators from true equilibrium statistics
and imperfect model approximation.
\end{itemize}

\item Model prediction stage:
\begin{itemize}
\item Use the optimally tuned parameter achieved from the previous step
in the reduced-order model to get statistical responses of the state
variables of interest in principal directions with all kinds of specific
external perturbations. 
\end{itemize}
\end{itemize}
\end{algorithm*}
Note that in the calibration step in the algorithm, only the unperturbed
statistics in equilibrium are required. Thus this offers the optimal
model parameters that are ideally valid for all kinds of specific
forcing perturbation forms. With the help of the linear response operator
we are able to find a unified way to tune the imperfect model parameters
and avoid the exhausting and impractical process to tune the models
each time with different kinds of perturbations.

\section{Reduced-Order Statistical Models Applied to a Suite of Stochastic Triad Models}

In this section, we illustrate the performance of the reduced-order
formulations by considering a simple but nevertheless instructive
model, namely the triad system with stochastic forcing \cite{gluhovsky1997interpretation,majda2002priori,majda2016introduction,sapsis2013statistically1}.
The triad systems where three modes interact through quadratic energy-conserving
nonlinear interactions form the building block for more general complex
turbulent flow, thus provide a nice simple test case for the mode
elimination strategy in the first stage. The nonlinear interaction
in triad systems is generic of nonlinear coupling between any three
modes in larger systems with quadratic nonlinearity. For a three-dimensional
system about state variables $\mathbf{u}=\left(u_{1},u_{2},u_{3}\right)^{T}\in\mathbb{R}^{3}$
with a quadratic part that is energy preserving, the triad system
possesses the general form\addtocounter{equation}{0}\begin{subequations}\label{triad}
\begin{eqnarray}
\frac{du_{1}}{dt} & = & L_{2}u_{3}-L_{3}u_{2}-d_{1}u_{1}+B_{1}u_{2}u_{3}+F_{1}+\sigma_{1}\dot{W}_{1},\label{eq:triad1}\\
\frac{du_{2}}{dt} & = & L_{3}u_{1}-L_{1}u_{3}-d_{2}u_{2}+B_{2}u_{3}u_{1}+F_{2}+\sigma_{2}\dot{W}_{2},\label{eq:triad2}\\
\frac{du_{3}}{dt} & = & L_{1}u_{2}-L_{2}u_{1}-d_{3}u_{3}+B_{3}u_{1}u_{2}+F_{3}+\sigma_{3}\dot{W_{3}}.\label{eq:triad3}
\end{eqnarray}
\end{subequations}The triad system (\ref{triad}) is easy to summarize
in the original abstract formulation (\ref{eq:abs_formu}), where
\[
L=\begin{bmatrix}0 & -L_{3} & L_{2}\\
L_{3} & 0 & -L_{1}\\
-L_{2} & L_{1} & 0
\end{bmatrix},\quad D=\begin{bmatrix}-d_{1}\\
 & -d_{2}\\
 &  & -d_{3}
\end{bmatrix},
\]
are the skew-symmetric and dissipation operator in (\ref{eq:prop1})
representing respectively the Coriolis forcing and dissipation; and
$B\left(\mathbf{u},\mathbf{u}\right)$ satisfies 
\[
B\left(\mathbf{u},\mathbf{u}\right)=\left(B_{1}u_{2}u_{3},B_{2}u_{3}u_{1},B_{3}u_{1}u_{2}\right)^{T},\quad B_{1}+B_{2}+B_{3}=0,
\]
which forms the nonlinear triad coupling that satisfies the energy
conservation in (\ref{eq:prop2}). The triad system can form the building
block of complex turbulent dynamical systems since it can be viewed
as a three-dimensional Galerkin truncation of many general dynamics.
One celebrated example is the famous Lorenz model \cite{lorenz1963deterministic}
that can be viewed as a special case of this procedure. An interpretation
of these low-order models with atmospheric problems and geoscience
is illustrated in \cite{gluhovsky1997interpretation}. Though simple
in appearance of this triad system, complex and interesting statistical
features can be generated through changing the model parameters.

\subsection{Statistical properties for the triad system}

First we can check the general statistical properties described in
Section 1.2 with the triad system. Typically we would like to investigate
the evolution of a smooth probability density function $p\left(\mathbf{u},t\right)$
due to the internal and external stochasticity. Associated with the
triad equations (\ref{triad}), the statistical solution satisfies
the Fokker-Planck equation
\begin{equation}
\begin{aligned}\frac{\partial p}{\partial t} & =-\left(B\left(\mathbf{u},\mathbf{u}\right)+\left(L+D\right)\mathbf{u}+\mathbf{F}\right)\cdot\nabla_{\mathbf{u}}p+\sum_{i=1}^{3}\left(d_{i}p+\frac{1}{2}\sigma_{i}^{2}\partial_{u_{i}}^{2}p\right),\\
p\left(\mathbf{u},t\right)\mid_{t=0} & =p_{0}\left(\mathbf{u}\right).
\end{aligned}
\label{eq:FP}
\end{equation}
Above we assume the forcing terms are only dependent on time, $\mathbf{F}\equiv\mathbf{F}\left(t\right),\sigma_{i}\equiv\sigma_{i}\left(t\right)$.
While the original triad system (\ref{triad}) is nonlinear, the statistical
dynamics in (\ref{eq:FP}) are linear equation for smooth functions
$p$. The Fokker-Planck equation will reduce to the \emph{Liouville
equation} in the case of zero stochastic noise $\sigma\equiv0$. However
the explicit solution of the Fokker-Planck equation (\ref{eq:FP})
is still difficult to get directly even with the triad model.

\subsubsection{Equilibrium invariant measure with equipartition of energy}

In general the explicit solutions for the Fokker-Planck equation above
(\ref{eq:FP}) is difficult to achieve due to the nonlinear interactions
in the triad system. Still under special arrangement about the damping
and noise coefficients, one special solution of a Gaussian invariant
measure, $p_{\mathrm{eq}}$, can be reached in the equilibrium. In
the absence of the deterministic forcing, $\mathbf{F=0}$, assume
the damping operator $d_{i}$ and random noise forcing $\sigma_{i}$
satisfy the following relation in each component
\begin{equation}
\sigma_{\mathrm{eq}}^{2}=\frac{\sigma_{1}^{2}}{2d_{1}}=\frac{\sigma_{2}^{2}}{2d_{2}}=\frac{\sigma_{3}^{2}}{2d_{3}}.\label{eq:equi_measure}
\end{equation}
Therefore, a Gaussian invariant measure as defined in (\ref{eq:inv_measure})
can be found with equipartition of energy in each component, that
is,
\begin{equation}
p_{\mathrm{eq}}=C^{-1}\exp\left(-\frac{1}{2}\sigma_{\mathrm{eq}}^{-1}\mathbf{u}\cdot\mathbf{u}\right).\label{eq:invar_m}
\end{equation}
Above $\sigma_{\mathrm{eq}}^{2}$ is the equilibrium variance in the
Gaussian invariant distribution $p_{\mathrm{eq}}$ that controls the
variability in each mode. To see this, we can substitute the invariant
measure (\ref{eq:invar_m}) back into the Fokker-Planck equation (\ref{eq:FP}).
It is a special case from the Theorem in \cite{majda2016introduction},
and detailed energy mechanism and stability for the triad system can
be found in \cite{majda2002priori}.

In the general case with deterministic external forcing and inhomogeneous
structure, energy is injected into the modes and transferred to each
other due to the nonlinear quadratic interaction through more complicated
mechanism, thus strong nonlinear non-Gaussian statistics with energy
cascade and internal instabilities can be generated.

\subsubsection{A link with quasi-geostrophic turbulence}

The triad model (\ref{triad}) is the building block of complex turbulent
dynamical systems since a three-dimensional Galerkin truncation of
many complex turbulent dynamics possesses the energy-conserving nonlinearity
as in (\ref{eq:abs_formu}). The random forcing together with the
damping term represents the inhomogeneous effect of the interaction
with other modes in a turbulent dynamical system that are not resolved
in the three dimensional subspace. Stochastic triad models are qualitative
models for a wide variety of turbulent phenomena regarding energy
exchange and cascades and supply important intuition for many effects.
They also provide elementary test models with subtle features for
prediction, UQ, and state estimation \cite{majda2002priori,majda2003systematic,sapsis2013blended}.

As a simple illustration about the link to more complex turbulent
systems, we can consider the quasi-geostrophic (QG) potential vorticity
equation with no external forcing and dissipation
\[
\frac{\partial q}{\partial t}+\nabla^{\bot}\psi\cdot\nabla q=0,\quad q=\nabla^{2}\psi.
\]
We have the barotropic triads of three barotropic components, $\psi_{\mathbf{k}},\psi_{\mathbf{m}},\psi_{\mathbf{n}}$,
obeying the selecting rule $\mathbf{k+m+n=0}$. Consider an initial
condition in which only these three components of a particular triad
are excited, then these three modes will only interact with each other
while no other modes will get excited due to the particular triad
relations as the system evolves in time. By projecting the above equation
to the active triad modes, we get the dynamical equations for the
selected modes
\begin{equation}
\frac{d\psi_{\mathbf{k}}}{dt}+A_{\mathbf{kmn}}\psi_{\mathbf{m}}\psi_{\mathbf{n}}=0,\quad\mathbf{k+m+n=0},\label{eq:baro_triad}
\end{equation}
where $A_{\mathbf{kmn}}=\frac{\left|\mathbf{n}\right|^{2}}{\left|\mathbf{k}\right|^{2}}\mathbf{m^{\bot}\cdot n}$
is the triad interaction coefficient with the detailed symmetry $A_{kmn}+A_{mnk}+A_{nkm}=0$,
showing the conservation of kinetic energy, 
\[
\frac{d}{dt}\left(\left|\mathbf{k}\right|^{2}\left|\psi_{\mathbf{k}}\right|^{2}+\left|\mathbf{m}\right|^{2}\left|\psi_{\mathbf{m}}\right|^{2}+\left|\mathbf{n}\right|^{2}\left|\psi_{\mathbf{n}}\right|^{2}\right)=0.
\]
The typical forward and backward cascades of energy and enstrophy
in turbulent flow are characterized by the triad interactions between
the three models. Hence from the above discussion, in the two-dimensional
QG turbulence, the nonlinear energy transfer is exactly governed by
the barotropic triads the same as (\ref{triad}) in the nonlinear
interaction part.

\subsection{Typical dynamical regimes in the triad system}

Though simple in appearance, the triad system (\ref{triad}) has representative
statistical features including energy cascade between modes and internal
instabilities that can be created in this simple set-up. A fundamental
factor in the triad system is the internal instabilities that make
the mean unstable over various directions in phase space as is typical
for anisotropic fully turbulent systems. Elementary intuition about
energy transfer in such models can be gained by looking at the special
situation with $L=D=F=\sigma\equiv0$ so that there are only the nonlinear
interactions in (\ref{triad}). We examine the linear stability of
the fixed point, $\bar{\mathbf{u}}=\left(\bar{u}_{1},0,0\right)^{T}$.
Elementary calculations show that the perturbation $\delta u_{1}$
satisfies $\frac{d\delta u_{1}}{dt}=0$ while the perturbations $\delta u_{2},\delta u_{3}$
satisfy the second-order equations
\[
\frac{d^{2}}{dt^{2}}\delta u_{2}=B_{2}B_{3}\bar{u}_{1}^{2}\delta u_{2},\;\frac{d^{2}}{dt^{2}}\delta u_{3}=B_{2}B_{3}\bar{u}_{1}^{2}\delta u_{3},
\]
so that 
\begin{eqnarray}
 & \mathrm{there\:is\:instability\:with}\:B_{2}B_{3}>0\:\mathrm{and}\nonumber \\
 & \mathrm{the\:energy\:of\:}\delta u_{2},\delta u_{3}\mathrm{\:grows\:provided\:}B_{1}\mathrm{\:has}\label{eq:triad_stability}\\
 & \mathrm{the\:opposite\:sign\:of\:}B_{2}\mathrm{\:and\:}B_{3}\mathrm{\:with\:}B_{1}+B_{2}+B_{3}=0.\nonumber 
\end{eqnarray}
The elementary analysis in (\ref{eq:triad_stability}) suggests that
we can expect a flow or cascade of energy from $u_{1}$ to $u_{2}$
and $u_{3}$ where it is dissipated provided the interaction coefficient
$B_{1}$ has the opposite sign from $B_{2}$ and $B_{3}$.

Then energy cascades can be induced from the strongly forced unstable
energetic mode to the stable less energetic modes with stronger damping
effects. Particularly, we can generate distinct statistical features
from Gaussian to highly skewed non-Gaussian PDFs in the following
dynamical regimes:
\begin{itemize}
\item \emph{Regime I: Equipartition of energy.} Set the equipartition of
energy in stationary steady state. That is, $\frac{\sigma_{1}^{2}}{2d_{1}}=\frac{\sigma_{2}^{2}}{2d_{2}}=\frac{\sigma_{3}^{2}}{2d_{3}}=\sigma_{\mathrm{eq}}^{2}$.
Gaussian distributions as in (\ref{eq:invar_m}), $p_{\mathrm{eq}}\sim\exp\left(-\frac{1}{2}\sigma_{\mathrm{eq}}^{-2}\mathbf{u}\cdot\mathbf{u}\right)$,
will be reached under this set-up in the equilibrium state. The parameters
are chosen as $d_{1}=0.2,d_{2}=0.1,d_{3}=0.1$, and $B_{1}=1,B_{2}=-0.6,B_{3}=-0.4$.
Skew-symmetric interactions are added as $L_{1}=3,L_{2}=2,L_{3}=-1$,
and there is no deterministic forcing $F_{1}=F_{2}=F_{3}=0$ added
for the unperturbed equilibrium;
\item \emph{Regime II: Nonlinear regime with forward energy cascade.} Consider
the system with one weakly damped strongly forced mode and two other
strongly damped weakly forced modes, that is, $d_{1}=1,d_{2}=2,d_{3}=2$,
and $\sigma_{1}^{2}=10,\sigma_{2}^{2}=0.01,\sigma_{3}^{2}=0.01$.
The nonlinear coupling is taken as $B_{1}=2,B_{2}=B_{3}=-1$. The
skew-symmetric interaction is set to be zero, $L_{1}=L_{2}=L_{3}=0$,
and there is no deterministic forcing $F_{1}=F_{2}=F_{3}=0$ added
in unperturbed equilibrium. In this case, the first mode is strongly
forced by the random forcing while the other two modes are much less
energetic. The values in the nonlinear coupling coefficients make
sure that the additional energy injected in $u_{1}$ cascades to the
other two less energetic modes $u_{2},u_{3}$, and then gets dissipated
by the strong damping;
\item \emph{Regime III: Nonlinear regime with dual energy cascade.} Use
the same damping and noise forcing parameters as in the energy cascade
case, that is, $d_{1}=1,d_{2}=2,d_{3}=2$, and $\sigma_{1}^{2}=10,\sigma_{2}^{2}=0.01,\sigma_{3}^{2}=0.01$.
The nonlinear coupling coefficients are also taken the same values
as before, $B_{1}=2,B_{2}=B_{3}=-1$. Skew-symmetric interactions
are added in this regime as $L_{1}=0.09,L_{2}=0.06,L_{3}=-0.03$ to
enhance the interactions between modes. Deterministic forcings are
applied in modes $u_{2},u_{3}$ as $F_{1}=0,F_{2}=-1,F_{3}=1$. In
addition to the forward energy cascade from $u_{1}$ to $u_{2},u_{3}$
as in the previous case, the additional forcing introduces energy
sources in modes $u_{2},u_{3}$ and leads to backward energy cascade
from modes $u_{2},u_{3}$ back to $u_{1}$. The linear skew-symmetric
operator further alters the equilibrium energy structure in the system,
resulting in skewed probability distribution functions in the steady
state. This regime is especially interesting because strong internal
instability can be generated here.
\end{itemize}
The first test case is the simplest but nevertheless representative
with equipartition of energy. The higher-order moment effects are
relatively small in this equipartition energy case, while the dynamics
are dominantly Gaussian with zero cross-covariances as the system
evolves in time. The second test case above is also relatively simple
without the skew-symmetric interaction between modes and most energy
will accumulate in the first dominant mode. Nevertheless important
third-order interactions will take place in this case, and large errors
will be introduced if the cross-covariances are ignored without care.
In the third test case the non-zero forcing in the unperturbed system
creates skewed equilibrium distributions with important third-order
moments. Also the non-zero linear skew-symmetric interaction terms
add extra emphasis on the cross-covariances. These induce stronger
interactions and energy cascades between the triad modes.

\subsubsection{Numerical results about the the unperturbed triad system in equilibrium}

The true statistical features of the triad system in the above dynamical
regimes are resolved through direct Monte-Carlo simulations. We run
an ensemble of $N=10000$ particles, which shall be enough for capturing
the statistics in a three-dimensional phase space. Forward Euler scheme
with small time step is used to integrate the system in time due to
its simplicity. The stochastic forcing is simulated through the standard
\emph{Euler-Maruyama} scheme. The initial ensemble is chosen from
a standard Gaussian random sampling.

\begin{figure}
\centering
\subfloat{\includegraphics[scale=0.26]{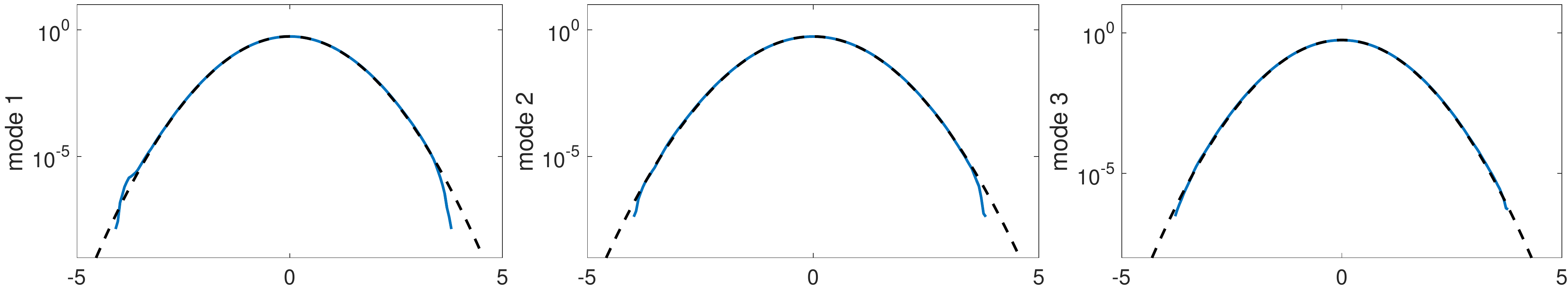}}

\subfloat[energy equipartition regime ]{\includegraphics[scale=0.26]{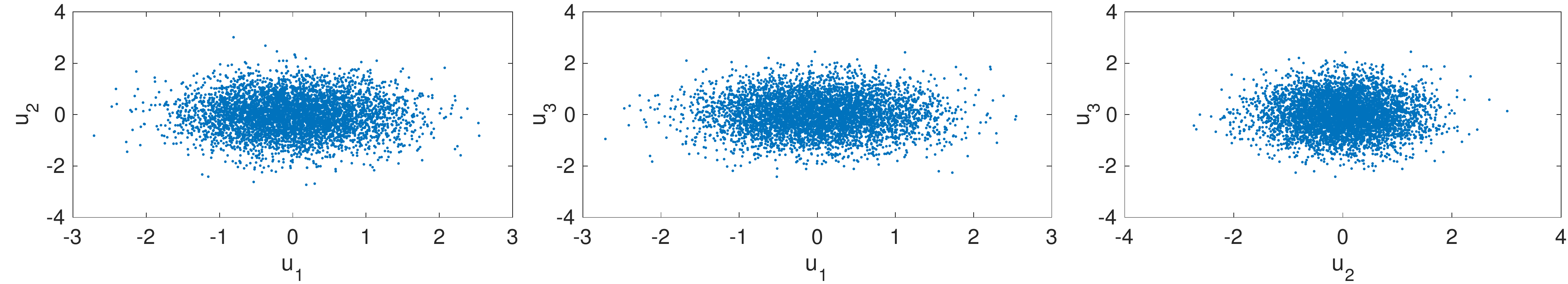}

}

\subfloat{\includegraphics[scale=0.26]{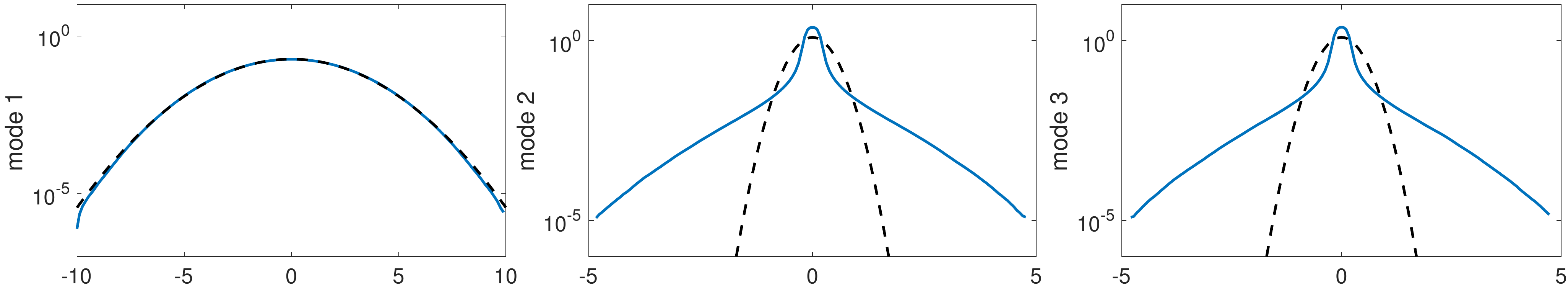}}

\subfloat[forward energy cascade regime]{\includegraphics[scale=0.26]{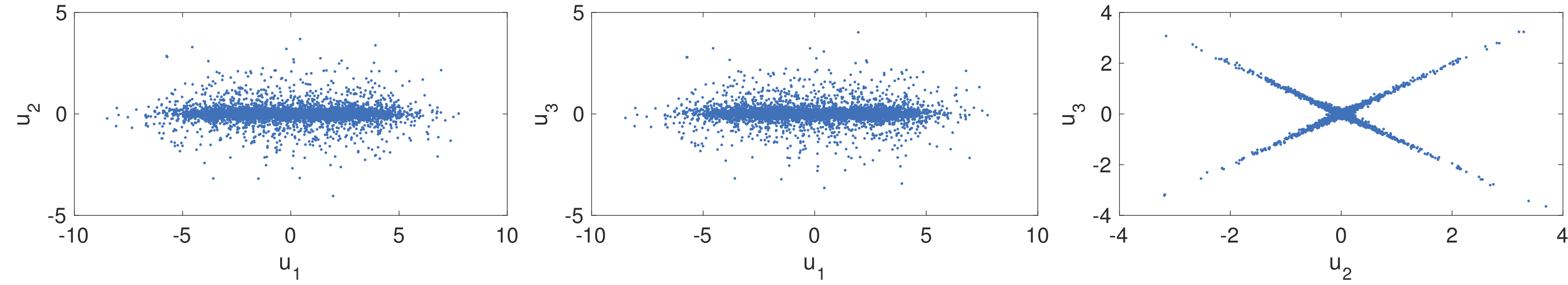}

}

\subfloat{\includegraphics[scale=0.26]{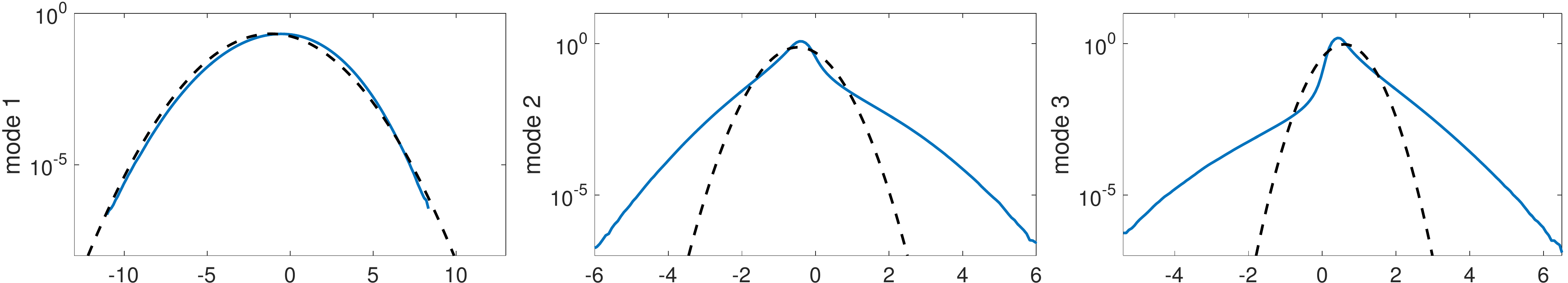}}

\subfloat[dual energy cascade regime]{\includegraphics[scale=0.26]{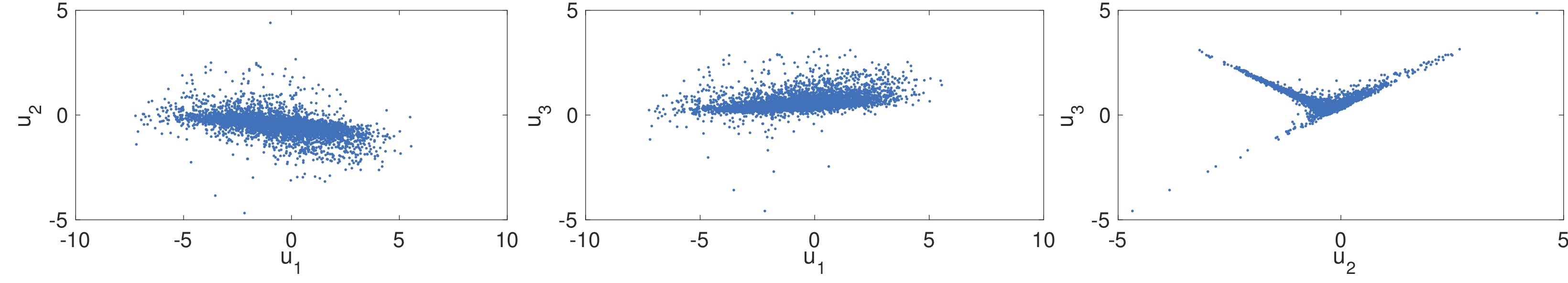}

}

\caption{Marginal PDFs in three modes $u_{1},u_{2},u_{3}$ and the joint distributions
in scatter plots in three regimes with distinct statistics. Gaussian
density functions with the same variance are shown in dashed black
lines. \label{fig:stat_pdf}}
\end{figure}

For more details about the model statistics in these regimes, Figure
\ref{fig:stat_pdf} displays the probability distributions in these
three test regimes. In mode $u_{1}$, Gaussian (or quasi-Gaussian)
distributions can always be observed in all the three regimes. Also
the equipartition of energy structure can be observed in the first
regime as predicted from analysis in the previous sections. Marginal
PDFs are consistent with the Gaussian fits from theory and the joint
2-dimensional distributions are also in Gaussian structure. On the
other hand, in the forward energy cascade case, fat tails can be observed
in the marginal PDFs in both $u_{2},u_{3}$ as well as the star-shaped
joint-distribution, showing the strong nonlinear effects in the modes.
Note that the non-Gaussianity in $u_{2},u_{3}$ can affect the final
structure in the dominant mode $u_{1}$ despite its near-Gaussian
marginal distribution. Furthermore, in the third test regime with
dual energy cascades, besides the fat tails, skewed PDFs appear in
both modes $u_{2},u_{3}$ due to the non-zero deterministic forcing
applied on them. The non-Gaussianity can be further confirmed by the
joint distributions where strong skewness is shown. Also note that
the mean states are not centered in zero in this case. These illustrate
important third-order moments in this case for accurate predictions.
Below we concentrate on this toughest regime.

\subsection{Reduced-order statistical models for the triad system}

In the development about reduced-order models, we focus on the accurate
estimation about the first two moments, that is, the statistical mean
state and covariance matrix. The major interest is to check the models'
skill in capturing sensitivity in response to external perturbations
besides equilibrium consistency in the models. Considering the relatively
simple structure in the triad dynamics, we focus on the GC1 and GC2
formulation in (\ref{eq:GC1}) and (\ref{eq:GC2}) respectively in
the following test cases and the parameter calibration strategy proposed
in (\ref{eq:model_nonflux}). In the model reduction procedure, we
first consider the \textbf{\emph{fully resolved model}}, where the
3-dimensional mean and $3\times3$ covariance matrix are resolved
entirely; and then the \textbf{\emph{diagonal model}}, where only
the mean and diagonal variances are calculated explicitly. In the
final step, a severely \textbf{\emph{reduced-order model}} is introduced
where only the variance in the principal mode is resolved.

\subsubsection{Reduced model formulation for the triad system (\emph{Model selection})}

Applying exactly the modeling procedure in Section 3 to the triad
system, the two-moment closure schemes replace the higher order moments
with additional damping and excitement containing only first two order
of moments. The first approach is to run the full system with mean
and covariance matrix where both diagonal variances and off-diagonal
components are resolved for $\bar{\mathbf{u}}\in\mathbb{R}^{3}$ and
$R\in\mathbb{R}^{3\times3}$\addtocounter{equation}{0}\begin{subequations}\label{GC}
\begin{eqnarray}
\frac{d\bar{\mathbf{u}}}{dt} & = & L\bar{\mathbf{u}}+B\left(\bar{\mathbf{u}},\bar{\mathbf{u}}\right)+\sum_{i,j}R_{ij}B\left(\mathbf{e}_{i},\mathbf{e}_{j}\right)+\mathbf{F},\label{eq:GC_mean}\\
\frac{dR}{dt} & = & L_{v}\left(\bar{\mathbf{u}}\right)R+RL_{v}^{*}\left(\bar{\mathbf{u}}\right)+Q_{\mathrm{GC}}+Q_{\sigma}.\label{eq:GC_var}
\end{eqnarray}
\end{subequations}As a test for reduced order methods, we also check
the models' skill by ignoring the cross-covariances. Especially for
the triad system, the off-diagonal cross-covariances play a crucial
role in the computation of mean and variance responses, while on the
other hand, are difficult and expensive to estimate with accuracy.
Then the unresolved covariances are approximated from steady state
information.
\begin{itemize}
\item \textbf{\emph{Fully Resolved Model}}: In this fully resolved model,
the entire mean and covariance matrix are calculated through the dynamical
equations (\ref{GC}). The nonlinear flux term is approximated by
additional damping and additional noise as in (\ref{eq:GC1}) and
(\ref{eq:GC2})
\[
Q_{\mathrm{GC}}=-\left(D_{M}R+RD_{M}^{*}\right)+Q_{M}.
\]

\item \textbf{\emph{Diagonal Model ignoring cross-covariances}}: In this
diagonal model, the cross-covariances between modes are ignored to
improve model efficiency. So the covariance matrix is replaced by
diagonal matrix $R=\mathrm{diag}\left(r_{k}\right)$. The correction
in $Q_{\mathrm{GC}}$ for higher-order moments calibration is kept
the same as the full model for GC1 or GC2 respectively. To correct
the error due to the neglect of cross-covariances, we add the cross-covariance
correction using only steady state information
\[
r_{ij}=\frac{\mathrm{tr}R}{\mathrm{tr}R_{\mathrm{eq}}}\left\langle u_{i}^{\prime}u_{j}^{\prime}\right\rangle _{\mathrm{eq}}.
\]
The diagonal model can be much more efficient compared with the fully
resolved model with a model reduction from $O\left(N^{2}\right)$
to $O\left(N\right)$.
\end{itemize}
Through construction using the unperturbed equilibrium information,
climate consistency is guaranteed in both GC1 and GC2 model. In GC1,
only constant damping and noise are added to approximate the unresolved
higher order moments, while these terms are further corrected with
the total variance in GC2. The exponents in the scaling factors are
designed to make them consistent in dimension with the estimated third-order
statistics. Actually, these scaling factor becomes quite crucial in
capturing model responses to perturbations and model performances
in transient state.

As a further reduction in the model we consider the two-moment closure
methods in reduced order subspace. In this case, we only resolve the
variance in the first mode $u_{1}$, and calculate the mean dynamics
for all three modes. Then \textbf{\emph{the reduced order model}}
can be expressed as\addtocounter{equation}{0}\begin{subequations}
\begin{eqnarray}
\frac{d\bar{\mathbf{u}}}{dt} & = & L\bar{\mathbf{u}}+B\left(\bar{\mathbf{u}},\bar{\mathbf{u}}\right)+\sum_{i,j\in\Lambda}C_{ij}^{\mathrm{red}}B\left(\mathbf{e}_{i},\mathbf{e}_{j}\right)+\mathbf{F}+\mathbf{G},\label{eq:GC_mean_red}\\
\frac{dC^{\mathrm{red}}}{dt} & = & L_{v}^{\mathrm{red}}\left(\bar{\mathbf{u}}\right)C^{\mathrm{red}}+C^{\mathrm{red}}L_{v}^{\mathrm{red}*}\left(\bar{\mathbf{u}}\right)+Q_{\mathrm{GC}}^{\mathrm{red}}+Q_{\sigma}^{\mathrm{red}}.\label{eq:GC_var_red}
\end{eqnarray}
\end{subequations}$\Lambda=\left\{ 1\right\} $ is the index set
that includes the resolved modes. Especially with only the leading
variance resolved, $C^{\mathrm{red}}=r_{1}=\left\langle u_{1}^{2}\right\rangle $
is only the variance in mode $u_{1}$. The construction about the
reduced-order parameters can follow the general strategy in (\ref{eq:mean_red1})
and (\ref{eq:cov_red1}) exactly.
\begin{itemize}
\item \textbf{\emph{Reduced Model with only the principal variance resolved}}:
In the variance equation (\ref{eq:GC_var_red}), the formulation is
similar as before and we only need to constrain the covariance matrix
$C^{\mathrm{red}}$ in the resolved subspace with the first mode $u_{1}$
resolved. The additional damping and noise can be applied in the same
way as the reduced-model correction as discussed in (\ref{red_model}).
In the mean dynamics, the unresolved second-order moments in $R_{ij}B\left(\mathbf{e}_{i},\mathbf{e}_{j}\right)$
are corrected following (\ref{eq:mean_corr})
\[
G=\frac{\mathrm{tr}R}{\mathrm{tr}R_{\mathrm{eq}}}\sum_{i,j\in\Lambda^{c}}R_{ij}B\left(\mathbf{e}_{i},\mathbf{e}_{j}\right).
\]
The total variance for model sensitivity correction can also be achieved
through the approximated statistical energy dynamics as in (\ref{eq:homo_ene})
\[
\frac{dE^{\mathrm{est}}}{dt}=-d_{\mathrm{eff}}E^{\mathrm{est}}+\sum_{i=1}^{3}\left(F_{i}\bar{u}_{i}+\frac{1}{2}\sigma_{i}^{2}\right).
\]
Note that we have calculated the mean in each mode, then the total
variance can be calculated through the total statistical energy $\mathrm{tr}R=2E-\sum_{i=1}^{3}\bar{u}_{i}^{2}$.
One additional difficulty in the inhomogeneous case is that different
damping rates $d_{j}$ are applied to different modes. So we introduce
the effective damping rate $d_{\mathrm{eff}}$ through the statistical
steady state information as
\[
d_{\mathrm{eff}}=\frac{\sum_{i=1}^{3}\left(F_{i}\bar{u}_{i,\mathrm{eq}}+\frac{1}{2}\sigma_{i}^{2}\right)}{E_{\mathrm{eq}}}.
\]

\end{itemize}
We also list the explicit statistical mean and variance dynamical
equations in Appendix B, as well as the explicit statistical energy
dynamics to achieve the total variance of the system when only the
leading mode is resolved.

\subsubsection{Model consistency and tuning parameters in the training phase (\emph{Model calibration})}

\begin{figure}
\centering
\subfloat[statistics in each mode]{\includegraphics[scale=0.36]{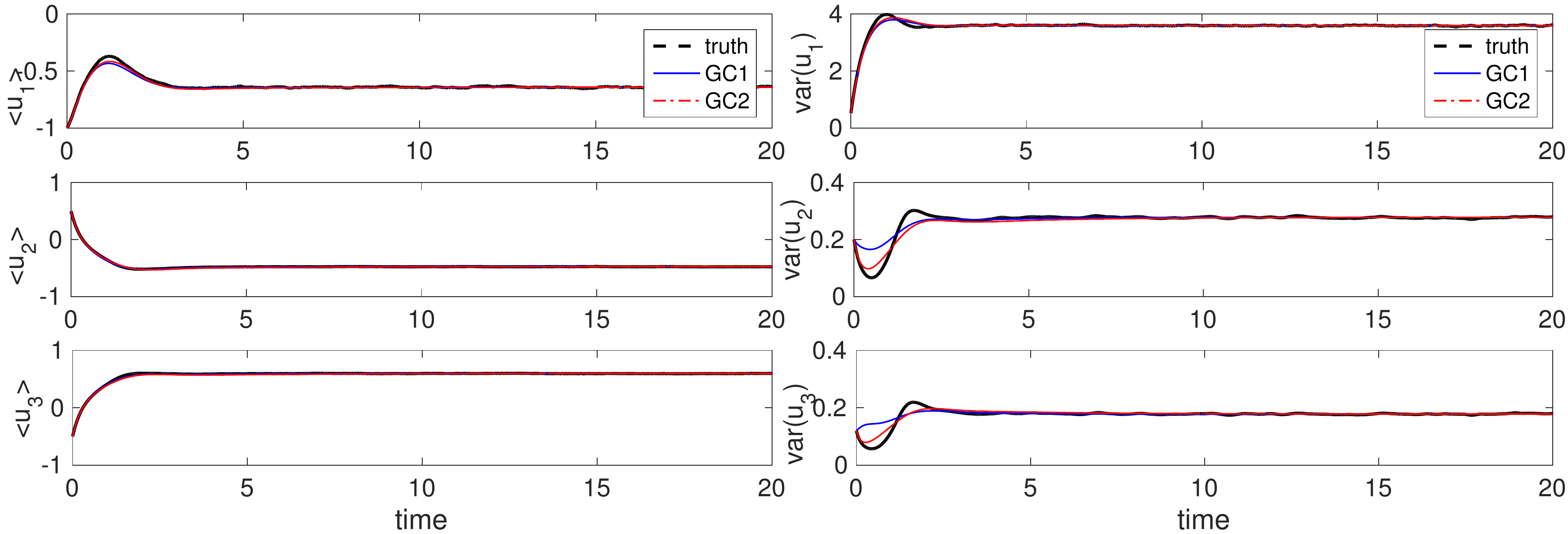}

}

\subfloat[total statistical energy]{\includegraphics[scale=0.36]{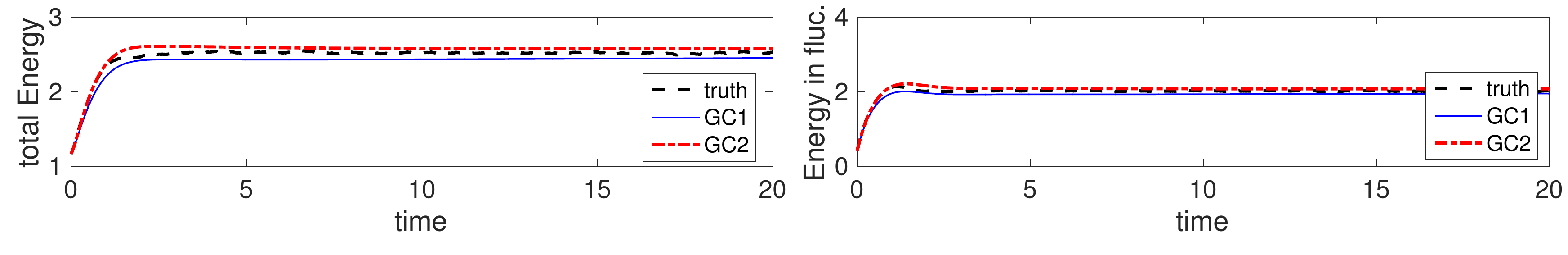}

}

\caption{Imperfect model performances in convergence to statistical equilibrium
state in dual energy cascade regime. Both GC1 (with constant damping
and noise) and GC2 (with correction from total variance) can recover
the steady state mean and variance but GC1 lacks the ability to accurately
capture the transient state in the beginning due to its lack of sensitivity.\label{fig:steady}}
\end{figure}

In the model calibration step, we check the imperfect models' consistency
with climatology when no perturbation is applied. As noticed before,
the imperfect models' skill in capturing the right statistics in transient
state is crucial for the convergence to the right fixed point in the
energy equation. As an example, Figure \ref{fig:steady} compares
the model performances in regimes with dual energy cascade. It can
be observed that the statistical steady state mean and variances are
recovered with accuracy due to the climate consistent choice of parameters
through (\ref{eq:GC1}) and (\ref{eq:GC2}), while GC1 lacks the ability
to accurately capture the transient state in the beginning due to
its lack of sensitivity.

\begin{figure}
\centering
\subfloat[information error with changing parameter values]{\includegraphics[scale=0.38]{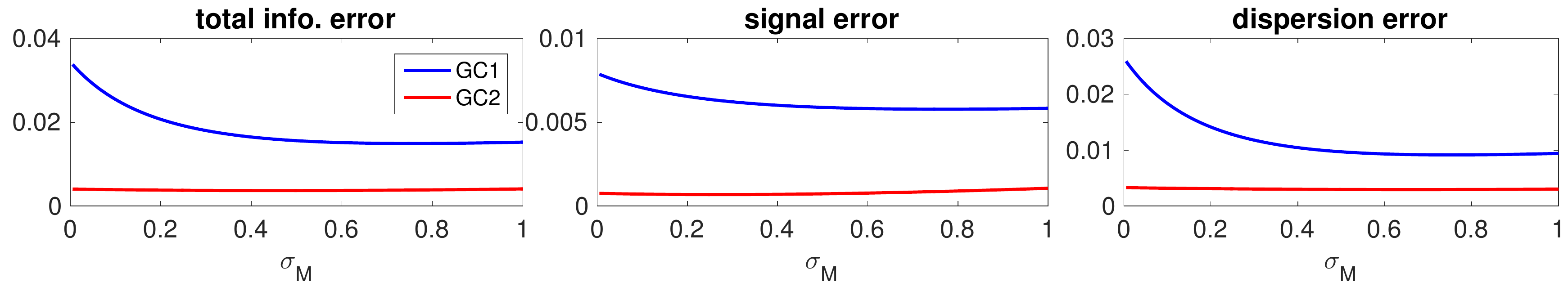}

}

\subfloat[linear responses with optimal parameter]{\includegraphics[scale=0.38]{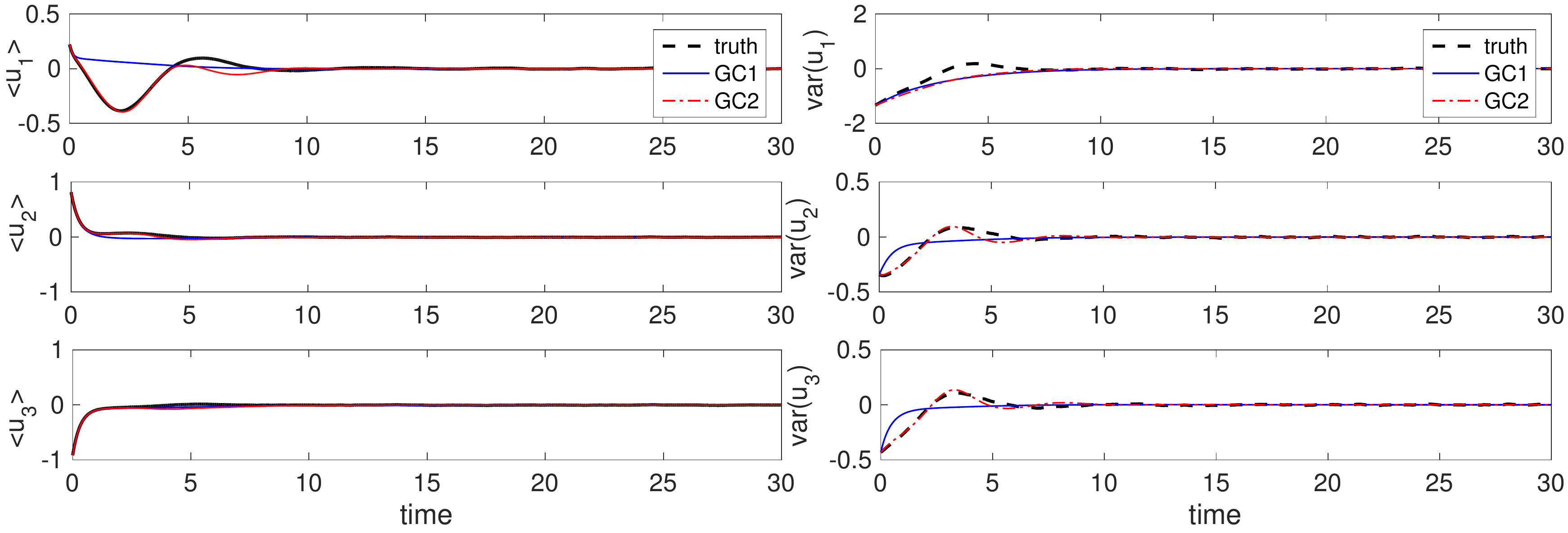}

}

\caption{Illustration about the training phase for tuning optimal imperfect
model parameters for GC1 and GC2. The first row is the information
errors with changing values of the tuning parameter $\sigma_{M}$.
The second row displays the approximation for the linear response
operators in GC1 and GC2 using optimal parameters from the tuning
process above.\label{fig:tune}}
\end{figure}

We illustrate the tuning process for optimal model parameters in the
training phase in Figure \ref{fig:tune}. Again we use the dual energy
cascade regime as an example. Since both the deterministic and stochastic
forcing might be perturbed in the external forcing, we consider a
kicked response in the initial value according to the perturbation
form described in (\ref{eq:pert1}) and (\ref{eq:pert2}). In the
first row, the information errors with changing model parameter values
are shown in total relative entropy from (\ref{eq:entro_pert}) as
well as the errors in signal and dispersion component. Note that the
errors in GC2 model results stay uniformly small in the entire parameter
regime, showing the robustness of the method; on the other hand, GC1
model displays larger information error no matter how well we tune
the model parameter. This illustrates the inherent information barrier
in the closure schemes if we do not consider proper statistical energy
scaling in the model damping and noise terms. In the second row, we
show the approximation about the linear response operators in the
mean and variance using optimal parameters. The transient structures
can be captured with accuracy in GC2 model, while GC1 lacks the skill
due to the insufficient calibration in the higher-order nonlinear
flux approximation.

\subsubsection{Model prediction skill to perturbations (\emph{Model prediction})}

\begin{figure}
\centering
\subfloat{\includegraphics[scale=0.35]{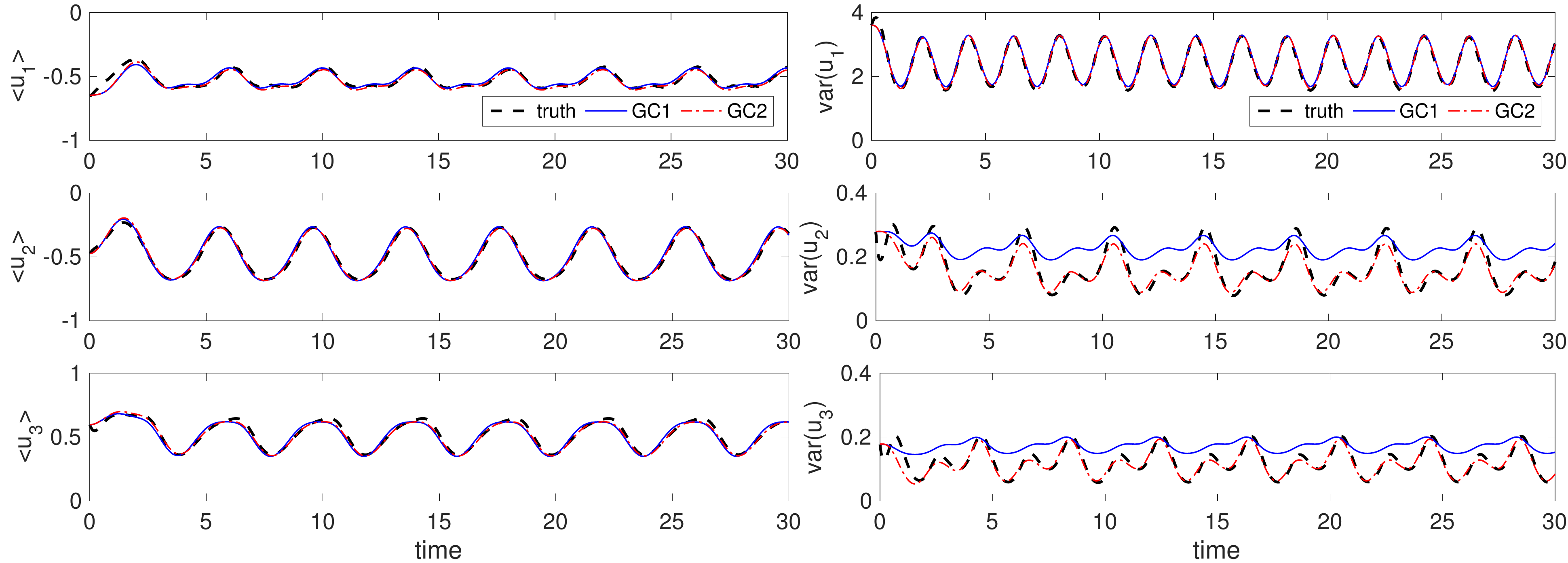}}

\subfloat{\includegraphics[scale=0.35]{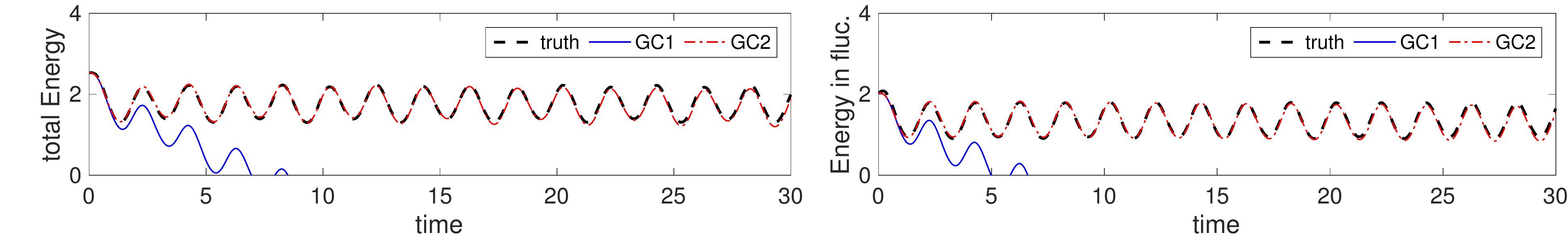}}

\subfloat{\includegraphics[scale=0.35]{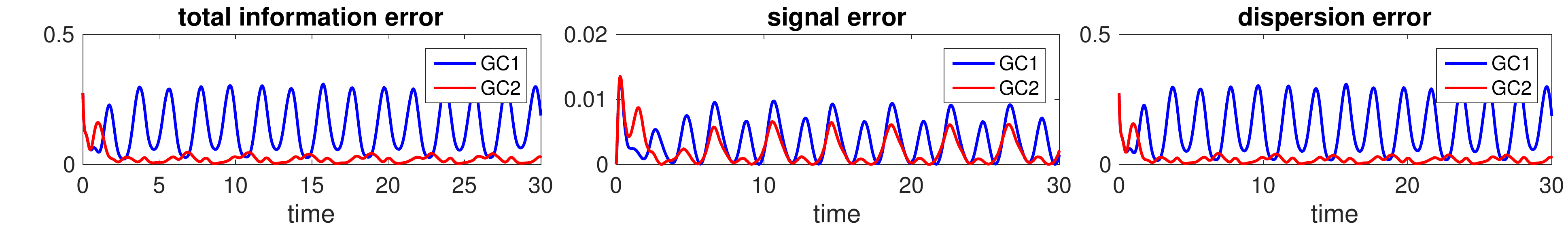}}

\caption{Full GC1 and GC2 model with optimal parameter in approximating responses
to periodic external forcing in regimes with dual cascade. The first
three rows show the GC1 and GC2 predictions for the statistical mean
and variances in each mode together with the truth from MC simulations
in thick black lines. The total energy $E$ from the energy equation
and the fluctuation energy $E^{\prime}=E-\frac{1}{2}\bar{\mathbf{u}}^{2}$
are compared in the following row. The last line shows the total information
error in the imperfect models together with the signal and dispersion
components.\label{fig:full}}
\end{figure}

Periodic perturbations added in both deterministic and random forcing
are representative in checking the imperfect models' prediction skill
in response to perturbations. The perturbations are introduced in
the following forms:
\begin{itemize}
\item \emph{periodic forcing perturbations in the deterministic forcing:}
The deterministic forcing perturbation is introduced by a periodic
addition to the mean forcing $F=\bar{F}+\delta F$. As one typical
test case, we add periodic forcing perturbations $\delta F$ to each
mode, that is,
\begin{equation}
\delta F_{i}=A_{i}\sin\left(\omega t\right),\label{eq:pert1}
\end{equation}
with $\omega$ taking the value $\pi/4$, and $A_{i}$ measures the
perturbation amplitude in each mode.
\item \emph{periodic perturbation in the stochastic random forcing:} Also
to add stronger time-dependent effect to the variances, we add periodic
random forcing to the system by setting
\begin{equation}
\sigma_{j}=\bar{\sigma}_{j}+\delta f^{2}\left(t\right)\left(\sigma_{T_{j}}-\bar{\sigma}_{j}\right),\label{eq:pert2}
\end{equation}
with $\delta f\left(t\right)=\sin\left(\omega t\right)$ also set
to be periodic. Here $\bar{\sigma}_{j}$ is the mean stochastic forcing
amplitude in the unperturbed case with $\sigma_{T_{j}}>\bar{\sigma}_{j}$
to add perturbations to this unperturbed mean.
\end{itemize}
We check the model performances in predicting response to the periodic
perturbations in both deterministic and stochastic components. Here
we display the imperfect model prediction skill in the toughest regime
with dual energy cascade. Thus strong nonlinear coupling is present
between the modes. In Figure \ref{fig:full}, the \emph{fully resolved
model} with mean and $3\times3$ covariance matrix is applied to the
typical regime in GC1 and GC2 closure methods. In this regime with
skewed distributions, higher order moments become crucial and need
more detailed calibration. As we can see from the results, the deficiency
of GC1 method appears in this regime due to the inaccurate approximation
about the third-order moments. Large deviation takes place in the
skewed modes $u_{2},u_{3}$ due to the errors from third-order energy
transfer. In contrast, GC2 model maintains the high skill in predicting
the mean and variances with the more careful calibration about the
nonlinear flux through the scaling factor using total statistical
energy.

\begin{figure}
\centering
\subfloat{\includegraphics[scale=0.35]{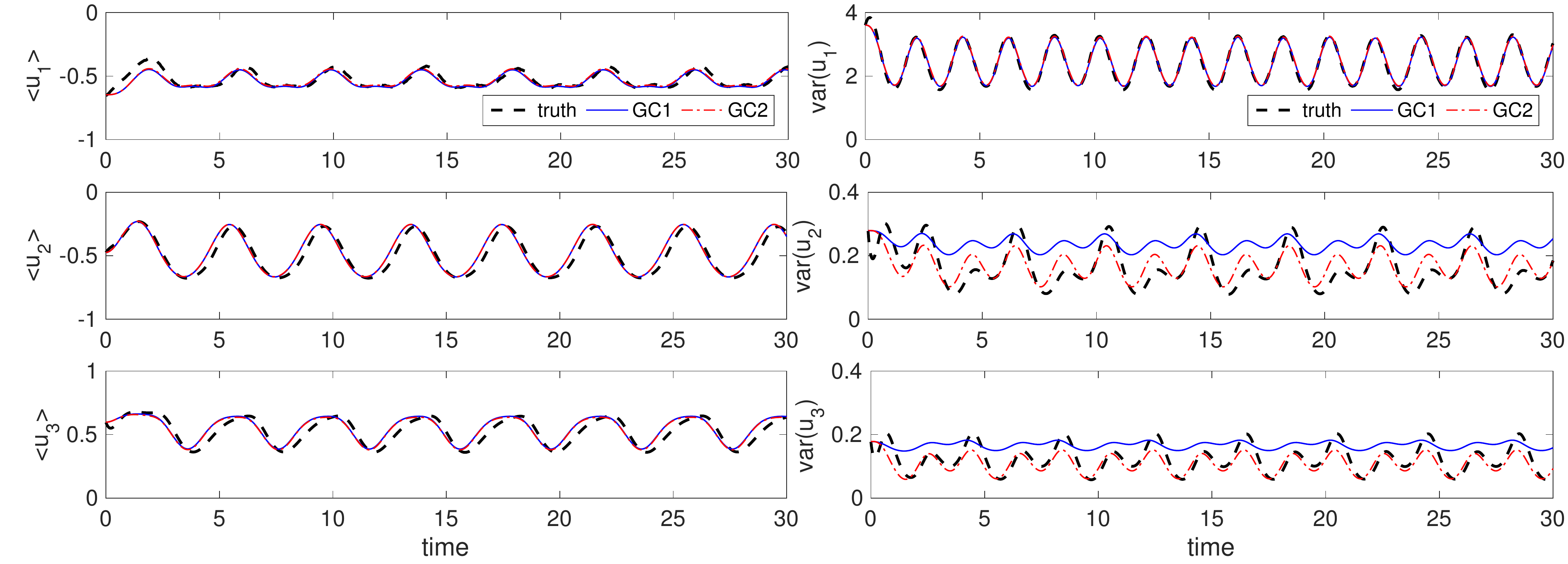}}

\subfloat{\includegraphics[scale=0.35]{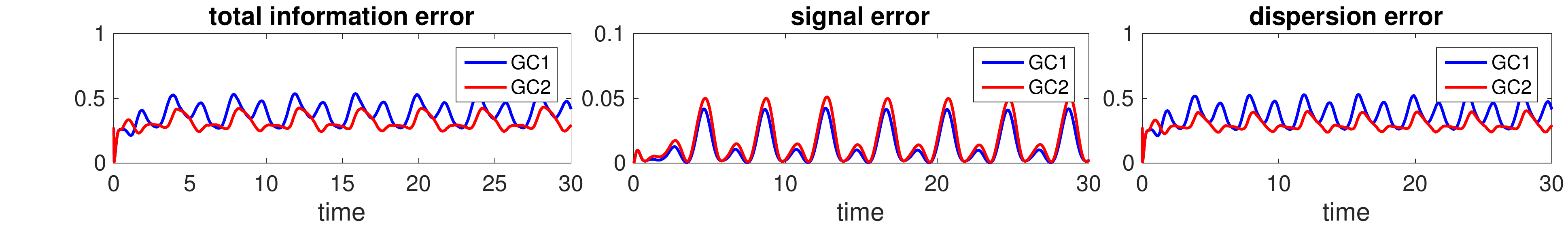}}

\caption{Diagonal GC1 and GC2 model with optimal parameter in approximating
responses to periodic external forcing in regimes with dual cascade.
The first three rows show the GC1 and GC2 predictions for the statistical
mean and variances in each mode together with the truth from MC simulations
in thick black lines. The last line shows the total information error
in the imperfect models together with the signal and dispersion components.\label{fig:diag}}
\end{figure}

In Figure \ref{fig:diag}, we check the \emph{diagonal models}\textbf{
}with only mean and variances in each mode resolved and ignoring the
off-diagonal cross-covariances. Like the previous case of full model,
GC1 loses the skill in predicting the responses in $u_{2},u_{3}$
due to the lack of information in the third-order interactions. The
errors in the beginning transient regime drive the statistical equation
to the wrong state or even blowing up. GC2 keeps the skill in capturing
the response structures of both the mean and variances. And again
most of the error takes place in the variance estimations.

\begin{figure}
\centering
\subfloat{\includegraphics[scale=0.35]{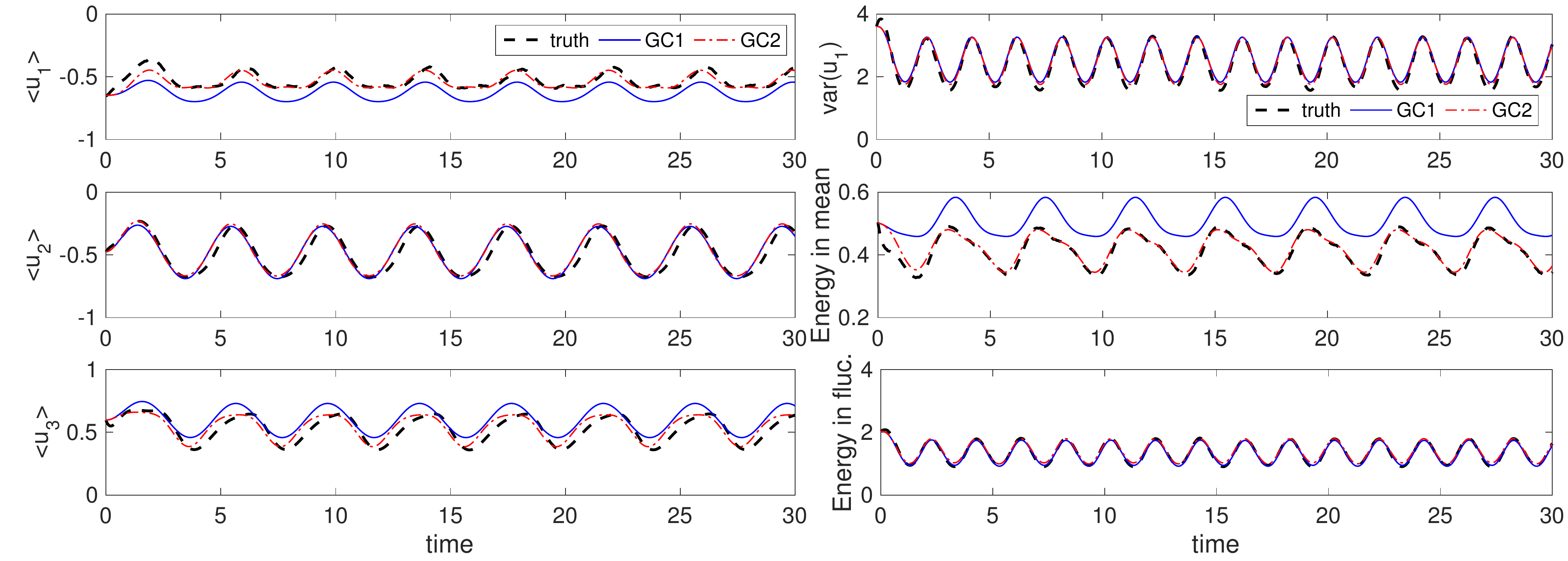}}

\subfloat{\includegraphics[scale=0.35]{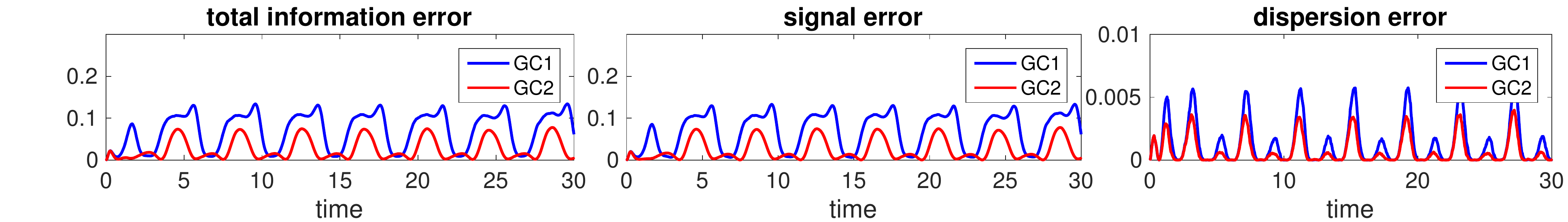}}

\caption{Reduced-order GC1 and GC2 model with the variance in first mode resolved
in approximating responses to periodic external forcing in dual energy
cascade regime. The first three rows show the GC1 and GC2 predictions
for the statistical mean and variances in each mode together with
the truth from MC simulations in thick black lines. The last line
shows the total information error in the imperfect models together
with the signal and dispersion components.\label{fig:redu}}
\end{figure}

Testing the feasibility of the \emph{reduced-order strategies} described
previous is important for further applications to general high dimensional
turbulent systems because the triad interactions always represent
the nonlinear interactions in different scales which usually are ignored
in realistic modeling. The same dynamical regime is tested here for
the reduced-order models. In this case, only the variance in the first
mode $u_{1}$ is resolved explicitly. Especially for this regime with
strong dual energy cascade, as we have seen in the true statistical
dynamics, strong coupling exists between the high energy mode $u_{1}$
and the less energetic modes, $u_{2},u_{3}$, in both third-order
moments and second-order cross-covariances. Thus this becomes a challenging
situation for the reduced-order models for capturing the responses
with accuracy. The model prediction results are shown in Figure \ref{fig:redu}.
With forward and backward energy cascade, strong nonlinear high-order
interactions become crucial here. GC1 loses its skill in this case
and ends with large errors especially in the mean state $u_{1}$.
This is no surprise considering the strong perturbed deviation from
the equilibrium state in this regime due to the nonlinear energy transfer
while GC1 model only uses unperturbed equilibrium information. On
the other hand, GC2 keeps its skill and can capture the responses
in both mean and variance with only a single low-order mode resolved.

\subsubsection{Additional results for the triad model with equipartition of energy}

\begin{figure}
\centering
\subfloat{\includegraphics[scale=0.35]{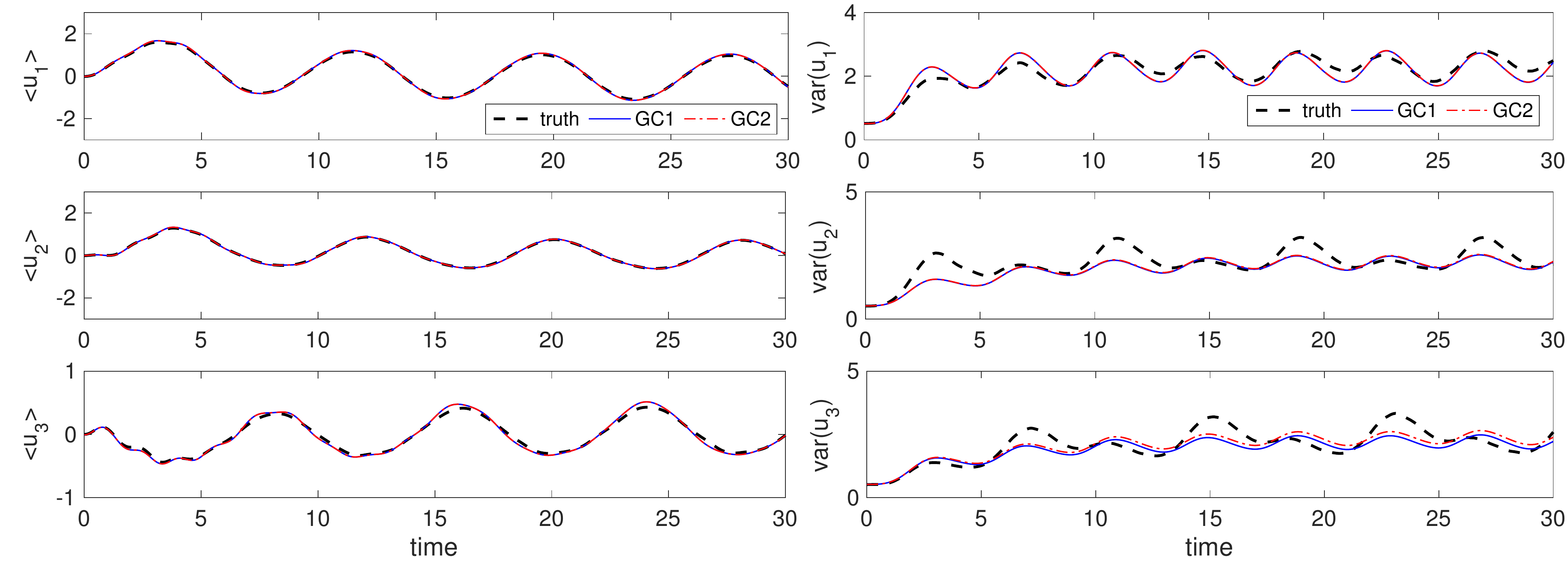}}

\subfloat{\includegraphics[scale=0.35]{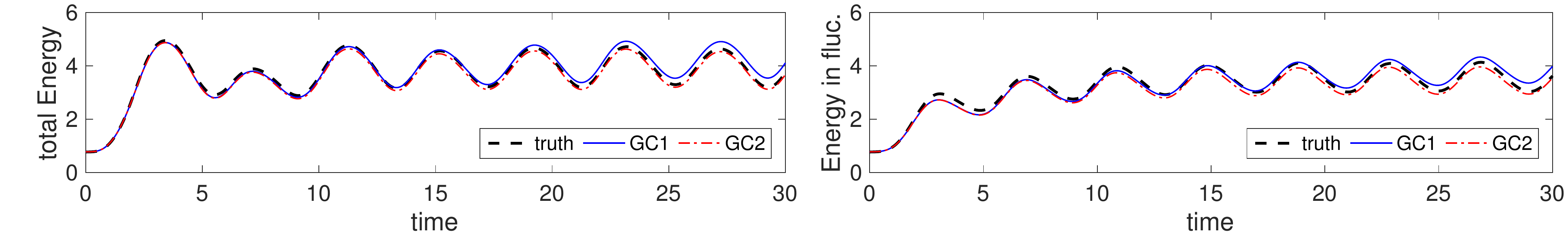}}

\subfloat{\includegraphics[scale=0.35]{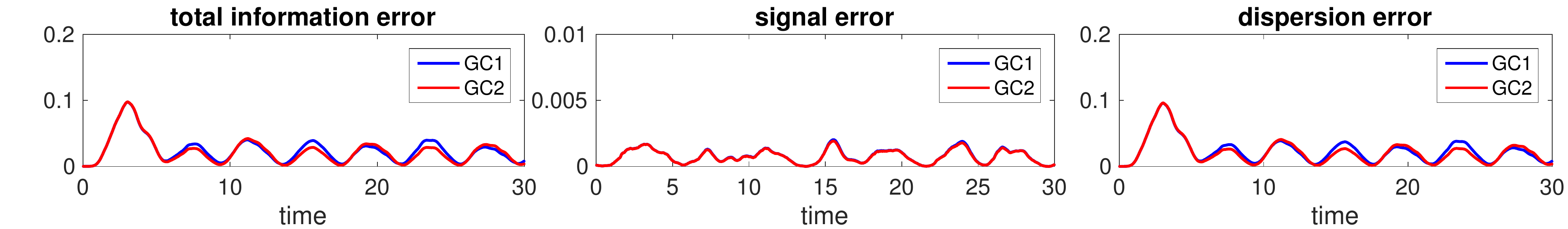}}

\caption{Diagonal GC1 and GC2 model with optimal parameter in approximating
responses to periodic external forcing in regimes with equipartition
of energy. The first three rows show the GC1 and GC2 predictions for
the statistical mean and variances in each mode together with the
truth from MC simulations in thick black lines. The total energy $E$
from the energy equation and the fluctuation energy $E^{\prime}=E-\frac{1}{2}\bar{\mathbf{u}}^{2}$
are compared in the following row. The last line shows the total information
error in the imperfect models together with the signal and dispersion
components.\label{fig:diag1}}
\end{figure}

\begin{figure}
\centering
\subfloat{\includegraphics[scale=0.35]{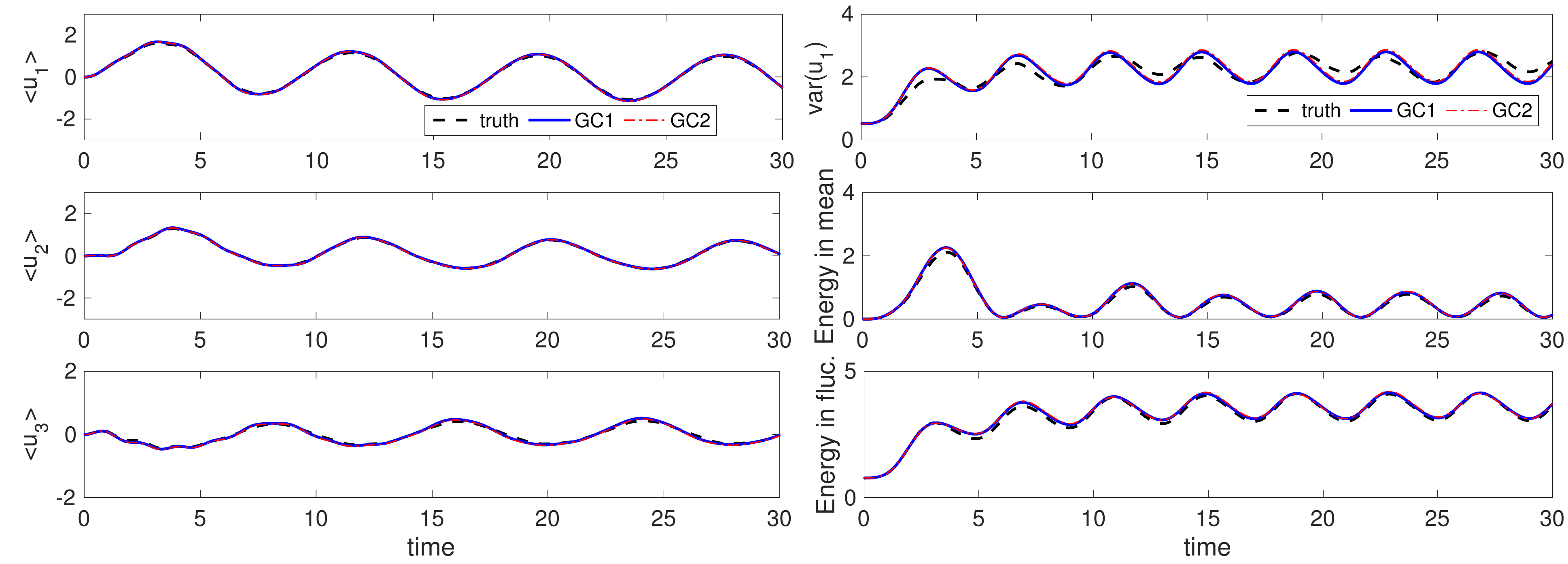}}

\subfloat{\includegraphics[scale=0.35]{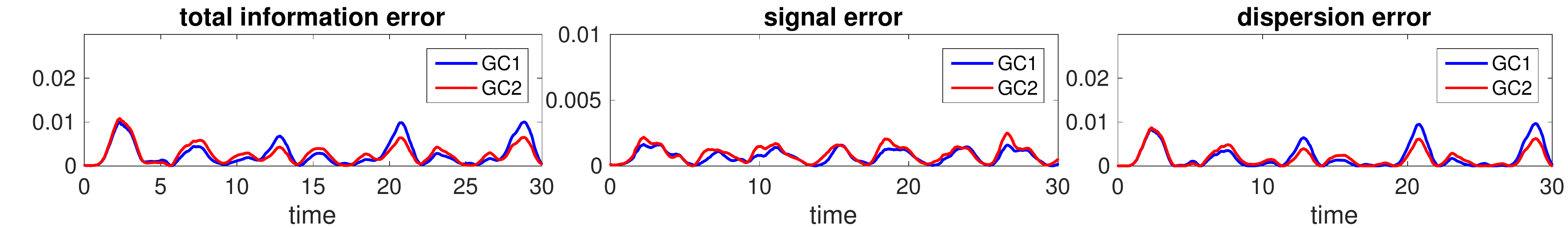}}

\caption{Reduced-order GC1 and GC2 model with the variance in first mode resolved
in approximating responses to periodic external forcing in equipartition
energy regime. \label{fig:red1}}
\end{figure}

In the final part of this section, we illustrate the imperfect model
prediction skills in the equipartition of energy regime with Gaussian
statistics. In this case, the three modes $\left(u_{1},u_{2},u_{3}\right)$
in the triad model possess same amount of energy in statistical equilibrium
state as in (\ref{eq:equi_measure}) and (\ref{eq:invar_m}). In Figure
\ref{fig:diag1} and \ref{fig:red1}, we show the prediction results
to the periodic perturbations from the \emph{diagonal model} with
off-diagonal cross-covariances neglected, and from the \emph{reduced
model} with only the variance in the first mode $u_{1}$ resolved.
Both GC1 and GC2 can capture the mean states in all three modes quite
accurately, and GC1 and GC2 results have little difference. This is
due to the relatively simple energy mechanism in this regime with
same amount of energy in each mode. On the other hand, in the prediction
of variances in the diagonal model larger errors appear especially
in the modes $u_{2}$ and $u_{3}$. This shows the important effects
of the off-diagonal covariances in predicting second order moments
in this equipartition energy regime. The reduced model gets good predictions
for the variance in the first resolved mode $u_{1}$. With the more
expensive full model, the prediction for the variances in all three
modes will become accurate with larger computational cost.

\section{Reduced-Order Statistical Models Applied to Two-Layer Baroclinic Turbulence}

In this section, we validate the performance of the reduced-order
models by the more complicated two-layer quasi-geostrophic system
with baroclinic instability. It is shown that the baroclinic model
is capable in capturing the essential physics of the relevant internal
variability despite its relatively simple dynamical structure. Two
dynamical regimes with typical statistical features are representative
in many applications \cite{srinivasan2012zonostrophic,bakas2011structural,qi2016low}.
The first one is the fully turbulent flow with homogeneous statistics
as a result of internal baroclinic instability corresponding to the
high-latitude ocean and atmosphere; the second one is the anisotropic
flow field with strong meandering zonal jets as in the low/mid-latitude
regime. Detailed discussions and comparisons of the construction and
test about the reduced-order models in various regimes with representative
zonal jets and vortices can be found in \cite{qi2016low}.

The governing two-layer quasi-geostrophic (QG) equations in a barotropic-baroclinic
mode formulation for potential vorticity anomalies $\left(q_{\psi},q_{\tau}\right)$
with periodic boundary condition in both $x,y$ directions are \cite{salmon1998lectures,vallis2006atmospheric}

\begin{equation}
\begin{aligned}\frac{\partial q_{\psi}}{\partial t}+J\left(\psi,q_{\psi}\right)+J\left(\tau,q_{\tau}\right)+\beta\frac{\partial\psi}{\partial x}+U\frac{\partial}{\partial x}\Delta\tau & =-\frac{\kappa}{2}\Delta\left(\psi-\tau\right)-\nu\Delta^{s}q_{\psi}+\mathcal{F}_{\psi}\left(\mathbf{x},t\right),\\
\frac{\partial q_{\tau}}{\partial t}+J\left(\psi,q_{\tau}\right)+J\left(\tau,q_{\psi}\right)+\beta\frac{\partial\tau}{\partial x}+U\frac{\partial}{\partial x}\left(\Delta\psi+k_{d}^{2}\psi\right) & =\frac{\kappa}{2}\Delta\left(\psi-\tau\right)-\nu\Delta^{s}q_{\tau}+\mathcal{F}_{\tau}\left(\mathbf{x},t\right).
\end{aligned}
\label{eq:dyn_twolayer}
\end{equation}
Above $q_{\psi}=\Delta\psi$, $q_{\tau}=\Delta\tau-k_{d}^{2}\tau$
are the \emph{disturbance} potential vorticities in barotropic and
baroclinic mode respectively, while $\psi,\tau$ are the corresponding
\emph{disturbance} barotropic and baroclinic stream functions. The
barotropic mode $\psi$ can be viewed as the vertically averaged effect
from the flow, and the baroclinic mode $\tau$ is usually related
with the thermal effect in heat transport. Besides, $J\left(A,B\right)=A_{x}B_{y}-A_{y}B_{x}$
represents the Jacobian operator. $k_{d}=\sqrt{8}/L_{d}=\left(2f_{0}/NH\right)^{2}$
is the baroclinic deformation wavenumber corresponding to the Rossby
radius of deformation $L_{d}$. A large-scale vertical shear $\left(U,-U\right)$
with the same strength and opposite directions is assumed in the background
to induce baroclinic instability. In the dissipation operators on
the right hand sides of the equations (\ref{eq:dyn_twolayer}), besides
the hyperviscosity, $\nu\Delta^{s}q_{i}$, we only use Ekman friction,
$\kappa\Delta\psi_{2}$, with strength $\kappa$ on the lower layer
of the flow.

\subsection{Representative dynamical regimes for the two-layer baroclinic turbulence}

The two-layer quasi-geostrophic system can display various dynamical
regimes with distinct statistical features as the parameters are changed.
Parameters for high and low/mid latitude dynamical regimes are shown
in Table \ref{tab:Model-parameters}. In numerical simulations, the
true statistics are calculated by a pseudo-spectra code by resolving
the two-layer equations (\ref{eq:dyn_twolayer}) with 128 spectral
modes zonally and meridionally, corresponding to $256\times256\times2$
grid points in total. In the reduced-order methods, only the large-scale
modes $\left|\mathbf{k}\right|\leq10$ are resolved, which is about
0.15\% of the full model resolution.

\begin{figure}
\centering
\includegraphics[scale=0.7]{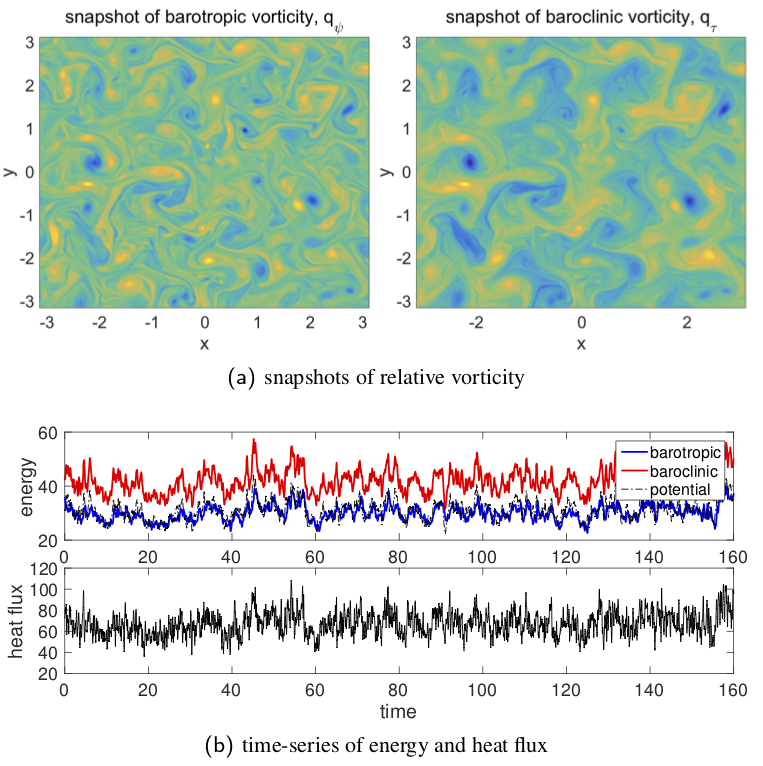}\caption{Snapshots of high-latitude ocean regime barotropic and baroclinic
vorticity in unperturbed system with no external forcing terms. Time-series
of energy in barotropic and baroclinic modes, as well as potential
energy, are also compared with the heat flux. \label{fig:snap_ocean}}
\end{figure}

\begin{figure}
\centering
\includegraphics[scale=0.7]{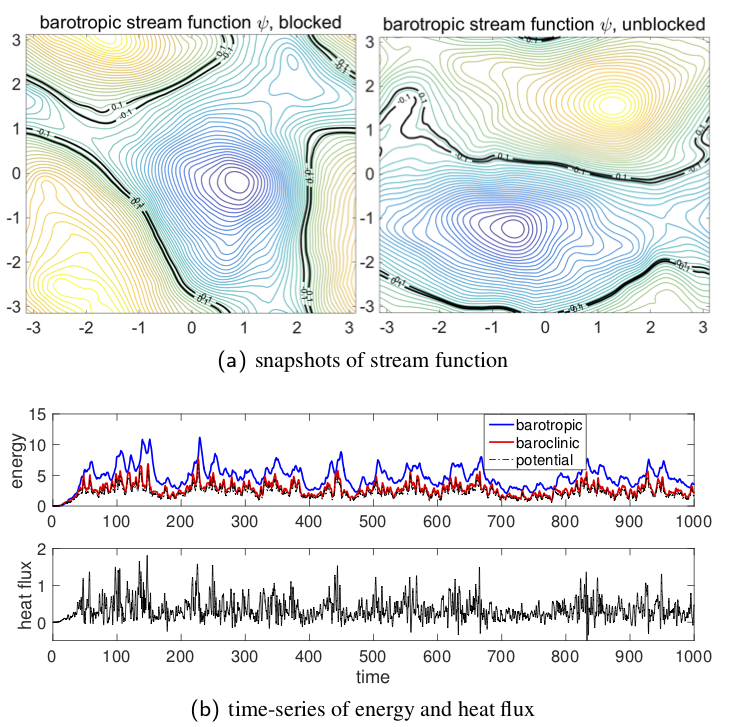}\caption{Snapshots of high-latitude atmosphere regime barotropic and baroclinic
stream function in unperturbed system with no external forcing terms.
Time-series of energy in barotropic and baroclinic modes, as well
as potential energy, are also compared with the heat flux. \label{fig:snap_atmos}}
\end{figure}

In the simulations for the unperturbed system in ocean and atmosphere
regimes, Figure \ref{fig:snap_ocean} displays the two-layer flow
structure in \emph{high-latitude ocean regime.} The first row is the
snapshots of the barotropic and baroclinic vorticity. Homogeneous
structure can be observed in both cases while larger scale structures
appear in the baroclinic mode. It is important to notice the strong
correlation in the coherent structures in the barotropic and baroclinic
field, illustrating the strong energy transfer between the two modes.
The following part shows time-series of energy in barotropic and baroclinic
mode, $-\fint\psi q_{\psi},-\fint\tau q_{\tau}$, as well as the potential
energy, $\fint k_{d}^{2}\tau^{2}$, compared with the meridional heat
flux, $k_{d}^{2}U\fint\psi_{x}\tau$. In Figure \ref{fig:snap_atmos}
the results for the two-layer flow in \emph{high-latitude atmosphere
regime} are compared. One important feature here is the flow field
alternating between blocked and unblocked regimes. In the stream functions,
it can be observed that in the blocked regime, zonal flow is blocked
and the field is restricted at separated regimes, while in the unblocked
regime strong zonal flow can be observed. Strong meridional heat flux
can be observed in the blocked regime while the flow is in state with
lower energy and low heat transfer in the zonal unblocked regime.

\begin{figure}
\centering
\includegraphics[scale=0.15]{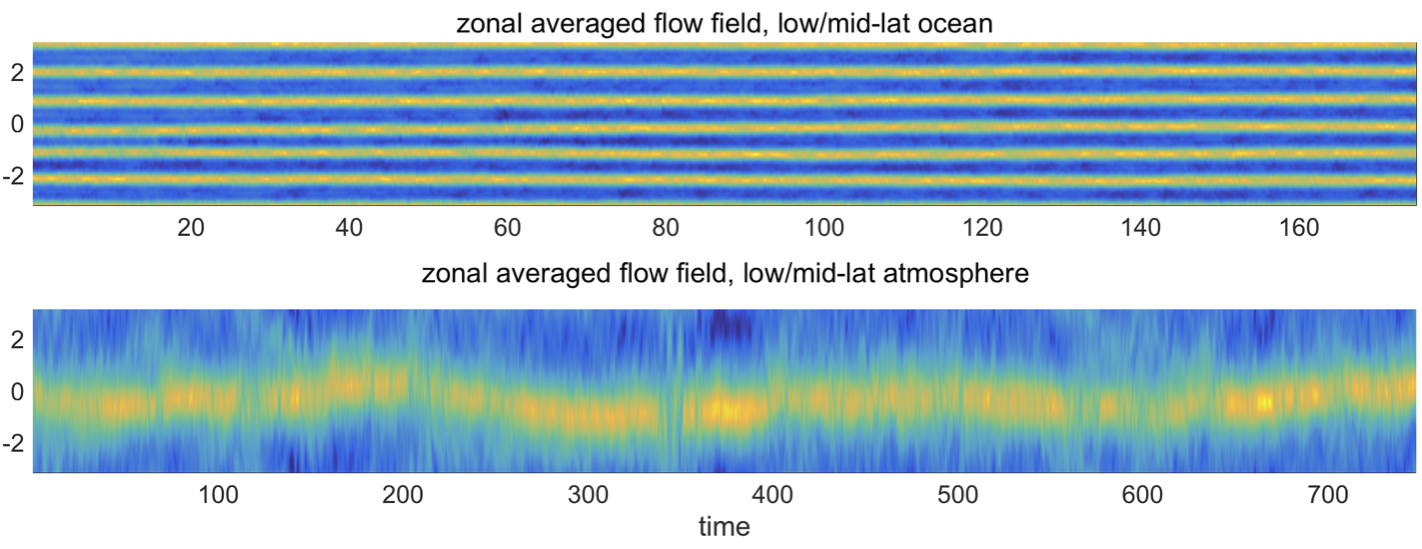}\caption{Time-series of zonal averaged mean flow with jets in low/mid-latitude
atmosphere and ocean regimes.\label{fig:U_low}}
\end{figure}

In mid/low latitude regimes, both the ocean and atmosphere are distinctly
inhomogeneous on large scales. The existence of large-amplitude meandering
zonal jets in these regimes suggests the regional metastable equilibria,
while the large-scale forced perturbations may lead to regular or
irregular fluctuations in some extent. The jet structures are illustrated
in more detail in Figure \ref{fig:U_low} for the time-series of the
zonally average mean flow, $u=-\partial_{y}\psi$.  In this low/mid
latitude case, especially for the ocean regime, due to the strong
zonal jets in wavenumber $k_{y}=6$, zonal modes with $k_{x}=5,6$
become active due to the nonlinear interactions.

\begin{table}
\caption{Model parameters for ocean and atmosphere dynamical regimes in high
and low/mid latitude. $N$ is the model resolution, $\beta,k_{d}$
are the rotation parameter and the deformation frequency, $U$ is
the background mean shear flow, $\kappa$ is the Ekman drag in the
bottom layer. The last three columns display the unstable waveband
from linear analysis. $\left(k_{\min},k_{\max}\right)$ shows the
range of unstable wavenumbers; $\sigma_{\mathrm{max}}$ is the largest
linear growth rate; and $\left(k_{x},k_{y}\right)_{\mathrm{max}}$
is the position of the mode with maximum growth rate.\label{tab:Model-parameters}}

\centering
\subfloat[high-latitude regime]{
\begin{tabular}{cccccccccc}
\toprule 
regime & $N$ & $\beta$ & $k_{d}$ & $U$ & $\kappa$ &  & $\left(k_{\min},k_{\max}\right)$ & $\sigma_{\mathrm{max}}$ & $\left(k_{x},k_{y}\right)_{\mathrm{max}}$\tabularnewline
\midrule
\midrule 
ocean regime, high lat. & 256 & 10 & 10 & 1 & 9 &  & $\left(2.25,14.61\right)$ & 0.411 & (4, 0)\tabularnewline
\midrule 
atmosphere regime, high lat. & 256 & 1 & 4 & 0.2 & 0.2 &  & $\left(1.58,6.78\right)$ & 0.099 & (2, 0)\tabularnewline
\bottomrule
\end{tabular}}

\subfloat[low/mid-latitude regime]{
\begin{tabular}{cccccccccc}
\toprule 
regime & $N$ & $\beta$ & $k_{d}$ & $U$ & $\kappa$ &  & $\left(k_{\min},k_{\max}\right)$ & $\sigma_{\mathrm{max}}$ & $\left(k_{x},k_{y}\right)_{\mathrm{max}}$\tabularnewline
\midrule
\midrule 
ocean regime, low/mid lat. & 256 & 100 & 10 & 1 & 1 &  & $\left(7.14,15.63\right)$ & 0.104 & (2, 8)\tabularnewline
\midrule 
atmosphere regime, low/mid lat. & 256 & 2.5 & 4 & 0.2 & 0.05 &  & $\left(2.51,7.06\right)$ & 0.053 & (3, 0)\tabularnewline
\bottomrule
\end{tabular}}
\end{table}

\begin{figure}
\centering
\subfloat[atmosphere regime]{\includegraphics[scale=0.3]{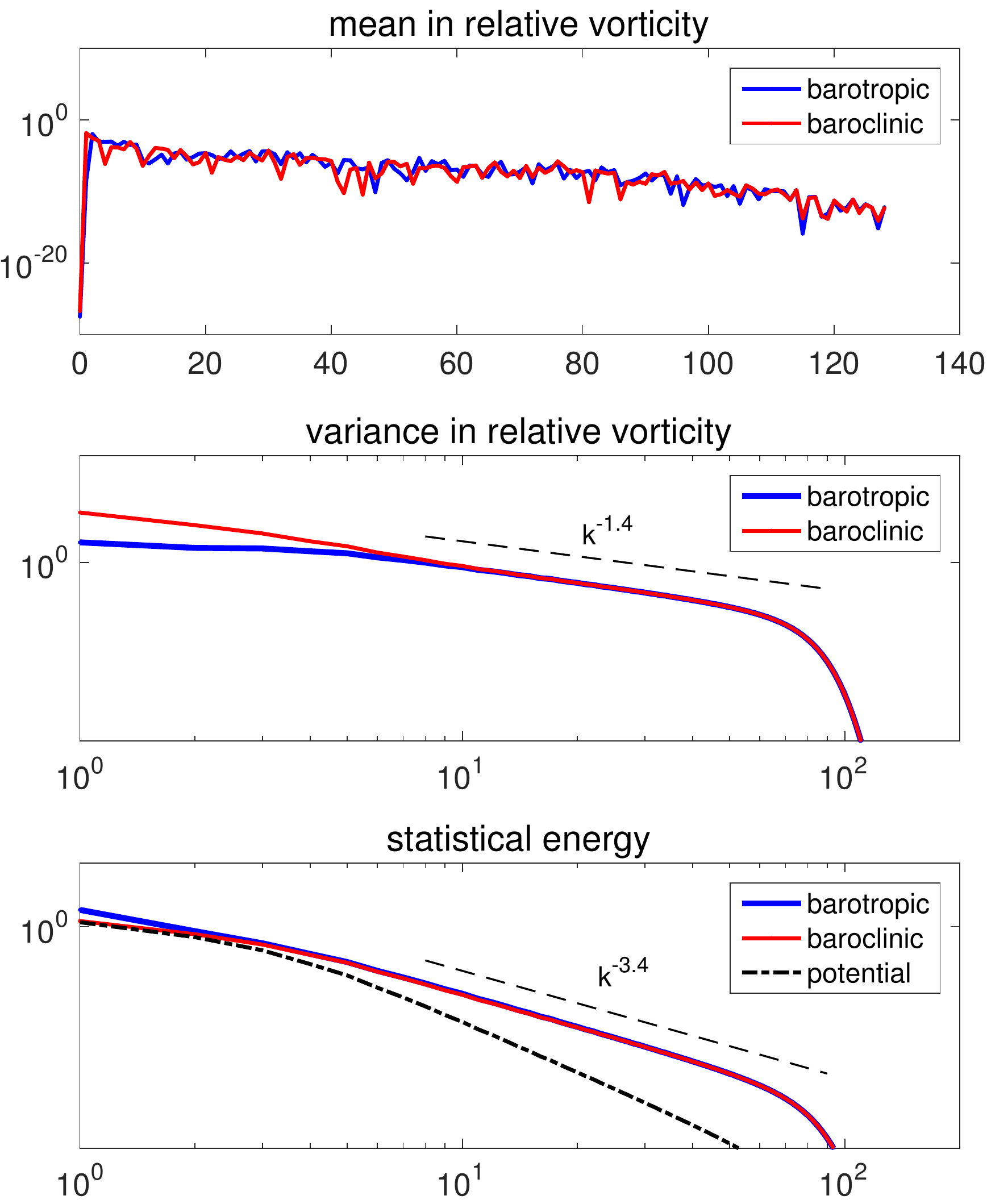}

}\subfloat[ocean regime]{\includegraphics[scale=0.3]{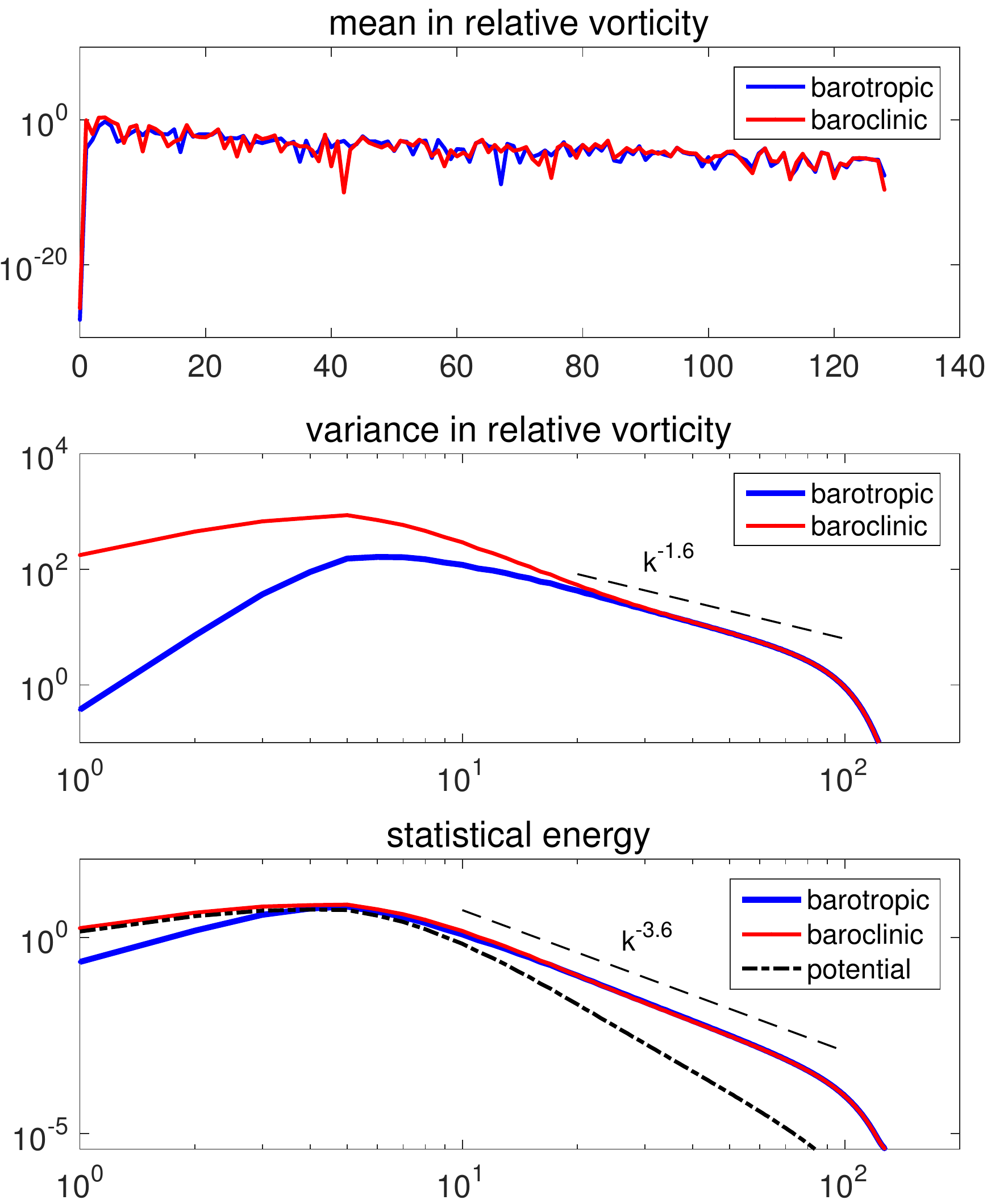}

}

\caption{Radial-averaged spectra in mean and second-order moments in both atmosphere
and ocean high-latitude regimes. The first row compares the statistical
mean states (in logarithmic coordinate). The following two rows show
the variances, and statistical energy, in barotropic and baroclinic
modes, as well as the potential energy.\label{fig:spectra-highlat}}
\end{figure}

The general steady state statistical structures in the spectral field
are shown in Figure \ref{fig:spectra-highlat}. As implied from the
homogeneous statistics in high-latitude, the mean states stay in small
values within fluctuation errors in both ocean and atmosphere regimes.
From the energy spectra, one observation is that the potential energy
is dominant in large scales in the baroclinic modes, and the kinetic
baroclinic energy becomes more important in small scales. For both
regimes, we observe wide and energetic spectra that exchange energy
between different scales, which indicates strong forward and backward
energy cascades along the entire spectral modes.

\subsection{Predicting quasi-geostrophic statistical responses in reduced-order models}

The quasi-geostrophic response to both stochastic and deterministic
perturbations is an important subject in understanding the earth\textquoteright s
atmospheric and oceanic interactions \cite{abramov2012low,lutsko2015applying}.
The same strategy developed in Section 3.3 and applied in the triad
system for the first mode in Section 4.3 can be directly generalized
to the statistical modeling of the two-layer QG system here.

\subsubsection{Statistical formulation about the two-layer baroclinic equations}

We formulate the two-layer QG system with Galerkin truncation to finite
number of spectral modes. In model simulations, consider a set of
rescaled normalized quantities with a high wavenumber truncation $N$
under standard Fourier basis $\mathbf{e_{k}}=\exp\left(i\mathbf{k\cdot x}\right)$
due to the periodic boundary condition, so that
\begin{equation}
\begin{aligned}p_{\psi,\mathbf{k}}= & q_{\psi,\mathbf{k}}/\left|\mathbf{k}\right|=-\left|\mathbf{k}\right|\psi_{\mathbf{k}},\\
p_{\tau,\mathbf{k}}= & q_{\tau,\mathbf{k}}/\sqrt{\left|\mathbf{k}\right|^{2}+k_{d}^{2}}=-\sqrt{\left|\mathbf{k}\right|^{2}+k_{d}^{2}}\tau_{\mathbf{k}}.
\end{aligned}
\label{eq:p_scale}
\end{equation}
The introduction of this new set of quantities (\ref{eq:p_scale})
offers convenience that the energy inner-product becomes the standard
Euclidean form. Under the above settings, the rescaled set of equations
of (\ref{eq:dyn_twolayer}) can be summarized in the abstract form
in the truncated subspace $\left|\mathbf{k}\right|\leq N$ as in (\ref{eq:abs_formu})
\begin{equation}
\frac{d\mathbf{p_{k}}}{dt}=B_{\mathbf{k}}\left(\mathbf{p_{k}},\mathbf{p_{k}}\right)+\left(\mathcal{L}_{\mathbf{k}}-\mathcal{D}_{\mathbf{k}}\right)\mathbf{p_{k}}+\mathcal{F}_{\mathbf{k}},\quad\mathbf{p_{k}}=\left(p_{\psi,\mathbf{k}},p_{\tau,\mathbf{k}}\right)^{T},\label{eq:dyn_p_scaled}
\end{equation}
where the linear operators are decomposed into the non-symmetric part
$\mathcal{L}_{\mathbf{k}}$ involving $\beta$-effect and vertical
shear flow $U$ and dissipation part $\mathcal{D}_{\mathbf{k}}$,
together with the forcing $\mathcal{F}_{\mathbf{k}}$ combining deterministic
component and stochastic component compared with (\ref{eq:dyn_twolayer}).
Most importantly, $B\left(\mathbf{p},\mathbf{p}\right)$ is the nonlinear
interactions that conserve both energy and enstrophy.

The same reduced-order modeling strategy then can be applied to the
two-layer model following the algorithm in Section 3.4. Therefore
the true dynamical equations for the statistical moment $R_{\mathbf{k}}=\left\langle \mathbf{p_{k}}^{*}\mathbf{p_{k}}\right\rangle $
in the form of a $2\times2$ matrix containing barotropic and baroclinic
mode in same wavenumber $\mathbf{k}$ become
\begin{equation}
\frac{dR_{\mathbf{k}}}{dt}=\left(\mathcal{L}_{\mathbf{k}}-\mathcal{D}_{\mathbf{k}}\right)R_{\mathbf{k}}+Q_{F,\mathbf{k}}+Q_{\sigma,\mathbf{k}}+c.c.,\quad\left|\mathbf{k}\right|\leq N,\label{eq:full_stat_model}
\end{equation}
where $c.c.$ represent the complex completion for the conjugate parts.
On the right hand side of the equation, $\mathcal{L}_{\mathbf{k}},\mathcal{D}_{\mathbf{k}}$
represent the linear interactions between modes, including $\beta$-effect
through the rotation of the earth, the effects from the mean shear
flow $U$, as well as the dissipations from Ekman drag and hyperviscosity.
$Q_{\boldsymbol{\sigma},\mathbf{k}}$ is the external forcing perturbations
represented by hypothetical stirring and heating forces. Importantly,
the nonlinear flux $Q_{F,\mathbf{k}}$ represents the nonlinear interactions
between different wavenumbers due to the advection term. Third-order
moments with triad modes $\mathbf{m+n=k}$ enter the first two order
moments dynamics representing the nonlinear energy transfer between
small and large scales. The nonlinear energy exchange mechanism is
crucial in the energy budget, and the conservation property is satisfied
due to the triad symmetry as $\sum_{\mathbf{k}}\mathrm{tr}Q_{F,\mathbf{k}}=0$.

\subsubsection{Reduced-order model predictions for responses in various dynamical regimes}

In constructing the reduced-order models, the same strategy is applied
to the crucial but expensive nonlinear flux term $Q_{F}$ as in (\ref{eq:model_nonflux})
\[
Q_{M,\mathbf{k}}=Q_{M,\mathbf{k}}^{-}+Q_{M,\mathbf{k}}^{+}=f_{1}\left(E\right)\left[-\left(N_{M,\mathbf{k},\mathrm{eq}}+d_{M}\right)R_{M,\mathbf{k}}\right]+f_{2}\left(E\right)\left[Q_{F,\mathbf{k},\mathrm{eq}}^{+}+\sigma_{M,\mathbf{k}}^{2}\right].
\]
Both equilibrium higher-order statistics and additional corrections
are combined, and statistical energy equation is important to provide
the scaling factor for optimal consistency and sensitivity. See \cite{qi2016low}
for more details.

\begin{figure}
\centering
\subfloat[$U=1.05$]{\includegraphics[scale=0.24]{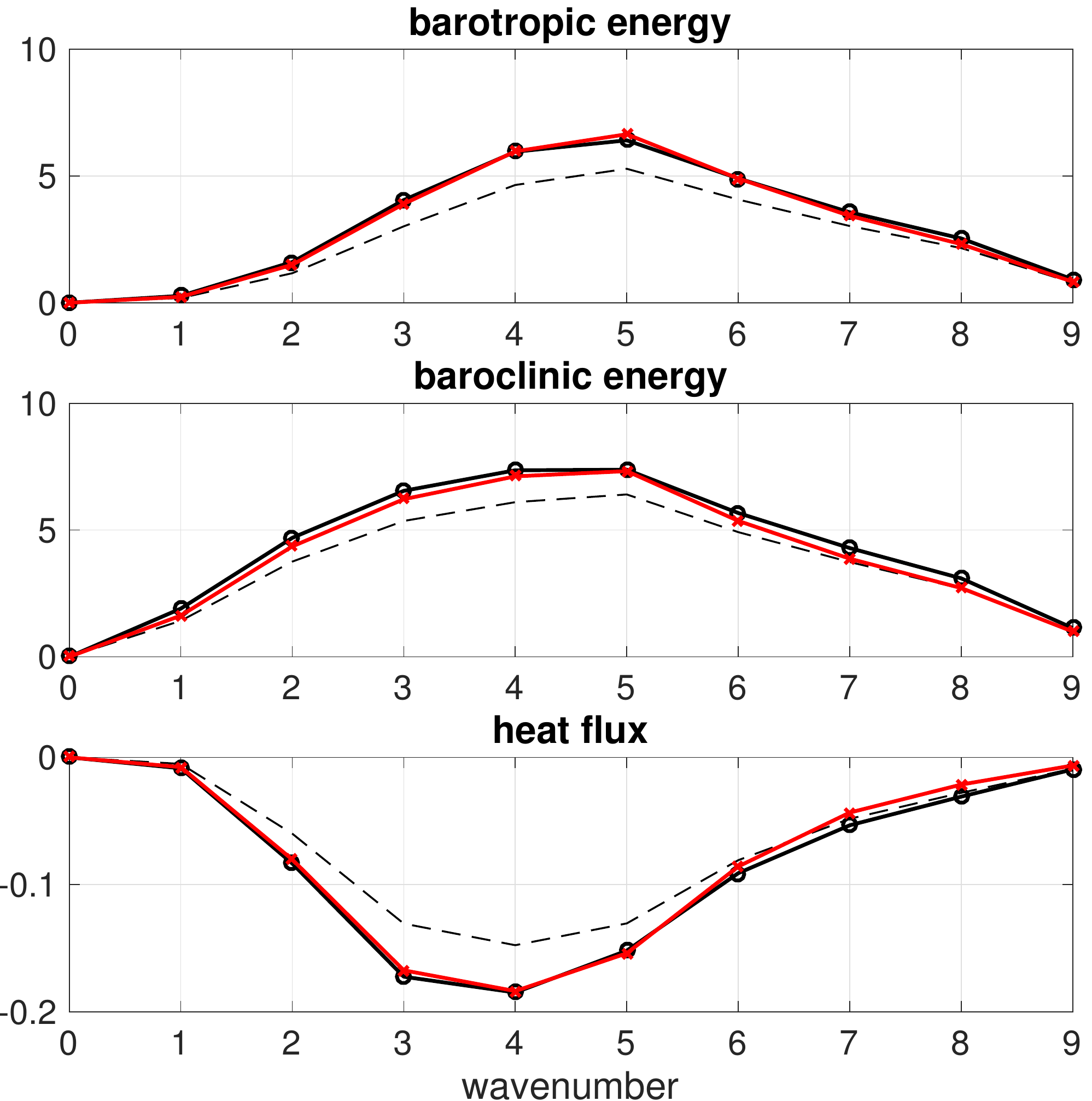}

}\subfloat[$U=0.95$]{\includegraphics[scale=0.24]{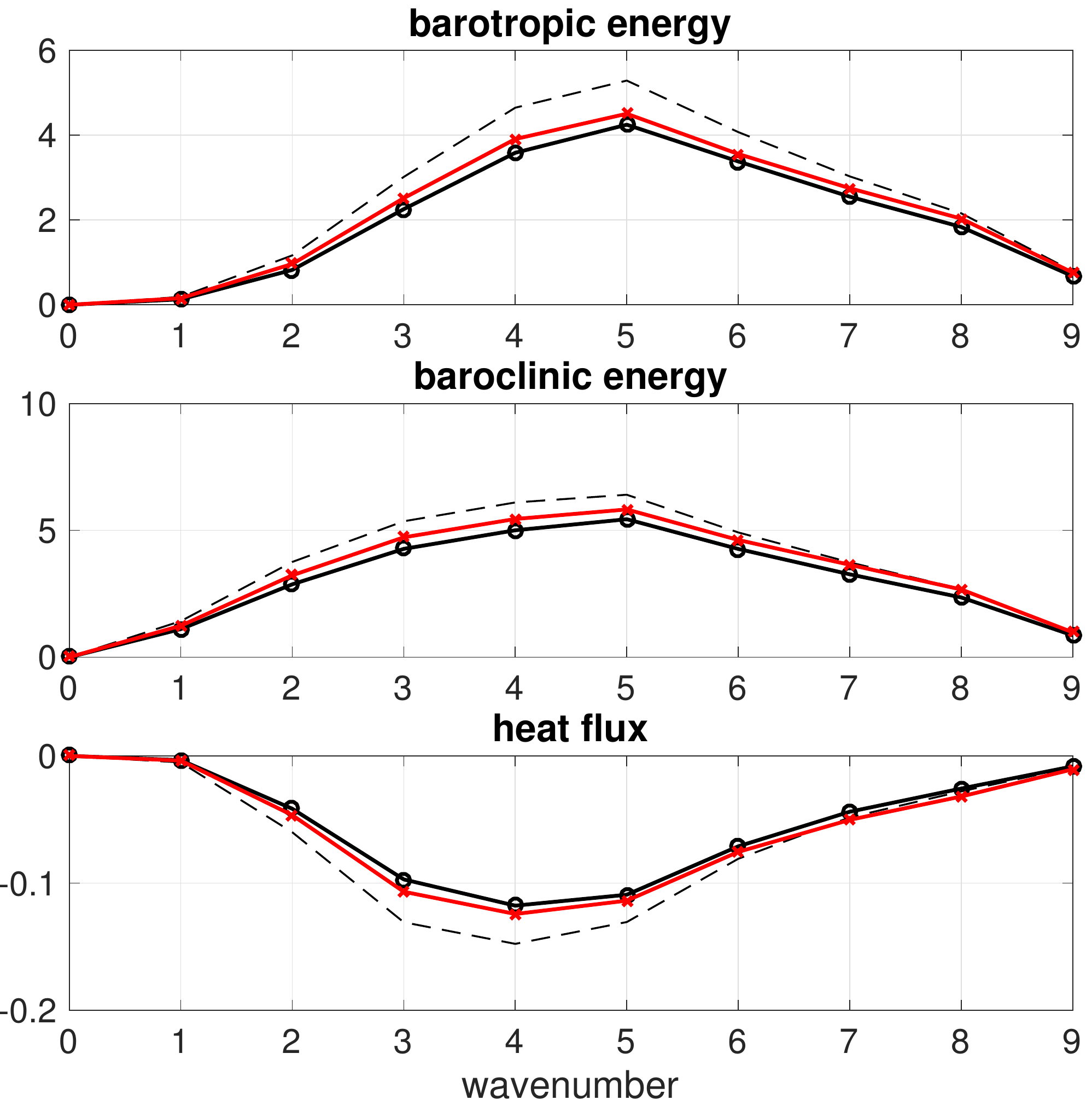}

}

\caption{Reduced-order model predictions to mean shear flow perturbation $\delta U=\pm0.05$
in high-latitude ocean regime. The reduced-order model predictions
for the spectra are compared with the truth. Black lines with circles
show the perturbed model responses in the barotropic energy, baroclinic
energy, and heat flux. The dashed black lines are the unperturbed
statistics, and the reduced order model predictions are in red lines.\label{fig:Reduced-order-model-predictions_pertU_oce}}
\end{figure}

\begin{figure}
\centering
\subfloat[$U=0.22$]{\includegraphics[scale=0.24]{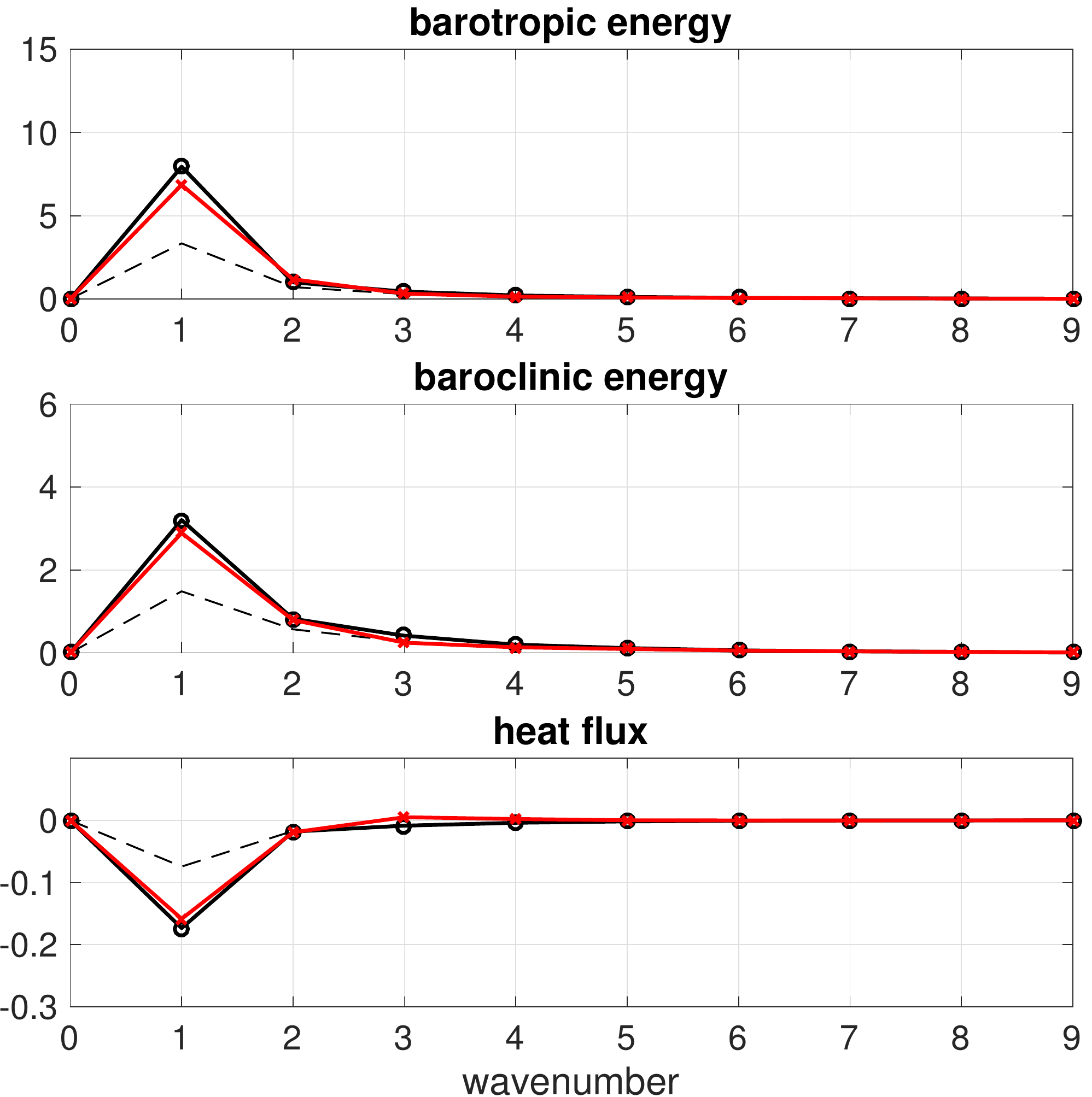}

}\subfloat[$U=0.19$]{\includegraphics[scale=0.24]{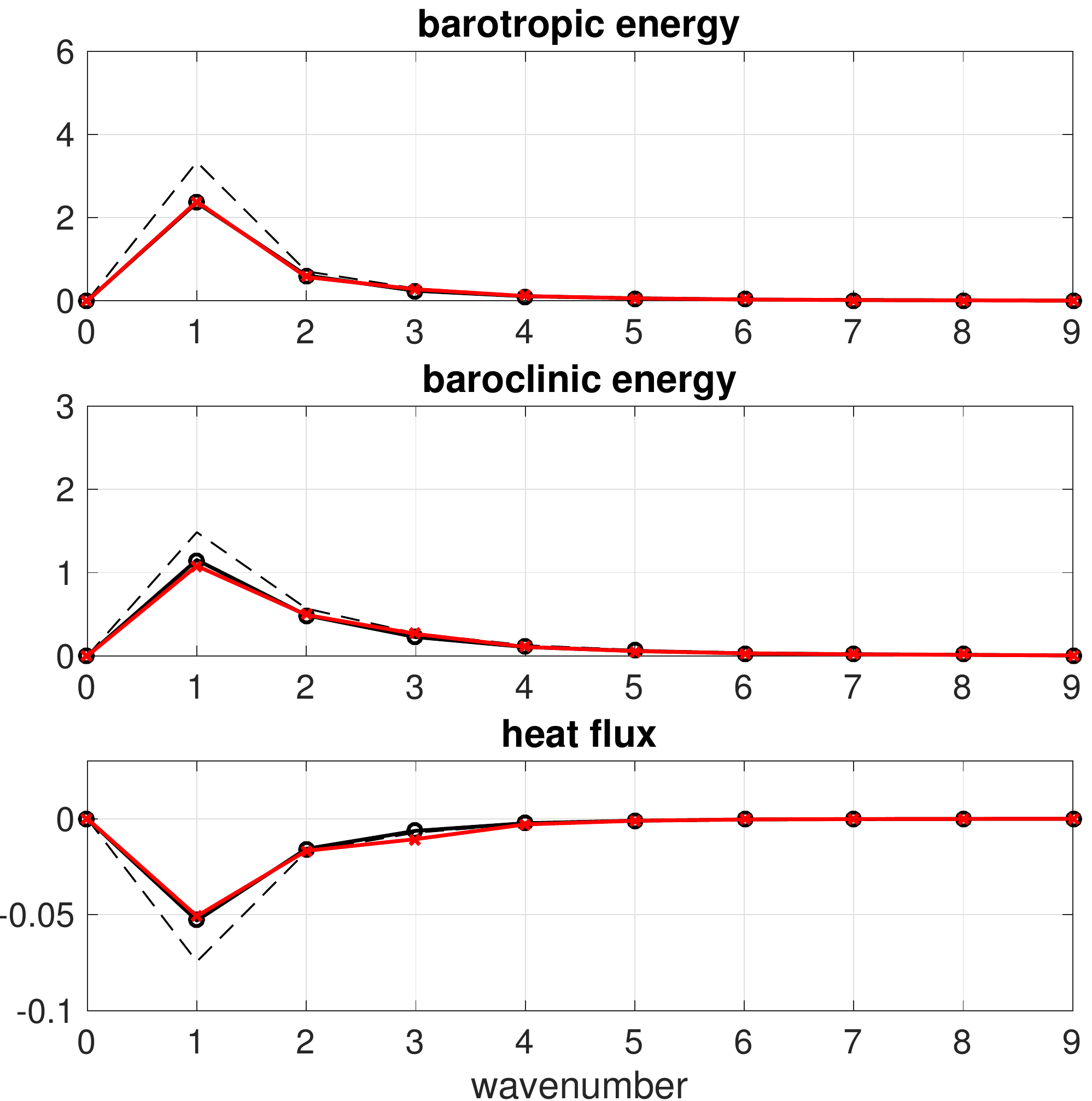}

}

\caption{Reduced-order model predictions to mean shear flow perturbation in
high-latitude atmosphere regime. The reduced-order model predictions
for the spectra are compared with the truth. Black lines with circles
show the perturbed model responses in barotropic energy, baroclinic
energy, and heat flux. The dashed black lines are the unperturbed
statistics. And the reduced order model predictions are in red lines.\label{fig:Reduced-order-model-predictions_pertU_atm}}
\end{figure}

In checking the model sensitivity in the homogeneous high-latitude
regimes, we introduce the forcing perturbation by changing the background
jet strength $U$. Note that the deterministic perturbation about
zonal mean flow advection forms a difficult test case because the
forcing is applied along all wavenumbers with stronger mean-fluctuation
interactions involved. On the other hand, for the reduced order methods,
only the perturbations at the limited resolved modes are quantified.
This gives the inherent difficulty for applying the reduced order
models to this kind of perturbations since we have no knowledge of
the unresolved modes where large amount of energy is contained. The
results with mean flow perturbations $\delta U=\pm0.05$ in the ocean
regime and perturbations $\delta U=0.02,-0.01$ in the atmosphere
regime are shown in Figure \ref{fig:Reduced-order-model-predictions_pertU_oce}
and \ref{fig:Reduced-order-model-predictions_pertU_atm} separately.
The perturbation accounts for about 5\%-10\% of the original shear
strength $U$, and the corresponding responses in both energy and
heat flux spectra are large due to this global perturbation at every
wavenumber and nonlinear energy cascade. In the ocean regime, a wide
waveband of modes $\left|\mathbf{k}\right|=3,4,5,6$ becomes sensitive
to the perturbations; while in the atmosphere regime, the first dominant
mode $\left|\mathbf{k}\right|=1$ is especially sensitive to even
small perturbations. This illustrates the strong nonlinear interactions
between the high and low wavenumber modes. The reduced-order method
displays uniform skill in capturing the sensitive responses in the
large-scale modes for both positive and negative perturbation cases
with only first $10\times10$ spectral modes resolved compared with
the $256\times256$ full resolution model. 

\begin{figure}
\centering
\subfloat{\includegraphics[scale=0.3]{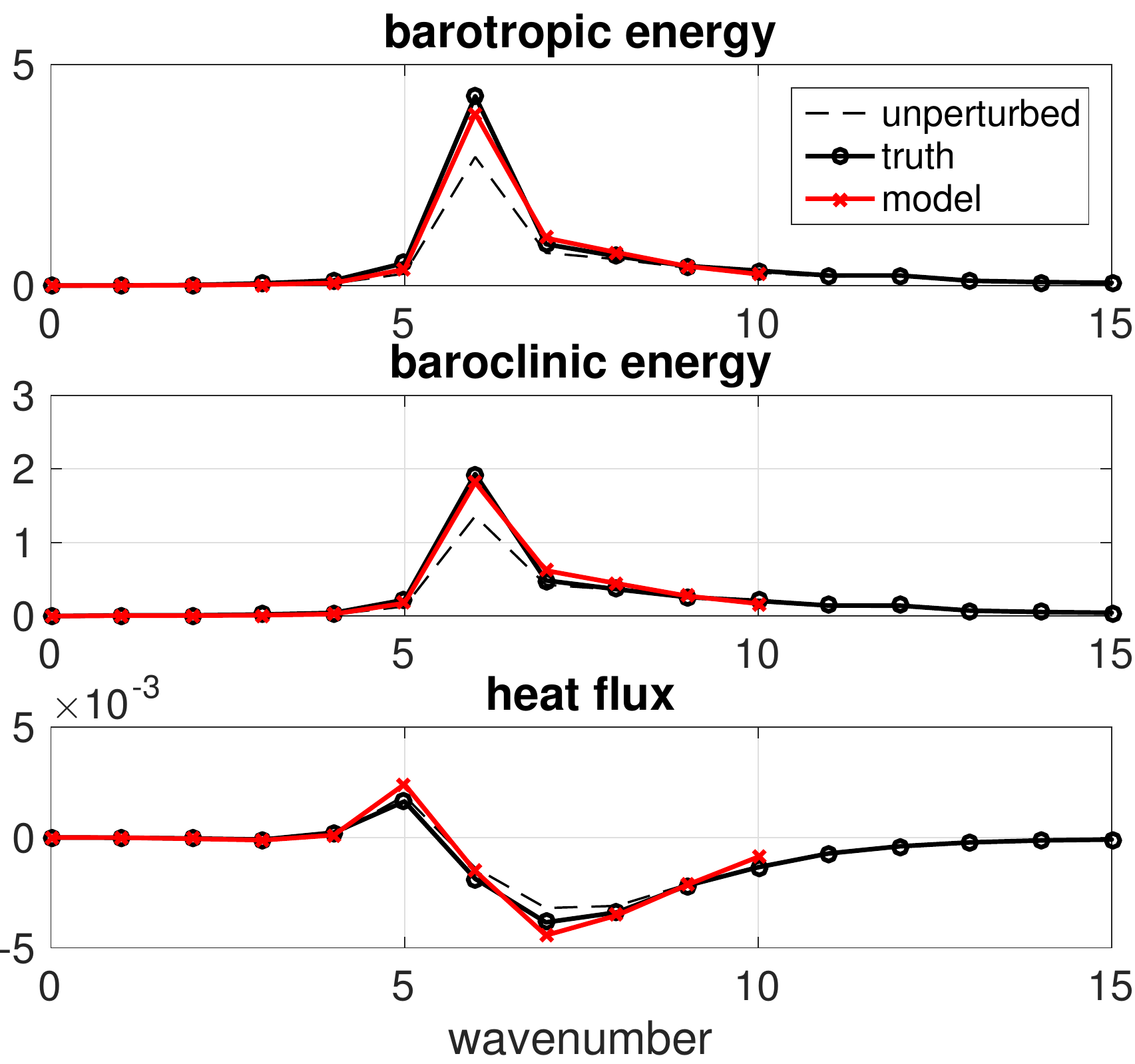}}\subfloat{\includegraphics[scale=0.3]{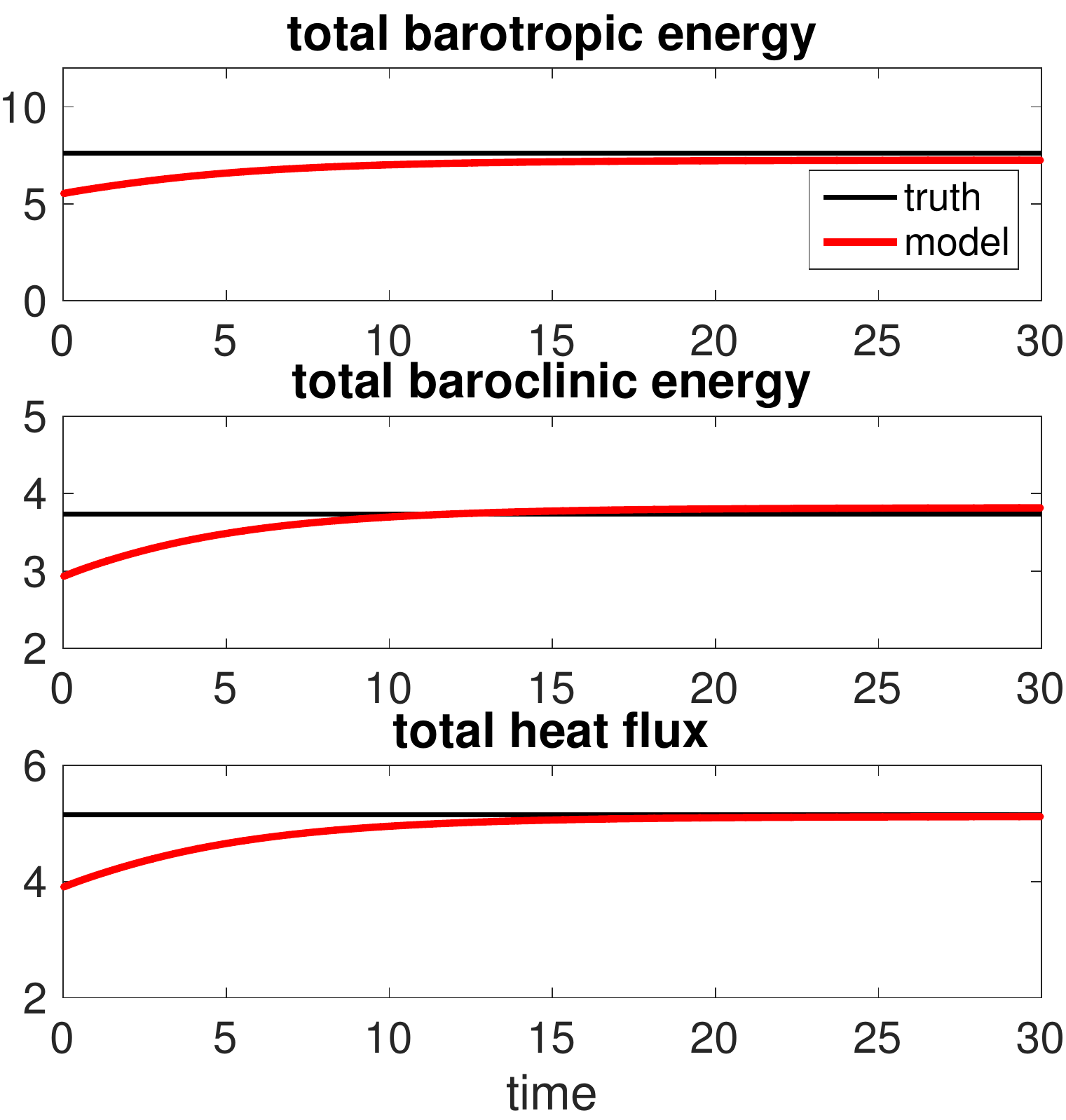}}

\caption{Model responses in low/mid-latitude ocean regime with random forcing
perturbation. The left panel shows the spectra for the barotropic
and baroclinic energy as well as the heat flux with first ten modes
resolved in the reduced-order method. The right panel is the time-series
of the total energy and heat flux. The truth is shown in black lines
while reduced-order model predictions are in red lines.\label{fig:Model-responses-ocean}}
\end{figure}

\begin{figure}
\centering
\subfloat{\includegraphics[scale=0.3]{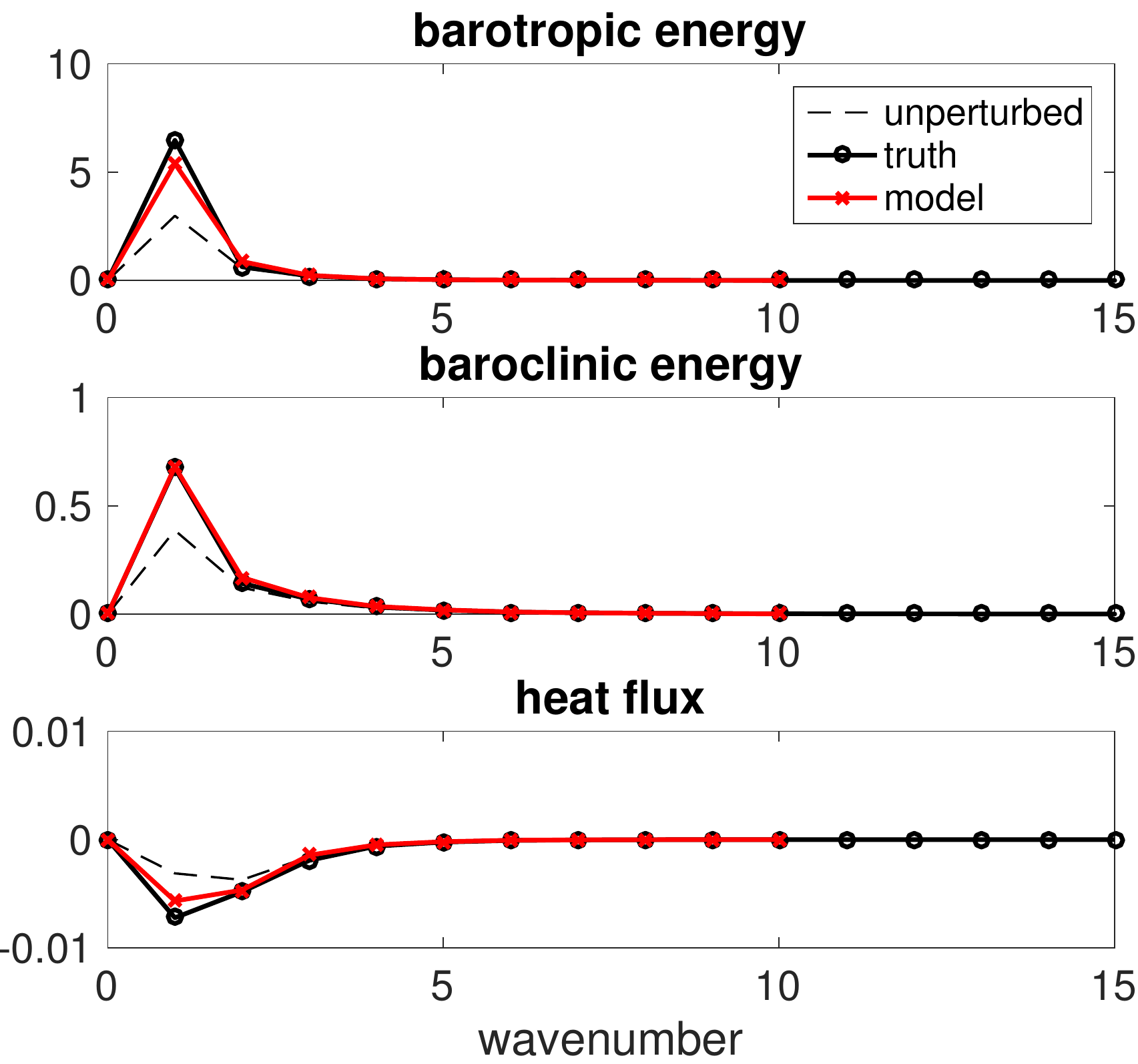}}\subfloat{\includegraphics[scale=0.3]{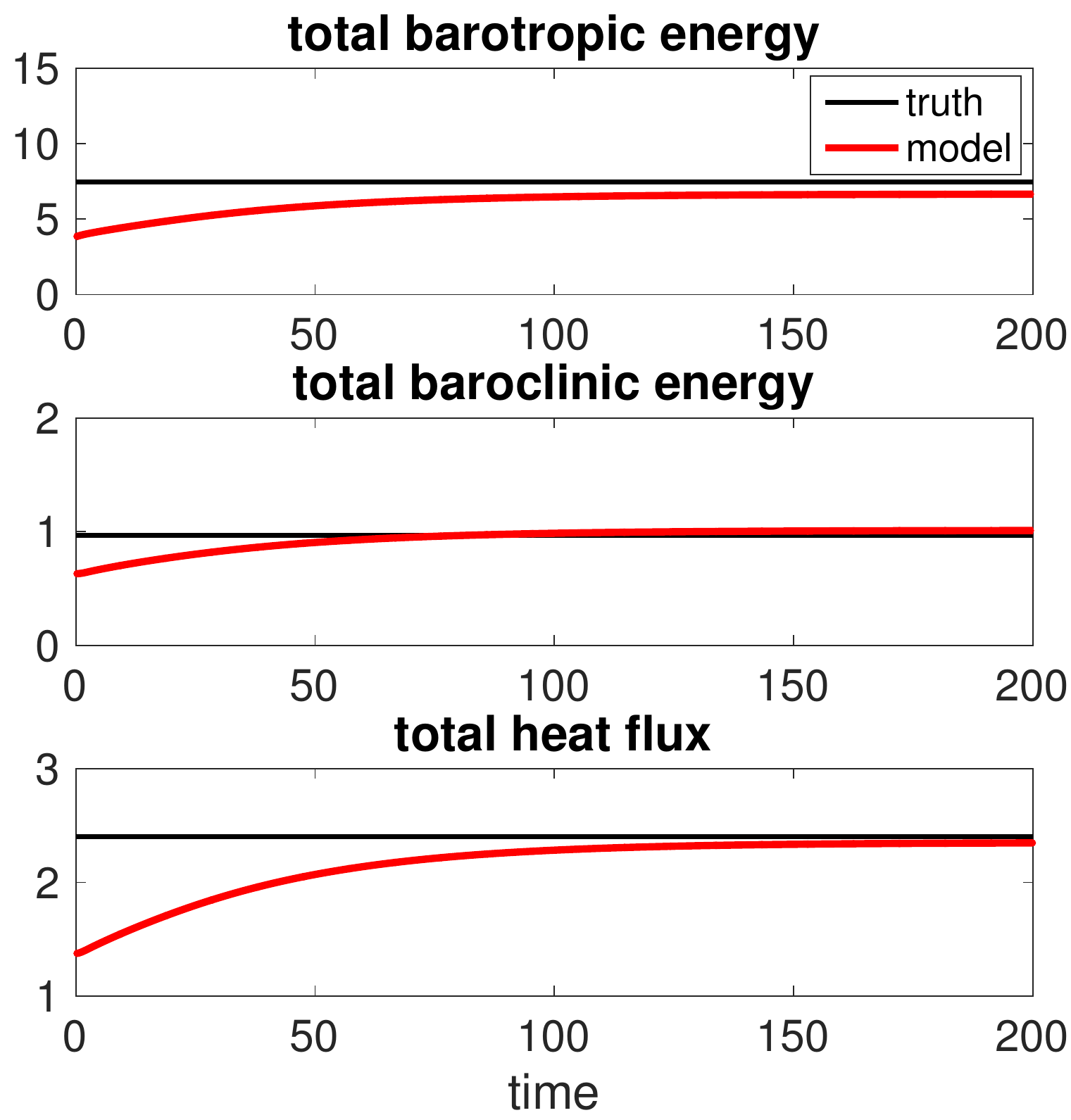}}

\caption{Model responses in low/mid-latitude atmosphere regime with random
forcing perturbation. The left panel shows the spectra for the barotropic
and baroclinic energy as well as the heat flux with first 10 modes
resolved in the reduced-order method. The right panel is the time-series
of the total energy and heat flux. The truth is shown in dashed black
lines while reduced-order model predictions are in red lines.\label{fig:Model-responses-atmos}}
\end{figure}

In Figure \ref{fig:Model-responses-ocean} and \ref{fig:Model-responses-atmos},
we compare the model responses in both low/mid-latitude ocean and
atmosphere regimes. In this inhomogeneous regime with anisotropic
jets, the statistical variables combine the responses in the mean
and variance, $\overline{p_{1,\mathbf{k}}^{*}p_{2,\mathbf{k}}^{\:}}=\bar{p}_{1,\mathbf{k}}^{*}\bar{p}_{2,\mathbf{k}}^{\:}+\overline{p_{1,\mathbf{k}}^{\prime*}p_{2,\mathbf{k}}^{\prime}}$,
to display the total effect from the perturbation. In the ocean regime,
the dominant mode with largest sensitivity is at wavenumber $\left|\mathbf{k}\right|=6$
due to the zonal jet structure. The sensitivity is captured with accuracy
in the reduced-order method. Also we compare the time evolvement of
the total resolved energy and heat flux. The prediction is also good
with small error. In the atmosphere regime, $\left|\mathbf{k}\right|=1$
mode gets the largest statistical energy and is most sensitive to
perturbations. One important feature is the large change in the heat
flux in the first two modes, representing the exchange of energy in
the dominant barotropic and baroclinic mode. Still the responses can
be captured with accuracy in each mode in the spectra as well as the
total energy and heat flux profile with only $10^{2}$ modes resolved.
Note that in both cases, the heat flux is weak due to the blocking
effect from strong zonal jets.

\section{Summary and Some Future Research Directions}

Understanding and improving the predictive skill of imperfect models for high-dimensional complex turbulent systems is a formidable and challenging problem and has been investigated through multiple approaches with various mathematical theories through the years \cite{salmon1998lectures,giannakis2012quantifying2,lesieur2012turbulence,majda2016improving,sapsis2013statistically,lutsko2015applying}.   Low-order truncation methods for statistical prediction can overcome the curse of dimensionality \cite{epstein1969stochastic,majda2012physics} by concentrating on the subspace containing largest variability. On the other hand, anisotropic turbulent processes are representative in many engineering and environmental fluid flows \cite{hinze1959turbulence,frisch1995turbulence} where energy transports intermittently from the smaller scales to impact the largest scales in these flows. Therefore significant model errors always occur due to the high wavenumber truncation in the imperfect model approximations. A systematic information-theoretic framework has been shown useful to improve model fidelity and sensitivity \cite{majda2011improving,majda2011link,branicki2015information} for complex systems including perturbation formulas and multimodel ensembles that can be utilized to improve model error. In many applications to complex systems with model error such as the climate change science \cite{emanuel2005improving,neelin2006tropical}, it is crucially important to provide guidelines to improve the predictive skill of imperfect models for their responses to changes in various external forcing perturbations.

We discuss the general framework of efficient low-dimensional
reduced-order models in this paper for turbulent dynamical systems with nonlinearity
to capture statistical responses to external perturbations. The validity
of the reduced-order modeling procedure is displayed via the simplest
3-dimensional triad model which is the building block of general turbulent
systems, and further on the more complicated two-layer barotropic
model with huge model reduction. The computational cost is reduced
through a systematic approximation about the expensive nonlinear higher-order
interactions. Additional damping and noise corrections are proposed
to replace the third-order moments. Model consistency in unperturbed
equilibrium and sensitivity to external perturbations are maintained
through careful calibration about imperfect model error in a training
phase before the prediction. The model errors are calibrated and reduced
effectively through a combination of linear response theory involving
only unperturbed equilibrium statistics and an information-theoretic
framework using information theory. The general framework has been
tested in detail for a series of dynamical systems with increasing
complexity \cite{majda2016improving,qi2016,qi2016low,majda2016introduction}.

For future development about the methods, there exist several interesting
and promising directions that are worth further investigation in the
next stage:
\begin{description}
\item [{A)}] Tracking the model fluctuation statistics about the perfect
statistical mean state. In many problems for turbulent systems, we
can assume the statistical mean state is known with reasonable accuracy
by averaging along the data trajectory, while the statistical fluctuations
about the mean state are the quantities of interest \cite{majda2006nonlinear,majda2003systematic}. It is useful
to consider accurate and efficient ways to quantify the model fluctuations
involving both uncertainties in the perfect system and errors due
to imperfect model approximation;
\item [{B)}] Design of a mitigation control strategy by using novel low-order
statistical models. There is need to combine control theory with the
statistical model reduction strategies for the principal large-scale
modes in the turbulent dynamical systems. For example, it is interesting
to consider the effects of climate change using control and statistical
modeling strategies;
\item [{C)}] Predicting passive scalar turbulence with complex flow field.
Besides the turbulent flow field, the dynamics of the passive tracer
advected by the turbulent flow has many interesting features with
practical implications and is worth investigating. One important feature
in the turbulent tracer field is the appearance of intermittency despite
the near-Gaussian statistics in the background advection flow. The
intermittency in time-series and fat-tails in the passive tracer distributions
have been observed in nature \cite{neelin2010long}, and have been
investigated under a simpler modeling framework both theoretically
and numerically \cite{majda2013elementary,majda2015intermittency,qi2015predicting}.
It is thus interesting to consider low-order stochastic and statistical
modeling about the  tracer field advected by complex turbulent flows;
\item [{D)}] More detailed consideration about the model nonlinearity.
In the reduced-order approximation here, the overall strategy does
not require explicit calculation of the inefficient quadratic forms
directly, but instead mimic the statistical symmetry in the nonlinearity
in simple and efficient forms. Still it is interesting to check the
improvement in the low-order models with more detailed approximation.
\end{description}

\section*{Acknowledgment }

This research of the Andrew Majda is partially supported by the Office
of Naval Research through MURI N00014-16-1-2161 and DARPA through
W911NF-15-1-0636. Di Qi is supported as a graduate research assistant
on these grant.

\appendix 
\numberwithin{equation}{section}

\section{Numerical strategies to calculate the kicked response operators}

In the calibration step of the reduced-order models, we use the statistical
kicked response theory to tune the imperfect model parameters in the
training phase. Here we describe the details about calculating the
kicked response operators for the mean and variance numerically. From
the formula in (\ref{eq:kicked_resp}), the response operators for
the mean and variance can be achieved from the perturbation part of
the probability density $\delta p^{\prime}$. And this density function
is also used to measure the information distance between the truth
and imperfect model result in the training phase. Below we describe
the numerical procedure to get this distribution function $\delta p^{\prime}$
for the true system and the imperfect closure model separately.
\begin{itemize}
\item \emph{Kicked response for the true model}: For the true system, we
want to achieve the most accurate possible estimation for the response
operators both for comparison with the imperfect model results and
for calculating the FDT linear prediction in (\ref{eq:LRT}). Therefore
we use a Monte-Carlo simulation with an large enough ensemble size
to capture the response in density. The initial equilibrium ensemble
is picked by sampling from a normal distribution with consistent equilibrium
mean and variance of the true system. For the kicked response to the
mean, a constant perturbation with ten percent of the equilibrium
state mean $\delta\mathbf{u}=0.1\bar{\mathbf{u}}_{\mathrm{eq}}$ is
added to each initial ensemble member (in fact, as observed in numerical
experiments, this perturbation amplitude has little effect on the
results in the response distribution as long as it's not too large);
and the initial variance of the ensemble is kept unchanged. The response
distribution $\delta p^{\prime}$ then is achieved by monitoring the
decay of the ensemble particles back to equilibrium under unperturbed
dynamics and uniformly perturbed initial value (and the length of
the time window that we need to monitor depends on the mixing property
of the turbulent system). See \cite{bell1980climate,majda2005information}
for similar version of this algorithm.
\item \emph{Kicked response for the imperfect model}: For the imperfect
model, we just need to run the closure equations to get the responses
for the mean and variance. In the same way as the true model, the
initial mean is taken from the equilibrium distribution and a perturbation
with amplitude $\delta\mathbf{u}=0.1\bar{\mathbf{u}}_{\mathrm{eq}}$
is added to the initial mean state. The initial value for the variance
is taken the same as the equilibrium state value and kept unperturbed.
Then using this initial mean and variance, the imperfect model with
specific closure strategies is applied to monitor the decay of the
mean and variance back to equilibrium.
\end{itemize}
One additional important point that requires attention is that even
if the unperturbed equilibrium initial conditions are applied, the
system will still deviate from the equilibrium state first and reapproach
equilibrium again after some relaxation time. This is due to the insufficient
characterization of the entire distribution of the true system with
a Gaussian approximation (note that nonlinearities are also included
in the imperfect closure methods). To eliminate this effect in computing
the kicked response in both the true and imperfect models, we subtract
the statistics computed using the unperturbed initial value from the
statistics computed using the perturbed Gaussian initial condition
to achieve more accurate characterization of the responses.

\section{Explicit statistical dynamical formulations for the triad system}

We can derive for the triad system (\ref{triad}) the \textbf{dynamical
equations for the mean state}\addtocounter{equation}{0}\begin{subequations}\label{triad_mean}
\begin{eqnarray}
\frac{d\bar{u}_{1}}{dt} & = & L_{2}\bar{u}_{3}-L_{3}\bar{u}_{2}-d_{1}\bar{u}_{1}+B_{1}\left(\bar{u}_{2}\bar{u}_{3}+\overline{u_{2}^{\prime}u_{3}^{\prime}}\right)+F_{1},\label{eq:mean1}\\
\frac{d\bar{u}_{2}}{dt} & = & L_{3}\bar{u}_{1}-L_{1}\bar{u}_{3}-d_{2}\bar{u}_{2}+B_{2}\left(\bar{u}_{3}\bar{u}_{1}+\overline{u_{3}^{\prime}u_{1}^{\prime}}\right)+F_{2},\label{eq:mean2}\\
\frac{d\bar{u}_{3}}{dt} & = & L_{1}\bar{u}_{2}-L_{2}\bar{u}_{1}-d_{3}\bar{u}_{3}+B_{3}\left(\bar{u}_{1}\bar{u}_{2}+\overline{u_{1}^{\prime}u_{2}^{\prime}}\right)+F_{3}.\label{eq:mean3}
\end{eqnarray}
\end{subequations}On the right hand sides of the above equations
(\ref{eq:mean1})-(\ref{eq:mean3}), the first parts include the skew-symmetric
interactions between modes as well as the linear damping for the mean.
The nonlinear interaction parts enter the mean dynamics both from
the interactions between the mean states, and more importantly from
the second-order moments of the fluctuations. Thus the mean dynamical
equations are not closed by themselves due to the inclusion of unresolved
higher-order statistics. Also note that in the second-order moments
in the mean equations, diagonal variances won't appear while the cross-diagonal
covariances take place as the role of transferring energy in the mean.
Next we consider the dynamics for the fluctuation parts of the state
variables\addtocounter{equation}{0}\begin{subequations}
\begin{eqnarray}
\frac{du_{1}^{\prime}}{dt} & = & L_{2}u_{3}^{\prime}-L_{3}u_{2}^{\prime}-d_{1}u_{1}^{\prime}+B_{1}\left(\bar{u}_{2}u_{3}^{\prime}+u_{2}^{\prime}\bar{u}_{3}+u_{2}^{\prime}u_{3}^{\prime}-\overline{u_{2}^{\prime}u_{3}^{\prime}}\right)+\sigma_{1}\dot{W}_{1},\label{eq:fluc1}\\
\frac{du_{2}^{\prime}}{dt} & = & L_{3}u_{1}^{\prime}-L_{1}u_{3}^{\prime}-d_{2}u_{2}^{\prime}+B_{2}\left(\bar{u}_{3}u_{1}^{\prime}+u_{3}^{\prime}\bar{u}_{1}+u_{3}^{\prime}u_{1}^{\prime}-\overline{u_{3}^{\prime}u_{1}^{\prime}}\right)+\sigma_{2}\dot{W}_{2},\label{eq:fluc2}\\
\frac{du_{3}^{\prime}}{dt} & = & L_{1}u_{2}^{\prime}-L_{2}u_{1}^{\prime}-d_{3}u_{3}^{\prime}+B_{3}\left(\bar{u}_{1}u_{2}^{\prime}+u_{1}^{\prime}\bar{u}_{2}+u_{1}^{\prime}u_{2}^{\prime}-\overline{u_{1}^{\prime}u_{2}^{\prime}}\right)+\sigma_{3}\dot{W_{3}}.\label{eq:fluc3}
\end{eqnarray}
\end{subequations}The above equations can be achieved by subtracting
the mean equations from the original triad system. Then the dynamics
for higher order moments can be achieved through the fluctuations
equations. Importantly, we can get the \textbf{dynamical equations
for the variances} in each mode\addtocounter{equation}{0}\begin{subequations}\label{triad_var}
\begin{eqnarray}
\frac{1}{2}\frac{d\overline{u_{1}^{\prime2}}}{dt} & = & L_{2}\overline{u_{1}^{\prime}u_{3}^{\prime}}-L_{3}\overline{u_{1}^{\prime}u_{2}^{\prime}}-d_{1}\overline{u_{1}^{\prime2}}+B_{1}\left(\bar{u}_{2}\overline{u_{1}^{\prime}u_{3}^{\prime}}+\overline{u_{1}^{\prime}u_{2}^{\prime}}\bar{u}_{3}\right)+B_{1}\overline{u_{1}^{\prime}u_{2}^{\prime}u_{3}^{\prime}}+\frac{1}{2}\sigma_{1}^{2},\label{eq:var1}\\
\frac{1}{2}\frac{d\overline{u_{2}^{\prime2}}}{dt} & = & L_{3}\overline{u_{1}^{\prime}u_{2}^{\prime}}-L_{1}\overline{u_{2}^{\prime}u_{3}^{\prime}}-d_{2}\overline{u_{2}^{\prime2}}+B_{2}\left(\bar{u}_{1}\overline{u_{2}^{\prime}u_{3}^{\prime}}+\overline{u_{1}^{\prime}u_{2}^{\prime}}\bar{u}_{3}\right)+B_{2}\overline{u_{1}^{\prime}u_{2}^{\prime}u_{3}^{\prime}}+\frac{1}{2}\sigma_{2}^{2},\label{eq:var2}\\
\frac{1}{2}\frac{d\overline{u_{3}^{\prime2}}}{dt} & = & L_{1}\overline{u_{2}^{\prime}u_{3}^{\prime}}-L_{2}\overline{u_{1}^{\prime}u_{3}^{\prime}}-d_{3}\overline{u_{3}^{\prime2}}+B_{3}\left(\bar{u}_{1}\overline{u_{2}^{\prime}u_{3}^{\prime}}+\overline{u_{1}^{\prime}u_{3}^{\prime}}\bar{u}_{2}\right)+B_{3}\overline{u_{1}^{\prime}u_{2}^{\prime}u_{3}^{\prime}}+\frac{1}{2}\sigma_{3}^{2}.\label{eq:var3}
\end{eqnarray}
\end{subequations}And the \textbf{dynamical equations for the cross-covariances}
between modes become\addtocounter{equation}{0}\begin{subequations}\label{triad_cov}
\begin{eqnarray}
\frac{d\overline{u_{1}^{\prime}u_{2}^{\prime}}}{dt} & = & L_{2}\overline{u_{2}^{\prime}u_{3}^{\prime}}-L_{3}\overline{u_{2}^{\prime2}}-d_{1}\overline{u_{1}^{\prime}u_{2}^{\prime}}+B_{1}\left(\bar{u}_{2}\overline{u_{2}^{\prime}u_{3}^{\prime}}+\overline{u_{2}^{\prime2}}\bar{u}_{3}\right)+B_{1}\overline{u_{2}^{\prime}u_{2}^{\prime}u_{3}^{\prime}}\nonumber \\
 & + & L_{3}\overline{u_{1}^{\prime2}}-L_{1}\overline{u_{1}^{\prime}u_{3}^{\prime}}-d_{2}\overline{u_{1}^{\prime}u_{2}^{\prime}}+B_{2}\left(\bar{u}_{1}\overline{u_{1}^{\prime}u_{3}^{\prime}}+\overline{u_{1}^{\prime2}}\bar{u}_{3}\right)+B_{2}\overline{u_{1}^{\prime}u_{1}^{\prime}u_{3}^{\prime}},\label{eq:cov1}\\
\frac{d\overline{u_{1}^{\prime}u_{3}^{\prime}}}{dt} & = & L_{2}\overline{u_{3}^{\prime2}}-L_{3}\overline{u_{2}^{\prime}u_{3}^{\prime}}-d_{1}\overline{u_{1}^{\prime}u_{3}^{\prime}}+B_{1}\left(\bar{u}_{2}\overline{u_{3}^{\prime2}}+\overline{u_{2}^{\prime}u_{3}^{\prime}}\bar{u}_{3}\right)+B_{1}\overline{u_{2}^{\prime}u_{3}^{\prime}u_{3}^{\prime}}\nonumber \\
 & + & L_{1}\overline{u_{1}^{\prime}u_{2}^{\prime}}-L_{2}\overline{u_{1}^{\prime2}}-d_{3}\overline{u_{1}^{\prime}u_{3}^{\prime}}+B_{3}\left(\bar{u}_{1}\overline{u_{1}^{\prime}u_{2}^{\prime}}+\overline{u_{1}^{\prime2}}\bar{u}_{2}\right)+B_{3}\overline{u_{1}^{\prime}u_{1}^{\prime}u_{2}^{\prime}},\label{eq:cov2}\\
\frac{d\overline{u_{2}^{\prime}u_{3}^{\prime}}}{dt} & = & L_{3}\overline{u_{1}^{\prime}u_{3}^{\prime}}-L_{1}\overline{u_{3}^{\prime2}}-d_{2}\overline{u_{2}^{\prime}u_{3}^{\prime}}+B_{2}\left(\bar{u}_{1}\overline{u_{3}^{\prime2}}+\overline{u_{1}^{\prime}u_{3}^{\prime}}\bar{u}_{3}\right)+B_{2}\overline{u_{1}^{\prime}u_{3}^{\prime}u_{3}^{\prime}}\nonumber \\
 & + & L_{1}\overline{u_{2}^{\prime2}}-L_{2}\overline{u_{1}^{\prime}u_{2}^{\prime}}-d_{3}\overline{u_{2}^{\prime}u_{3}^{\prime}}+B_{3}\left(\bar{u}_{1}\overline{u_{2}^{\prime2}}+\overline{u_{1}^{\prime}u_{2}^{\prime}}\bar{u}_{2}\right)+B_{3}\overline{u_{1}^{\prime}u_{2}^{\prime}u_{2}^{\prime}}.\label{eq:cov3}
\end{eqnarray}
\end{subequations}For most situations, it is the diagonal variances
in (\ref{eq:var1})-(\ref{eq:var3}) that we are more interested in,
while the off-diagonal covariances (\ref{eq:cov1})-(\ref{eq:cov3})
are less important and expensive to resolve. On the other hand, it
is noticed that only the cross-covariance terms take place in the
central variance dynamics in (\ref{eq:var1})-(\ref{eq:var3}) for
the linear and quasi-linear interaction. In this typical case, if
we only consider the diagonal model and ignore the off-diagonal terms
in the statistical closure dynamics, huge errors could be introduced
in the variance dynamical equations. Thus in the development of reduced
order statistical models, careful calibration about the unresolved
components becomes crucial in the accuracy of the prediction results.

With the dynamical equations for the mean (\ref{triad_mean}) and
for the variances in each mode (\ref{triad_var}) and (\ref{triad_cov}),
we can derive the statistical energy dynamics following the general
framework proposed in (\ref{eq:energy_conservation}) of Theorem \ref{thm:thm1} and also \cite{majda2015statistical}. The statistical energy can
be defined as the combination of the mean energy and the fluctuation
energy as
\[
E=\frac{1}{2}\sum_{i=1}^{3}\left(\bar{u}_{i}^{2}+\overline{u_{i}^{\prime2}}\right).
\]
In the dynamics for the mean and covariance, the major difficulty
in resolving the equations explicitly comes from the complex third-order
moments in $Q_{F}$ as well as the covariance interactions through
$R_{ij}B\left(\mathbf{e}_{i},\mathbf{e}_{j}\right)$. However due
to the conservation of energy and the detailed triad symmetry in the
triad system, nonlinear interactions cancel in the mean and variance
equations and the \textbf{statistical energy equation} becomes
\begin{equation}
\frac{dE}{dt}=\frac{d}{dt}\left(\frac{1}{2}\sum_{i=1}^{3}\left(\bar{u}_{i}^{2}+\overline{u_{i}^{\prime2}}\right)\right)=-\sum_{i=1}^{3}d_{i}\left(\bar{u}_{i}^{2}+\overline{u_{i}^{\prime2}}\right)+\sum_{i=1}^{3}\left(F_{i}\bar{u}_{i}+\frac{1}{2}\sigma_{i}^{2}\right).\label{eq:stat_ene}
\end{equation}
Therefore the total statistical structure can be calculated from (\ref{eq:stat_ene})
without knowing the higher-order moments as well as cross-covariances
which are in general difficult to resolve exactly without error. Furthermore,
considering the special case with homogeneous damping in each mode,
$d_{j}\equiv d$, the dissipation term on the right hand side of (\ref{eq:stat_ene})
becomes $-dE$. Thus we can get the total second-order statistics
from only the information in the first-order moments with the help
of the statistical energy dynamics.

\bibliographystyle{plain}
\bibliography{ref_theory}

\end{document}